\documentclass[a4paper,12pt]{JHEP3}

\usepackage{caption,graphicx,color}
\usepackage{amsmath,epsfig}
\usepackage{amssymb,amsfonts}
\usepackage{latexsym}
\usepackage[latin1]{inputenc}
\usepackage{float}
\usepackage{overpic}
\usepackage{slashed}
\usepackage{xcolor}
\let\normalcolor\relax
\usepackage{graphicx}
\usepackage{longtable}

\relax
\renewcommand{\theequation}{\arabic{section}.\arabic{equation}}

\def\be{\begin{equation}}
\def\ee{\end{equation}}

\newcommand{\de}{\partial}
\newcommand{\bear}{\begin{eqnarray}}
\newcommand{\bea}{\begin{eqnarray}}
\newcommand{\eear}{\end{eqnarray}}
\newcommand{\eea}{\end{eqnarray}}

\def\hri#1#2{\href{http://arxiv.org/abs/#1}{[ArXiv:#1]#2}}
\def\hre#1#2{\href{http://arxiv.org/abs/#1/#2}{[ArXiv:#1/#2]}}

\def\hrj#1#2{\href{www.doi.org/#1}{#2}}
\newbox\pippobox

\def\II{\relax{\rm I\kern-.18em I}}

\def\m{\mu}
\def\n{\nu}

\def\g{\gamma}

\def\pa{\partial}

\def\sp{\;\;\;,\;\;\;}

\def\p{\partial}

\def\f{\varphi}
\def\z{\zeta}

\def\ss{{\cal S}}

\title{Holographic theories at finite $\theta$-angle,  CP-violation, glueball spectra and strong-coupling instabilities }

\author{Yuta Hamada$^\natural$, Elias Kiritsis$^\natural$$^\flat$, Francesco Nitti$^\natural$
~\\
~\\
~\\
$^\natural$ \href{http://www.apc.univ-paris7.fr}{Universit\'e de Paris, CNRS, Astroparticule et Cosmologie,  F-75006 Paris, France}\\
~\\
$^\flat$ \href{http://hep.physics.uoc.gr}{Crete Center for Theoretical Physics}, Institute for Theoretical and Computational Physics,
Department of Physics, Voutes University Campus,\\
GR-70013, Heraklion, GREECE.
}

\abstract{A general class of holographic theories with a nontrivial $\theta$-angle are analyzed. The instanton density operator is dual to a bulk axion field. We calculate the ground-state solutions with nontrivial source, $a_{UV}$,  for the axion, for both steep and soft dilaton potentials in the IR, and both in $d=3$ and $d=4$. We find all cases to be qualitatively similar.  We also calculate the spin$=2,0$ glueball spectra and show that the glueball masses monotonically decrease as functions of $a_{UV}$ (or $\theta$-angle).
The slopes of glueball masses are different, generically, in different potentials.
In the case of steep dilaton potentials, the glueball (masses)$^2$  turn negative before the maximum of $a_{UV}$ is attained. We interpret this as a signal for a favored instanton condensation in the bulk.
We also investigate strong CP-violation in the effective glueball action.}

\preprint{CCTP-2020-9\\ITCP-IPP-2020/9}

\begin{document}

\maketitle 

\section{Introduction}

The YM topological density and topological densities more generally, are sources of CP-violation in the associated theories, as well as mediators of interesting topological dynamics. The role of instantons in QCD has been analyzed since a long time, and the large N analysis has indicated that their effects are not exponentially suppressed due to strong coupling effects,   \cite{Witten0,Witten1,Veneziano,Witten2}. This fact has implications for the U(1)$_A$ problem in QCD.
The presence of a new CP-odd coupling in YM theory highlights a possible source of CP-violation in the strong force. Experimental data, however, indicate that such an angle must be tiny, \cite{edm}. This is known as the strong-CP problem. This fact has motivated the introduction of the axion, \cite{PQ}, in order to render the solution of the strong-CP problem natural.

The dynamics of the $\theta$-angle and the associated operator of the instanton density, $Tr[F\wedge F]$,  is special as well as notoriously difficult to determine.
The $\theta$-term in the action is trivial in perturbation theory, as it is a total derivative. It is however non-trivial non-perturbatively. Its correlation functions are very special,  as they contain important contact terms, but also they lack the generic UV divergences that affect other operators in the gauge theory, \cite{panago,DelDebbio:2002xa,DelDebbio:2004ns,DelDebbio:2006yuf,Giusti:2007tu,Mazur}.  The correlators of the topological density control the diffusion of topological charge at finite temperature, via the Chern-Simons diffusion rate, \cite{BM,cs,BC3}. This is a transport coefficient obtained from  the infrared limit of the retarded two point-function of the instanton density. It is expected to be important for studying of charge separation in heavy-ion collisions, mediated by the chiral anomaly.

The special properties of instanton densities in QFT have also an impact in composite axion theories. As holography suggests, a bulk axion can be interpreted as the propagating effective field of states generated by the instanton density out of the QFT vacuum. Moreover, as argued in \cite{smgrav} such states have special properties, the most important being that their effective masses are insensitive to UV effects.
It is on the basis of this, that we expect that hidden holographic theories may generate, beyond emergent gravity, also emergent axions, with properties that are distinct from conventional composite axion models, \cite{axion}.
In particular, in this framework, the emergent axion masses may not be connected necessarily to the QCD scale. The brane-world picture of such emergent axions is reminiscent of earlier work in \cite{dudas}.
Composite axions  may also be instrumental in connecting the self-tuning mechanism of the cosmological constant, \cite{self}, to the gauge hierarchy problem, \cite{hknw}.

Many aspects of the dynamics of the instanton density have been formulated and calculated, in the context of the holographic description of large-N strongly-coupled gauge theories, \cite{malda}. In the top-down black D$_4$ theory, \cite{d4}, the finite $\theta$-angle solution and its properties has been discussed in \cite{witten,Dubovsky:2011tu,BC1}. The $\theta$-dynamics in the extension of this theory with flavor, the Witten-Sakai-Sugimoto model, \cite{ss}, has been studied in \cite{BC2,BC3}. The associated non-derivative couplings that are induced for the PQ axion, by the instanton density dynamics, have been computed recently in \cite{BCJK}.

The dynamics of instanton density has been studied also, \cite{data}, in bottom up holographic models of YM, like Improved Holographic QCD, \cite{iQCD}.
In this context,  the ground state was determined at finite $\theta$-angle and  the spectrum of $0^{\pm}$ glueballs calculated. The two-point function of the instanton density was also calculated, \cite{cs}, allowing the calculation of the Chern-Simons diffusion rate.
Upon the addition of back-reacting flavor degrees of freedom, one obtains the VQCD  model for QCD in the Veneziano limit, \cite{VQCD}. The solutions in this theory, in the presence of a non-trivial $\theta$-angle, were analyzed in \cite{tVQCD} and exhibit a highly non-trivial structure in the space of coupling constants and vevs, including a complex generalization of the Efimov spiral.

Holographic theories with a space dependent $\theta$-angle have been considered with the goal of generating Lifshitz critical points in the IR, \cite{tada}. Such theories, being anisotropic, they have been used as laboratories to study the (breakdown of) the universality of shear viscosity at strong coupling, \cite{s0}-\cite{s3}.

In all holographic contexts, the instanton density is dual to a (string theory) axion.  The prototypical example of this is the IIB ten-dimensional RR axion field, dual to the $N=4$ super Yang-Mills (sYM) instanton density operator. By the holographic dictionary,  the leading (i.e. non-normalizable) term in the near-boundary  expansion of the axion  field, is the field theory $\theta$-angle\footnote{Modulo an integer number of $2\pi$ shifts.}.

Seen as a coupling constant,  the $\theta$-angle is usually not considered to run under the renormalization group (RG), as the instanton density operator is not perturbatively renormalized. However, in holography, a non-trivial vacuum expectation value (vev) for the instanton density operator drives a flow corresponding to a non-constant bulk axion field, and can be interpreted as a non-trivial RG flow of the associated  $\theta$-angle, driven by non-perturbative effects. On the gravity side, this is captured by a holographic RG flow solution, in which the axion has a non-trivial bulk profile.
The RG flow of  $\theta$-angles has also been considered in QFT in several works, \cite{Knizhnik:1984kn,Levine:1984pv,Latorre:1997af,Pruisken:2000my,Apenko:2007wk,Nakamura:2019ind}.
It should be stressed that the renormalization flow of the $\theta$-angle is not driven by short distance divergences. It is a finite renormalization that is known to occur in solvable effective gauge theories like the Seiberg-Witten theory, \cite{SW} and many of its generalizations.

The dynamics of YM-like holographic theories, can be described by including the dual of the most important relevant (scalar) operator in the gravitational bulk action. In the YM case, this is $Tr[F^2]$. The relevant gravitational theory therefore is an Einstein-scalar theory (typically called Einstein-dilaton theory). It is considered at the two-derivative level, as it is expected to be valid at strong coupling. If the dual quantum field theory is a $d$-dimensional QFT, then the gravitational dual will live in a bulk space-time, that has at least $d+1$ space-time dimensions. The presence of more than $d+1$ dimensions is associated with symmetries that are realized in the adjoint sector of the theory, \cite{diss,ap}. Here,  we will consider theories without adjoint global symmetries and therefore our holographic theory will live in $d+1$ bulk space-time dimensions.\footnote{This assumption is without an important loss of generality. Many higher-dimensional solutions can be dimensionally reduced to this form.}

The instanton density is dual to a bulk axion (pseudo)scalar, with no potential, reflecting the perturbative shift symmetry that is the holographic avatar of the fact that a shift in the $\theta$-angle does not change the theory in perturbation theory. (Dilute) instanton effects are exponentially suppressed at large $N$.

A rather general holographic description of the dynamics of the instanton densities, in a $d$-dimensional theory, can be captured therefore by the class of Einstein-axion-dilaton gravitational theories in $d+1$ bulk dimensions.
This description also includes theories that are defined in higher dimensions but are subsequently dimensionally reduced to a lower dimension.
The holographic renormalization of such theories has been fully described in \cite{Papa}.

In \cite{axion-rg} the bulk axion RG-flows in a generic $(d+1)$-dimensional Einstein-axion-dilaton theory were studied in full generality.
The Einstein-axion-dilaton Lagrangian enjoys an exact axion shift-symmetry (i.e.~the axion enters neither in the potential nor in the metric in field space).  These theories serve as bottom-up phenomenological models, or may be considered as proxies for low-energy effective supergravities emerging from top-down string theories\footnote{What is not included in our setup are RG flows/solutions where the fields depend on more than one internal coordinates. In the picture in which we only keep the holographic coordinate, such flows can be represented only upon the inclusion of an infinite number of KK-generated bulk fields.}.
Their general form after field redefinitions is
\be
S= M_p^{d-1} \int d^{d+1}x \sqrt{-g} \left[
R - \frac{1}{2} (\pa\f)^2-{Y(\f)\over 2}(\pa a)^2
- V(\f)
\right],
\label{A1}\ee
and depend on two functions, $V(\f)$ and $Y(\f)$ that are not completely arbitrary. They are constrained by several properties of string theory, and such constraints have been studied in \cite{Gubser,iQCD,thermo,cgkkm,axion-rg}. Moreover, we will be studying theories whose potential $V$, leads to confinement, \cite{iQCD}.

The holographic RG flow geometries, which display $d$-dimensional Poincar\'e invariance (and correspond therefore  to vacuum states of the dual QFT), are of the general form
\be \label{intro5}
ds^2 = du^2 + e^{2A(u)} \eta_{\mu\nu}dx^\mu dx^\nu, \quad \f = \f(u), \quad a = a(u)
\ee
characterized by a scale factor $A(u)$, dilaton profile $\f(u)$ and axion profile $a(u)$, where $u$ is the holographic coordinate. The  solutions have an asymptotic AdS boundary, and as such,  they are dual to field theories with a UV conformal fixed point\footnote{This can be relaxed to include holographic theories which in the UV match the logarithmic running of asymptotically free QFTs, with no significant change in the qualitative picture \cite{iQCD}}, deformed by a relevant operator dual to the dilaton and have generically a $\theta$-angle if the axion is non-trivial in the solution.
In \cite{axion-rg}, the fully backreacted system was studied, in which the effect of the axion on the metric and dilaton dynamics was fully taken into account.

The issue of backreaction however deserves some comments. When axion running is considered in holography, this is most often discussed in the probe limit, i.e.~ignoring the backreaction of the axion on other bulk fields such as the metric, dilaton, etc. This is because, in known string theory examples, the axion backreaction is suppressed by $1/N_c^2$. It is therefore subleading in the large-$N_c$ limit. This corresponds to the gauge theory expectation, since the $\theta$-term in the action, with $\theta\sim \mathcal{O}(1)$, is subleading in the large-$N_c$ limit, \cite{Witten0}.  The axion running can still give $\mathcal{O}(1)$ contributions to quantities which are vanishing at leading order, such as the topological susceptibility, which indeed can be matched to lattice results in phenomenological models \cite{iQCD,data}. However, the implicit assumption is that this does not lead to significant effects in the other sectors of the theory (e.g.~the dynamics of the running coupling and the associated Yang-Mills Lagrangian operator) whose free energy is $\mathcal{O}(N_c^2)$.

There are important exceptions, where the axion backreaction becomes relevant. This is the case when the bulk axion becomes effectively  non-compact, and is allowed to take on arbitrarily large values, e.g.~$\mathcal{O}(N_c)$. In this case, the axion contribution to the bulk Einstein  equations  is unsuppressed. This is the case, for example, in axi-dilaton black holes with a linear axion profile \cite{tada,s0,s1,s2}. In string theory, this can also occur in models with axion monodromy \cite{monodromy, monodromy2}, in which the axion decompactifies due to the coupling with extended objects. For large axion values, the axion backreaction is indeed important and hence cannot simply be ignored \cite{1011.4521, 1411.2032, 1602.06517, 1611.00394}.

The axion bulk profile is characterized by two parameters, $a_{UV}$ and $Q$,  which enter as the two integration constants of the second order axion equation of motion and control the leading and subleading terms in the near-boundary expansion. This expansion has schematically the form,
\be \label{intro1}
a(u) = a_{UV}  + Q \, e^{d u /\ell} + \ldots \qquad u \to -\infty,
\ee
where $u\to -\infty$ corresponds to the AdS boundary,  and $\ell$ is the (UV) AdS length. In the holographic dictionary,  $a_{UV}$ is related to the value of the $\theta$-angle in the UV field theory (modulo 2$\pi$ shifts)  and $Q$ is proportional to the vacuum expectation value of the corresponding instanton density. The precise expressions have been presented in \cite{axion-rg}. Roughly speaking,  the probe limit corresponds to small $Q$.
 Due to the exact axion shift symmetry of the bulk Lagrangian, of the two parameters, only $Q$ enters non-trivially in the non-linear equations and solutions  for the metric and dilaton. This does not mean however that the value of $a_{UV}$ does not affect the solution, as a relation between $Q$ and $a_{UV}$ arises due to boundary conditions in the far interior.

In \cite{axion-rg} it was argued  rather generally, that the correct regularity condition for axion flows in the interior of the bulk is that
\be \label{intro2}
\text{axion regularity:} \qquad a_{IR} \equiv a (u_{IR}) = 0.
\ee
where $u_{IR}$ is the IR endpoint of the bulk geometry, to be defined precisely in later sections.
This was motivated by  top-down string theory constructions, where the axion is a form field component along an  internal cycle, which shrinks to zero-size in the IR as in \cite{witten}. Single-valuedness then demands that the axion field vanishes at such points. Assuming this notion of axion regularity to hold in general, it leads to a consistent holographic interpretation of axion RG flows. In the probe limit, imposing equation (\ref{intro2}) results in a linear relation on the UV coefficients in (\ref{intro1}) of the type
\be \label{intro4}
Q = c\,  a_{UV}
\ee
where $c$ is a constant that  depends only on the metric and dilaton profiles. However, as shown in \cite{axion-rg},  backreaction will turn (\ref{intro4}) into a  non-linear  relation. Interestingly,  the condition (\ref{intro2}) will also lead us to discard as unphysical a full class of solutions, in which $Q$ is fixed independently of $a_{UV}$.

In this paper we continue the analysis of \cite{axion-rg} towards understanding the physics of CP-violation in holographic theories with a non-trivial $\theta$ angle.

The first question we address is motivated by the fact that the axion must always vanish in the IR part of the geometry according to (\ref{intro2}). As the running axions can be  considered as the effective running $theta$-angle, this may seem to suggest that CP-invariance will be restored in some sense in the IR. One can ask  whether this can alleviate the strong-CP problem.

In \cite{axion-rg} it was found that the range of values of the source $a_{UV}$ for which a regular axion solutions exists is always bounded: $|a_{UV}|\in [0,a_{UV}^{max})$. The maximum value $a_{UV}^{max}$ depends on the dimension $d$ as well as the bulk potential functions $V(\f)$ and $Y(\f)$. Here, we shall investigate the stability of the ground state solution of the theory for all allowed values of $a_{UV}$.

To characterize and classify the bulk models,  it is convenient to use a general  parametrization of  the asymptotic form of the bulk dilaton potentials for large values of $\f\to \infty$, of the form:
\be
V(\f)\sim e^{b\f}\sp b\geq 0\sp  \f\to \infty
\label{a}\ee
The behavior of the solution in the IR is classified according to the value of the parameter $b$ \cite{iQCD}.
For a confining (and gapped) theory,
\be
\sqrt{\frac{2}{d-1}}< b<\sqrt{\frac{2d}{d-1}}.
\label{aa}\ee
The upper bound is the well known Gubser bound, \cite{Gubser,iQCD}, and is imposed so that mild bulk singularity is resolvable.
If $b$ is smaller than the lower bound in (\ref{aa}) then the theory is gapless and non-confining.
Theories with asymptotics as in (\ref{aa}) have a hyperscaling-violating, scaling regime in the IR, \cite{GK}. This is explained by the fact that these asymptotics are obtained by compactifying a  higher dimensional AdS solution on a sphere, \cite{GK}.
In the range (\ref{aa}),   the end-of space in the IR is at a finite value of the conformal radial coordinate\footnote{This is the coordinate $r$ defined by the relation:
$$
du = e^A dr,
$$
where $u$ is radial coordinate in which the  metric takes the form (\ref{intro5}) }. 
The glueball masses scale as
\be
m_n\sim n\sp n\to \infty
\label{aaaa}\ee
which is the scaling one obtains in cutoff AdS space.

The lower bound value $b=\sqrt{\frac{2}{d-1}}$ is special, and for this value we can refine the large-field asymptotics of the potential as follows,
\be
V(\f)\sim e^{\sqrt{\frac{2}{d-1}}\f}\f^{P}\sp P\geq 0\sp  \f\to \infty .
\label{aaa}\ee
Again this describes confining gapped theories but in this case the potential is softer in the IR. Such solutions do not have a scaling symmetry in the IR. If $P>1$, the end of space is again at a finite  value of the conformal radial coordinate, and the glueball masses have the same asymptotic behavior as in  (\ref{aaaa}).\footnote{Strictly speaking, $m_n\sim n/\log n$ for $P=1$.}
If $0<P<1$ the end of space is at an infinite value of the conformal radial coordinate. The glueball masses behave as
\be
m_n\sim n^{P}\sp n\to \infty\;.
\ee
 This case contains the linear trajectories for $P={1\over 2}$ which is the choice in Improved Holographic QCD, \cite{iQCD}.

\subsection{Results and Outlook}

We have analyzed the holography of Einstein-axion-dilaton theories in $d=3$ and $d=4$, where $d$ is the space-time dimension of the dual QFT.
We have also analyzed potentials that are in class (\ref{aa}) that we call in the sequel  ``steep potentials"
and potentials that are in class (\ref{aaa}) that we will call in the sequel ``soft potentials".
In the case of steep potentials, we have analyzed various (allowed) asymptotic behaviors for the function $Y(\f)$ that controls the kinetic term of the axion. In the case of the soft potentials, we have fixed $P={1\over 2}$ to have a Regge-like glueball spectrum. Moreover,  we have fixed the large-$\f$ asymptotic behavior of $Y(\f)$ requiring {\em glueball universality}, i.e. the requiring that (in the CP-symmetric limit) the $0^{+-}$ glueball trajectory have the same slope as the  $0^{++}$ and $2^{++}$  trajectories.\footnote{As shown in appendix \ref{universality}, for steep potentials or for soft potentials with $P\geq 1$  glueball universality is automatic and independent of the specific large-$\f$ behavior of $Y(\f)$.}

We have also analyzed several values of possible parameters in these potentials. Although in the rest of the paper we exhibit concretes example in each case, we expect, based on our calculations, that their behavior is generic.

\begin{itemize}

\item We find the background solutions with non-trivial dilaton and axions  both in $d=3$ and $d=4$ and with both steep and soft potentials. We also determine the maximum values of $a_{UV}$ in each case.
    We observe that such solutions, in all cases,  have qualitatively similar features.

\item We compute that glueball spectra of spin-2 glueballs, arising from the transverse-traceless part of the bulk metric, as well as the two spin-0 towers that arise from the axion and the dilaton. The spin-0 problem can be mapped to a coupled system of Schr\"odinger equations. This system factorises only when $a_{UV}=0$ and the background axion field is trivial.

    When $a_{UV}=0$, we have CP-symmetry and  the eigenstates are the towers of the $0^{++}$ and $0^{+-}$ glueballs.

We find that, generically, the glueball masses decrease as the $\theta$ angle (or $|a_{UV}|$) increases. This is similar to what was observed in \cite{Dubovsky:2011tu,BC1}  using   Witten's black D$_4$ holographic model.
All of the above are valid both at $d=3$ and $d=4$ and for both soft and steep potentials.

\item For steep potentials, and for sufficiently large $a_{UV}<a_{UV}^{max}$ the lowest glueball mass becomes tachyonic signaling an instability of the saddle-point solution.
    This happens for both $d=3$ and $d=4$.

\item We diagonalise the quadratic action for glueballs and we then compute the four cubic couplings of the two lightest spin-0 glueballs. This is done in $d=3$ only, as we use results, appropriately adapted from the study of non-gaussianities in similar theories in the context of cosmology. For this action, we calculate the CP-violating effects of the interactions in detail. In the limit of zero $\theta$-angle, two such couplings vanish because of the CP-symmetry.

 \item  We find that, at finite $\theta$,  no particular suppression exists for the CP-violating effects. We expect, from previous experience that these results are qualitatively correct also in $d=4$.

\end{itemize}

There are two clear puzzles that emerge from our analysis.

\begin{enumerate}

 \item Why there are no regular solutions in the theory for $a_{UV}>a_{UV}^{max}$?

 \item Why are there instabilities for steep potentials at sufficiently large $a_{UV}$, and in such a case what is the dominant and stable solution?

 \end{enumerate}

We believe, that the answer to both of the questions above is related, and is also similar to a phenomenon seen in other classes of holographic solutions, \cite{cc}.
In such cases, the resolution was correlated with the fate of the cosmic censorship conjecture, and tied interestingly with the weak gravity conjecture, \cite{wgc} and its generalizations.

The analogous resolution in this case relies on the fact that axions in string theory are generalized gauge fields and there are D-instantons that are charged minimally under them. Such instantons are solitonic but may condense in the bulk solution generating a novel setup analogous to the condensation of scalar fields in RN black holes, triggering the appearance of a new phase.
Such novel solutions must be examined in order to ascertain as to whether ``instanton condensation"  can describe the stable ground states of the theories in question.\footnote{An instanton-related domain wall was proposed with a different motivation in \cite{Shu}.}
This investigation is left for the future. 

The structure of this paper is as follows:
In section \ref{EAD} we introduce our Einstein-axion-dilaton theory and discuss its holographic interpretation. We also present numerical solutions for the background.
In section \ref{Quadratic} we compute spectra of spin-0 and spin-2 glueballs.
In section \ref{cubic} we compute CP-violating cubic couplings among the spin-0 glueballs.

In appendix \ref{geometry} we summarize the geometry of the field space. In appendix \ref{conformal} we provide the definition and equations of motion in the conformal coordinate system. In appendix \ref{wavefunction} we describe asymptotics of the wave-function of Schrodinger equations. In appendix \ref{universality} we discuss the universality of glueballs. In appendix \ref{IR} we derive IR asymptotic solutions with soft potentials. In appendix \ref{quadratic} we present a derivation of the quadratic fluctuation equations.
In appendix \ref{transformation} we provide a transformation law from the conformal radial coordinate to a coordinate using the scale factor.
Finally appendix \ref{continuation} we describe an analytic continuation which is used to compute the CP-violating cubic couplings.

\section{Einstein-axion-dilaton theory and holography}\label{EAD}

We consider  an Einstein-axion-dilaton theory in a $(d+1)$-dimensional bulk space-time parametrized by coordinates $x^a\equiv (u, x^\mu)$.
The most general two-derivative action with the axion shift symmetry is
\be
S= M_p^{d-1} \int d^{d+1}x \sqrt{-g} \left[
R - \frac{1}{2} G_{IJ} g^{ab}\de_a \phi^I \de_b \phi^J
- V(\f)
\right] + S_{GHY},
\label{A2}\ee
where $g_{ab}$ is the bulk metric, $R$ is its associated Ricci scalar, $V(\f)$ is the bulk scalar potential, and $S_{GHY}$ is the Gibbons-Hawking-York term.
The field space metric $G_{IJ}$ and vector $\phi^I$ are defined as
\be
G_{IJ}=
\begin{pmatrix}
G_{\f\f}&G_{\f a}\\
G_{a\f}&G_{aa}
\end{pmatrix}=
\begin{pmatrix}
1&0\\
0&Y(\f)
\end{pmatrix},
\quad
\phi^I =
\begin{pmatrix}
\f \\
a
\end{pmatrix}.
\label{thi3}\ee
In appendix \ref{geometry}, we summarize the geometry of the field space specified by the metric \eqref{thi3}. This would be useful in the calculations in the later sections.

The scalar field $\f$ is dual to  a relevant operator of the UV field theory. In a YM-like theory it is expected to correspond to $Tr[F^2]$, but we keep its interpretation open for the rest. The massless scalar field $a$ is expected to be dual to the instanton density operator. The metric, $g_{\m\n}$, as usual, it is dual to the stress tensor of the theory.

The bulk field equations, stemming from the action (\ref{A2}) are given by
\be
R_{ab} -{1\over 2} g_{ab} R = {1\over 2}\de_a\f\de_b \f +\frac{Y}{2}\de_a a\de_b a-  {1\over 2}g_{ab}\left( {1\over 2}g^{cd}\de_c\f\de_d \f+{Y\over 2}(\pa a)^2 + V \right),
\label{FE1}\ee
\be
\frac{1}{\sqrt{-g}}\de_a \left(\sqrt{-g} g^{ab}\de_b \f \right)- \frac{d V}{d \f}-{1\over 2}\frac{d Y}{d\f}(\pa a)^2 =0,
\label{FE2}\ee
\be
  \de_a \left(\sqrt{-g} \,Y\,g^{ab}\de_b a \right)=0.
\label{FE3}\ee

We consider the bulk space-time solution to  have $d$-dimensional Poincar\'e invariance, so that the solution would be dual to the ground state of a Lorentz-Invariant QFT$_d$ defined on Minkowski space-time. With these symmetries, the  solution can be put in the form (up to diffeomorphisms):
\be\label{FE7}
ds^2 = du^2  + e^{2A(u)} \eta_{\mu\nu}dx^\mu dx^\nu
\, , \qquad \f = \f(u) \, , \qquad  a=a(u),
\ee
where  $u$ is the (holographic) domain-wall coordinate. We also
  use a conformal radial coortdinate $r$, in which the solution takes the form
\be\label{FE7-ii}
ds^2 = e^{2A(r)} \left( dr^2 + \eta_{\mu\nu}dx^\mu dx^\nu \right)
\, , \qquad \f = \f(r) \, , \qquad  a=a(r).
\ee
The $u$ and $r$ coordinates are related by
\be
{du \over dr} = e^A.
\label{th2}\ee
The domain world coordinate $u$ is more convenient when we discuss the solutions of the equations of motion as RG flows, while the conformal coordinate $r$ is more useful when we study fluctuations around the solutions.

The UV AdS boundary and the IR endpoint correspond to $u=u_{UV}=-\infty$ ($r=r_{UV}=0$) and $u=u_{IR}=\infty$ ($r=r_{IR}$), respectively.
As we shall soon see,  $r_{IR}$ can be  either finite or infinite, \cite{iQCD}.

The bulk field equations for the  ansatz (\ref{FE7}) are
\be
d(d-1)\dot A^2-{1\over 2}\dot \f^2 -{Y\over 2}\dot a^2+V=0\sp 2(d-1)\ddot A+\dot\f^2+Y \dot a^2=0,
\label{a1}\ee
\be
\ddot\f+d\dot A\dot \f- \p_\f V - \frac{\p_\f Y}{2}\dot a^2=0,
\quad
\pa_u(Y\,e^{dA}\,\dot a)=0,
\label{a3}\ee
where a dot stands for a $u$ derivative while $\p_\f$ stands for a $\f$ derivative. The second equation in (\ref{a1}) is redundant as it can be obtained from the other equations.
The expressions in the conformal coordinate  system (\ref{FE7-ii})  are presented in appendix \ref{conformal}.

The axion equation of motion integrates to
\be
\dot a =\ell^{d-1}{Q\over Ye^{dA}},
\label{a5}\ee
with $Q$ being an integration constant. The mass dimension of $Q$ is $d$.
The system can be written as a first order system by introducing the scalar functions $W, S$ and $T$ as
\be
\dot A=-{W(\f)\over 2(d-1)},
\quad
 \dot \f=S(\f),
\quad
\dot{a}^2 = \frac{T(\f)}{Y^2},
\quad
T(\f)=(\ell^{d-1}Q)^2~e^{-2dA}
\label{a6}\ee
with
 \be
 S^2-\frac{dW}{d\f}S+{T\over Y}=0,
 \quad  {1\over T}\frac{dT}{d\f}={d\over d-1}{W\over S},
 \quad {d W^2\over 4(d-1)}-{S^2\over 2}-{T\over 2Y}+V=0.
 \label{a8}
 \ee

A bookkeeping  of the constants of integration of the above system is important. The original system of equations in (\ref{a1}) and (\ref{a3}) has 5 integration constants. This is the same number one finds in the first order fiormulation: the three first order equations in (\ref{a6}) have three integration constants, and the system of  equations in (\ref{a8}) has  two more.
The interpretation of the three integration constants in (\ref{a6}) are as sources, or alternatively, as couplings in the dual QFT. The additive integration constant of the $A$ equation sets the overall scale of the solution, and it can be fixed in the UV.   TYhe integration constants of the $\f$ and $a$ equations are the (UV) relevant coupling of the operator $O$, dual to $\f$, as well as the $\theta$ angle\footnote{The precise correspondence of the source of $a$ and the $\theta$-angle can be found in \cite{axion-rg} and will be discussed later on.}.
The two further integration constants in the system (\ref{a8}) correspond to the vevs of the operators $O$ and the instanton density.

\subsection{The near-boundary asymptotic solutions}

A UV fixed point generically corresponds to a maximum of the bulk scalar potential $V(\f)$. By an appropriate shift of $\f$, we can set the maximum to occur at $\f=0$.
Around the UV fixed point, the bulk functions $V(\f)$ and $Y(\f)$ are expanded as
\be
V=-{d(d-1)\over\ell^2}-{1\over2}{m^2\over\ell^2}\f^2  + \mathcal{O}(\f^3) \, ,
\quad
Y= Y_0  + \mathcal{O}(\f) ,
\label{UV1}\ee
with
\be
\Delta_{\pm}={d\over 2}\pm \sqrt{{d^2\over 4}-m^2\ell^2}.
\label{UV29}\ee
For a maximum,  $m^2>0$,  ${d\over 2}<\Delta_{+}<{d}$ and $0<\Delta_{-}<{d\over 2}$.

The UV expansion of  $\f$, $a$ and $A$, can be found by solving the equations near the maximum\footnote{Here the minus branch solution in \cite{exotic} is chosen.}:
\be
\f=
\f_- \ell^{\Delta_-}e^{\Delta_- u/\ell}+{C d\left(\f_-\,\ell^{\Delta_-}\right)^{{\Delta_+\over\Delta_-}}\over(\Delta_+-\Delta_-)\Delta_-}e^{\Delta_+{u/\ell}}+\cdots,
\quad
a= a_{UV} + \frac{Q\ell^d}{dY_0} e^{\frac{du}{\ell}}+\ldots,
\label{UV23}\ee
$$
e^{A}=e^{-{u\over\ell}}+\ldots,
$$
where $\f_-, \,C, \,a_{UV}$ and $Q$ are  integration constants.

Holographically, the integration constant $\f_-$ is the source of the relevant operator corresponding to $\f$ in the dual QFT. The holographic map of $a_{UV}$ to the UV $\theta$-angle is, \cite{witten,iQCD,data,axion-rg}
\be
a_{UV}= c \frac{\theta_{UV}+2\pi k}{N_c},
\label{UV9}\ee
where $\theta_{UV}\in [0,2\pi)$, $k\in \mathbb{Z}$ and $c$ is a dimensionless number depending on the precise setting of the bulk-boundary correspondence. 
The expectation values of the operator dual to $\f$ and $a$ (instanton density) are given by
\be
\langle O_\f \rangle = C \, (M_p \ell)^{d-1}{d\over \Delta_-}|\f_-|^{\Delta_+\over\Delta_-},
\quad
\langle O_a \rangle= c\, Q \frac{(M_p \ell)^{d-1}}{N_c}.
\label{UV27}\ee

Because of the axion shift symmetry in the action \eqref{A2}, the axion source $a_{UV}$ is a free parameter and is not related to $Q$.
However, as argued in \cite{axion-rg}, and in analogy with regular examples in string theory, the appropriate IR regularity condition for the bulk axion is
\be
a(u_{IR})=0.
\label{a10}\ee
This condition gives the relation between the vev and source of the axion, as expected in holography.

\subsection{IR asymptotic solutions}

The dilaton potential $V(\f)$ is an important part of the bulk action and controls the physics of the theory in the absence of the $\theta$-angle.
A general analysis of confining potentials and their properties has been done in \cite{iQCD}.  They are of two types, which we call {\em steep} and {\em soft} potentials, depending on their large-$\f$  asymptotics, which determines the IR properties of the solution.

For the  first class, the physics in IR region is similar to a higher-dimensional  AdS theory compactified on a (internal) sphere, \cite{GK}. In such a case, the conformal coordinate has a finite range.

The second class contains the mildest potentials in the IR, with a conformal coordinate having infinite range. Improved holographic QCD is in this class, \cite{iQCD}.

\begin{enumerate}
\item \noindent \textbf{Steep potentials.}

The  leading large-$\f$ behavior is conveniently parametrized by exponential functions,
\be
V\overset{\f \rightarrow +\infty}{\longrightarrow}
-V_\infty e^{b\f},
\quad \quad
Y\overset{\f \rightarrow +\infty}{\longrightarrow}
Y_\infty e^{\g \f}
\label{a32}\ee
with $V_\infty,Y_\infty$ positive, and
\be
\sqrt{\frac{2}{d-1}}<b<\sqrt{\frac{2d}{d-1}}.
\label{Num1-2}\ee
The mass dimension of $V_\infty$ and $Y_\infty$ are $2$ and $0$, respectively.
 The lower bound  on $b$ comes from the requirement that the theory is  confinning, whereas  the upper bound comes from  Gubser's bound \cite{Gubser,thermo}, which can be interpreted as a condition for the (mild) IR singularity to be resolvable\footnote{If however we would like the correlation functions of the theory not to depend on the resolution of the (mild) IR singularity then there is a more stringent upper bound on $b$, \cite{thermo,cgkkm}. This happens because otherwise, both fluctuation solutions near the singularity are normalizable and an extra condition is needed to choose the correct solution. Requiring that only one of the two solutions of the fluctuations is normalizable we obtain  \cite{iQCD}
\be
b < \sqrt{\frac{2(d+2)}{3(d-1)}},
\label{Num1-4}\ee
as described in appendix \ref{wavefunction}.}.

For steep potentials, we restrict the large-$\f$ behavior of the axion kinetic function by requiring that
 \cite{iQCD}.
\be\label{Num1-2-i}
\g > \frac{2d}{(d-1)b}-b \equiv \g_\text{min}
\ee
The lower bound on $\g$ was derived in \cite{axion-rg} and is required for overall regularity and consistency of solutions.

The IR asymptotics (\ref{a32}, \ref{Num1-2}) give a  confining geometry: this means that the holographic Wilson loop obeys an area law, and that the spectrum of bulk excitations  is gapped and made up of a discrete tower of states (interpreted as glueballs in the dual field theory \cite{iQCD}.

The IR endpoint is at a finite value of the conformal coordinate $r$,  $r=r_{IR}=\text{(finite)}\equiv r_0$ \cite{iQCD}. For $r\to r_0$, the scalar field diverges to $+\infty$, and  the  behavior of $W, S$ and $T$ is given by\footnote{The derivation of the IR asymptotic forms is given in \cite{axion-rg}.}
\be
W=\left(W_\infty
	-{D\over2}\frac{e^{-(\g-\g_\text{min})\f}}{{b\over2}+\g-{d\over (d-1)b}}
	\right)e^{{b\over2}\f}
+\ldots,
\label{sub8-2}\ee
$$
S=\left({b\over2}W_\infty
	-{D\over2}\frac{{b\over2}+\g}{{b\over2}+\g-{d\over (d-1)b}}
	e^{-(\g-\g_\text{min})\f}
	\right)e^{{b\over2}\f}+\ldots,
$$
$$
T={b\over2}D\, W_\infty Y_\infty  \,e^{{2d\over(d-1)b}\f}+\ldots,
$$
In the above equations,  $D$ is an integration constant which is the IR avatar of the vev parameter $Q$ appearing in the UV expansion (\ref{UV23}), and
\be
W_\infty \equiv 2\sqrt{\frac{V_\infty}{1-\frac{1}{(d-1)\delta}}},
\quad
\delta \equiv {1\over \frac{d-1}{2}b^2 - 1}.
\label{a34}\ee
Note that by requiring \eqref{Num1-2} and \eqref{Num1-4} we obtain:
\be
\delta > \delta_\text{min},
\quad
\delta_\text{min} \equiv \frac{3}{d-1},
\label{a34-2}\ee

The integration constant $D$ is related to the integration constant $Q$ in \eqref{a5}:
\be
\lim_{\f_{UV}\to0,\f_{IR}\to\infty} \frac{Q^2\ell^{2(d-1)} \left|\frac{\f_{UV}}{\f_- \ell^\Delta_-}\right|^{\frac{2d}{\Delta_-}} e^{{d\over d-1}\int^{\f_{IR}}_{\f_{UV}}d\f\,{W\over S}}}{{b\over2} D W_\infty Y_\infty  e^{{2d\over(d-1)b}\f_{IR}}} = 1.
\label{sub14}\ee

The bulk fields $\f, a$ and the warp factor $A$ behave near the IR end-point as
\be
\f = - b (d-1) \delta \log\left(r_0-r\over \ell\right) +\ldots,
\quad
A = \delta \log\left(r_0-r\over \ell\right) + \ldots,
\label{th20-2}\ee
\be
a = - {\rm sign} (Q) \sqrt{2D \over bW_\infty Y_\infty }{1\over {b\over 2}+\gamma-{d\over (d-1)b}}
e^{\left((d-1) \big(1+b(\g-\g_\text{min})\big) - \frac{1}{\delta} \right) A}+\ldots.
\label{th20-2-ii}\ee
Holographically, the energy scale of the boundary field theory is measured by the scale factor,
\be
\mu \sim e^{A}.
\label{th11}\ee
The identification \eqref{th11} does not determine the absolute units of the energy scale.
From the identifications \eqref{UV9} and \eqref{th11}, \eqref{th20-2-ii} can be written schematically as
\be
a(\mu)=c\,{\theta(\mu) + 2\pi k \over N_c} \sim \mu^{  (d-1) \big(1+b(\g-\g_\text{min})\big) - \frac{1}{\delta}}.
\label{th14}\ee
From \eqref{Num1-2}, \eqref{Num1-2-i} and \eqref{a34-2}, the exponent of $\mu$ in \eqref{th14} is positive.
Therefore, in the IR, as $\mu \to 0$, $a(\mu) \to 0$ by construction.

The spectrum of linear fluctuations around the vacuum solution  (which will be discussed extensively in Section 4) is gapped and discrete, and corresponds to the gauge-invariant coposite particle states  (glueballs) in the field theory.  One can show\footnote{See appendix \ref{universality} for the derivation.} that,  asymptiotically (i.e. for large mass quantum number), the  glueball spectra for steep potentials  behave as
\be
m_{n_+} = \frac{\pi}{r_0} n_+ + \ldots,
\quad m_{n_-} = \frac{\pi}{r_0} n_-  + \ldots, \qquad n_{\pm} \to +\infty
\label{w63}\ee
where $n_\pm$ are the quantum numbers.

As explained in appendix \ref{universality}, there are two distinct towers of bound states.
The masses $m_{n_+}$ and $m_{n_-}$ correspond to $0^{-+}$ and $0^{++}$ glueball masses for $a_{UV}=0$, respectively. Note that, in the presence of non-trivial axion source, $a_{UV}\neq0$, there is no invariant distinction between $0^{-+}$ and $0^{++}$ glueballs.
The ratio of slopes of the two towers
\be
\lim_{n\to\infty} \frac{m_{n_-}}{m_{n_+}}=1.
\label{w65}\ee

\item \noindent \textbf{Soft potentials.}

Confining potentials in the ``soft'' class correspond to setting $b$ to saturate the lower bound in equation (\ref{Num1-2}), and refining the large $\f$ asymptotic behavior by a power-law 
\be
V \overset{\f \rightarrow +\infty}{\longrightarrow}
- V_\infty \f^{P} e^{\sqrt{\frac{2}{d-1}} \f},
\label{qc1}\ee
The solution is confining for $P>0$, and for $P\geq 1$ things are qualitatively the same as in the previous case of steep potentials (IR endpoint at finite value of the conformal coordinate, discrete spectrum with asymptotics behavior (\ref{w63})). For this reason we include the asymptotic (\ref{qc1})  in the ``steep potential'' class, and define soft potentials by:
\be \label{soft}
\text{\em soft potentials} = \text{(\ref{qc1}) with } \; 0<P<1.
\ee
For this class of asymptotics, the IR end of space is at infinity in conformal coordinates.
For $d=4$, the choice $P=1/2$ realizes linear glueball asymptotics ($m_n^2 \sim n$) and is the choice for Improved Holographic QCD  \cite{iQCD}, while for general $P$ we have:
\be \label{qc1-iii}
m_{n_{\pm}} \sim n_{\pm}^P \qquad n_{\pm} \to +\infty .
\ee

For the axion kinetic function, we take:
\be \label{qc1-ii}
Y \overset{\f \rightarrow +\infty}{\longrightarrow}
Y_\infty e^{\sqrt{2(d-1)}\f}
\ee
We have taken the exponent of $Y$ to correspond to $\gamma_\text{min}$ in (\ref{Num1-2-i}) for $b={\sqrt{2\over d-1}}$.
This exponent  is the same as the one chosen in Improved Holographic QCD,  \cite{iQCD,cs} for $d=4$.\footnote{The relation between $\lambda$ in \cite{cs} and $\f$ in \eqref{A2} is $\log \lambda = \sqrt{d-1\over 8} \f$.}
 and it is determined by requiring that, for large mass quantum numbers, the glueballs have a spectrum whose slope is independent of their spin and parity\footnote{
Indeed, as described in appendix \ref{universality}, for general $\gamma$ the asymptotic spectrum with $\g\geq\g_\text{min}=\sqrt{2(d-1)}$ is
\be
\lim_{n\to\infty} \frac{m_{n_-}}{m_{n_+}} =
\left(\sqrt{\frac{2}{d-1}}\g-1\right)^{1-P\over P},
\label{w9}\ee
From \eqref{w9}, we observe that we have glueball universality if
\be
\lim_{n\to\infty} \frac{m_{n_-}}{m_{n_+}} =1 \quad \text{for} \quad \g=\sqrt{2(d-1)}.
\label{w11}\ee
}.

Next, we summarize the IR asymptotic solutions. The derivation is provided in appendix \ref{IR}. The   expressions below hold for  the asymptotics   (\ref{qc1-ii}), i.e. we take $\gamma=\sqrt{2(d-1)}$. At large $\f$, we find: 
\be
W =
\left(
2 \sqrt{V_\infty} \f^{P\over2}
+  P \sqrt{2V_\infty \over d-1} \f^{{P\over2}-1}
- {D \over 2\sqrt{V_\infty} Y_\infty} \f^{-\left(d+{1\over2}\right)P}
\right) e^{{1\over\sqrt{2(d-1)}}\f}+\ldots,
\label{s27}\ee
$$ 
S = \left(
\sqrt{\frac{2V_\infty}{d-1}} \f^{P\over2}
+ \frac{\sqrt{V_\infty} d P}{d-1} \f^{{P\over2}-1}
 - \frac{D(2d-1)}{2\sqrt{2(d-1)V_\infty} Y_\infty} \f^{-\left(d+\frac{1}{2}\right)P}
\right) e^{\frac{1}{\sqrt{2(d-1)}}\f} +\ldots,
$$
$$
T = D \,\f^{-d P} e^{\sqrt{\frac{2}{d-1}}\,d\f} +\ldots
$$

From the last equation of \eqref{a6} and \eqref{s27}, the relation between the integration constants $D$ and $Q$ in this background is 
\be
\frac{\ell^{2d} Q^2}{\ell^2D}=\lim_{\f_{IR}\to\infty} \f_{IR}^{-d P} e^{\sqrt{\frac{2}{d-1}}\,d\f_{IR}}e^{2dA(\f_{IR})}.
\label{sub14-2-i}\ee
Using the first two equations in \eqref{a6} and \eqref{UV23}, $e^{2dA(\f_{IR})}$ is expressed as
\be
e^{2dA(\f_{IR})}=
\lim_{\f_{UV}\to0} \left|\frac{\f_{UV}}{\f_- \ell^\Delta_-}\right|^{-\frac{2d}{\Delta_-}} e^{-\frac{d}{d-1}\int^{\f_{IR}}_{\f_{UV}} d\f \frac{W}{S}}.
\label{sub14-2-ii}\ee
Combining \eqref{sub14-2-i} and \eqref{sub14-2-ii}, we obtain
\be
\frac{\ell^{2d} Q^2}{\ell^2 D}=
\lim_{\f_{UV}\to0,\f_{IR}\to\infty} \frac{\f_{IR}^{-d P} e^{\sqrt{\frac{2}{d-1}}d\f_{IR}}}{\left|\frac{\f_{UV}}{\f_- \ell^\Delta_-}\right|^{\frac{2d}{\Delta_-}} e^{{d\over d-1}\int^{\f_{IR}}_{\f_{UV}}d\f\,{W\over S}}}.
\label{sub14-2}\ee

In the IR, $A, \f$ and $a$ behave as:
\be
A=- {G\over\sqrt{2(d-1)}} \left(r\over\ell\right)^{1\over 1-P} + \ldots,
\quad
\f= G \left(r\over\ell\right)^{1\over 1-P} + \ldots,
\label{s14}\ee
$$
a=- \frac{{\rm sign} (Q)}{Y_\infty} \sqrt{D\over V_\infty} \f^{-{d+1\over2}P} e^{-\sqrt{d-1\over2}\f}+\ldots,
$$
where we have defined
\be
G\equiv  \left(
(1-P) \sqrt{\dfrac{2\ell^2V_\infty}{d-1}}  
\left(\frac{\ell^{2d}Q^2}{\ell^2D}\right)^{\frac{1}{2d}} 
\,
\right)^{\frac{1}{1-P}}.
\label{s16-2}\ee

By using the identifications \eqref{UV9} and \eqref{th11}, we obtain the RG scale  dependence of the axion field in the IR:
\be
a=c\,{\theta(\mu) + 2\pi k \over N_c}
\sim \mu^{d-1} \left(\log \mu \right)^{-d P}.
\label{s17}\ee
Again, in the IR the running theta parameter decays to zero as a power-law of the energy scale $\mu$.

\end{enumerate}

\subsection{Numerical solutions for the background}\label{background_numerical}

In this subsection we will find explicit solutions to the bulk equations using numerical techniques. We will do so for both four-dimensional holographic QFTs ($d=4$) and three-dimensional holographic QFTs ($d=3$).

Without loss of generality, we set
\be
\ell=1.
\label{tn1-2}\ee
All the dimensionful quantities are evaluated in units of the UV AdS length $\ell$.
In the numerical calculations, we use \eqref{tn1-2} throughout the paper.

\subsubsection{Steep potentials}

 \begin{figure}[t]
 \begin{center}
  \includegraphics[width=.49\textwidth]{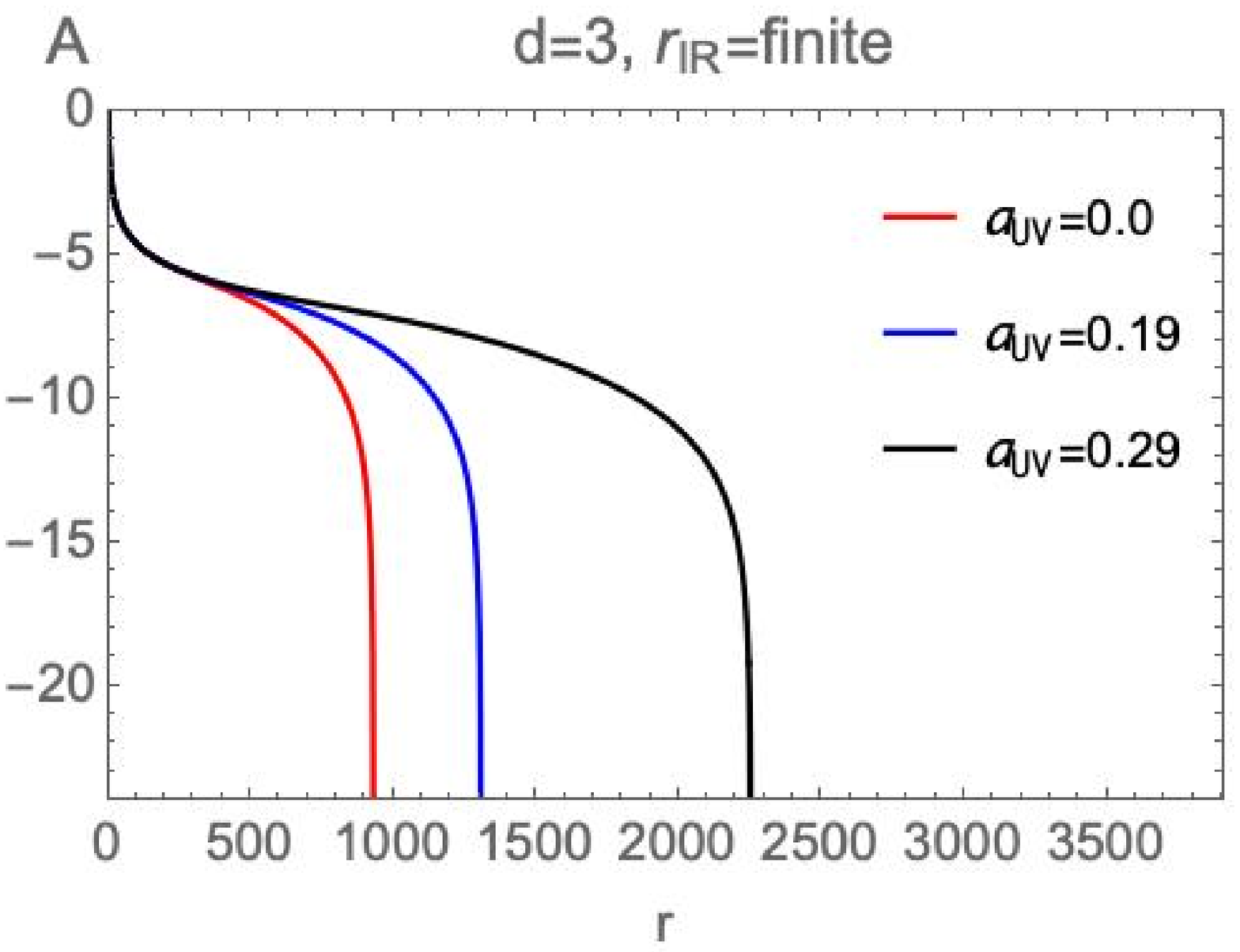}\hfil\hfil
   \includegraphics[width=.49\textwidth]{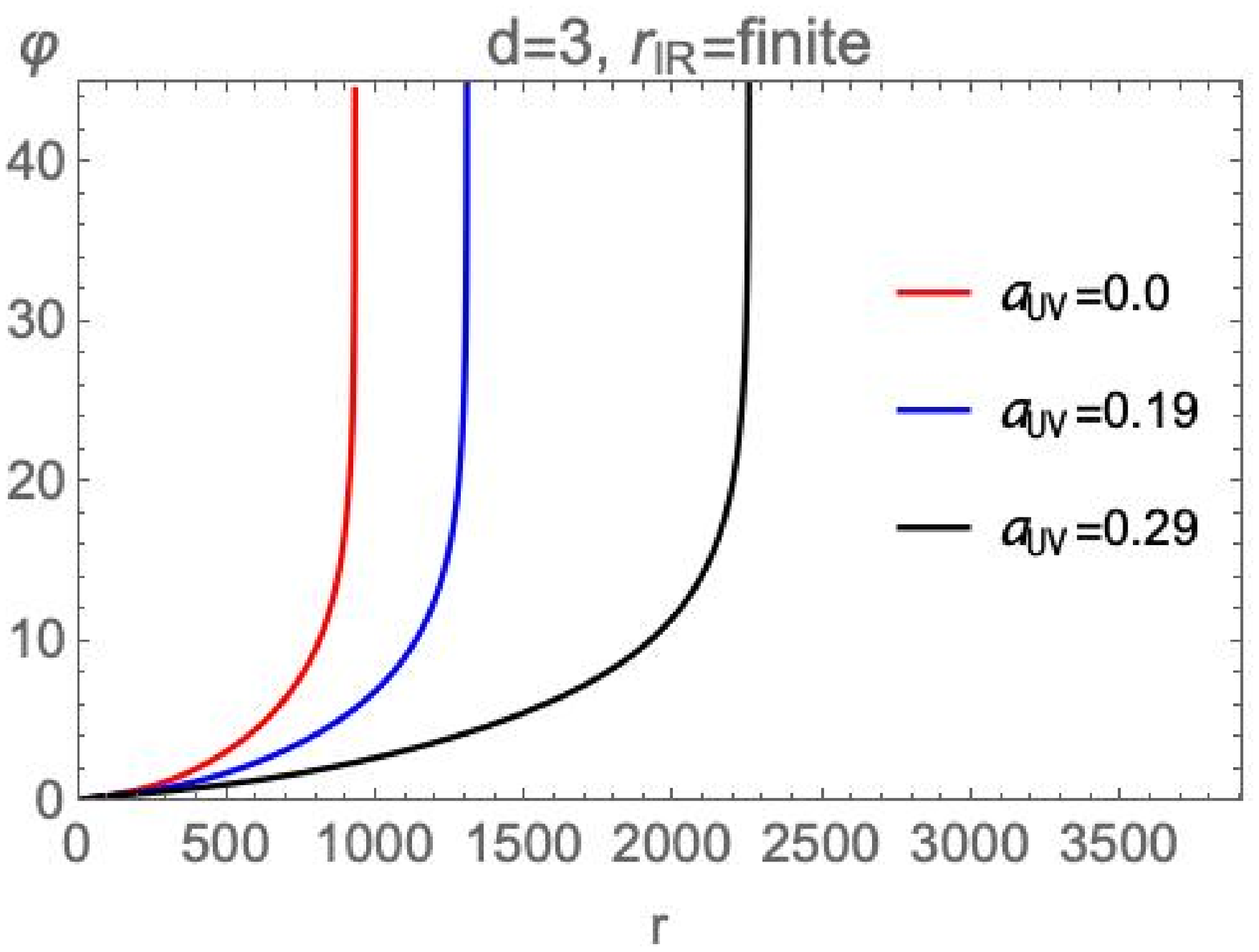}\hfil\hfil
 \end{center}
 \caption{Plots of $A$ (\textbf{left}) and $\f$ (\textbf{right}) as functions of $r$ with the bulk functions \protect\eqref{Num1}. The model parameters are \protect\eqref{4d2}.
 }
  \label{fig42}
 \end{figure}

 \begin{figure}[t]
 \begin{center}
  \includegraphics[width=.49\textwidth]{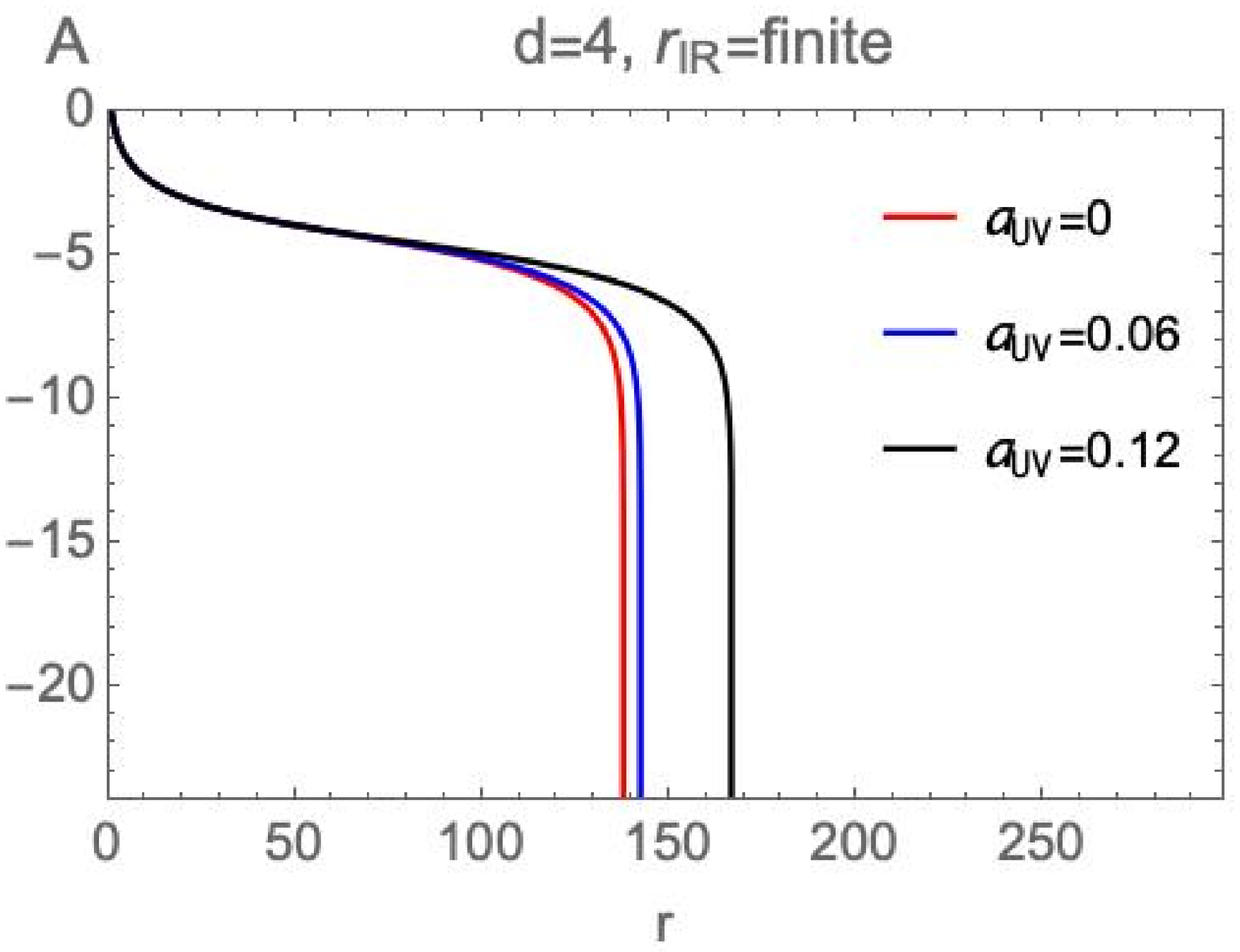}\hfil\hfil
   \includegraphics[width=.49\textwidth]{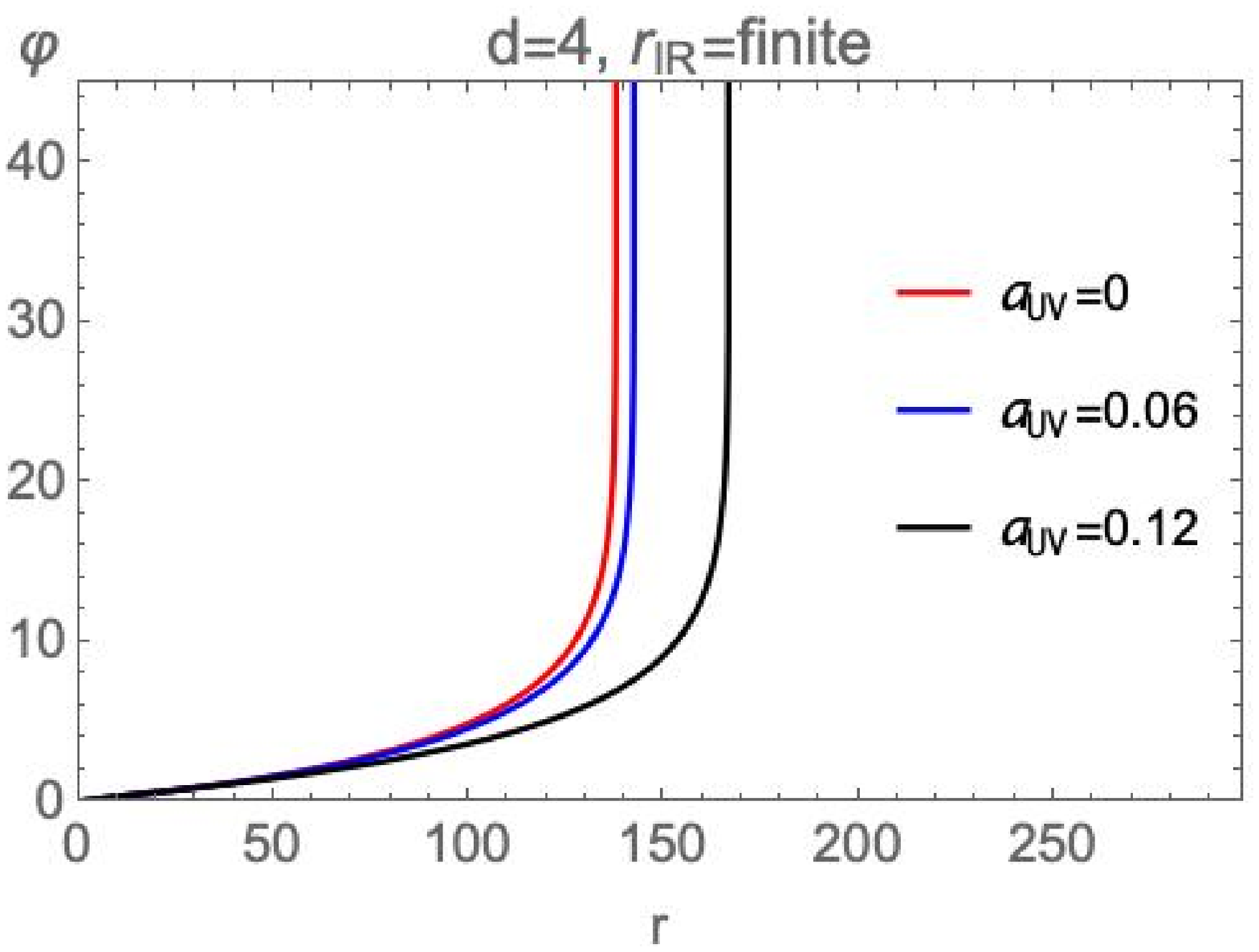}\hfil\hfil
 \end{center}
 \caption{Plots of $A$ (\textbf{left}) and $\f$ (\textbf{right}) as functions of $r$ with the bulk functions \protect\eqref{Num1}. The model parameters are \protect\eqref{4d2-2}.
 }
  \label{fig42-2}
 \end{figure}

  \begin{figure}[t]
 \begin{center}
  \includegraphics[width=.49\textwidth]{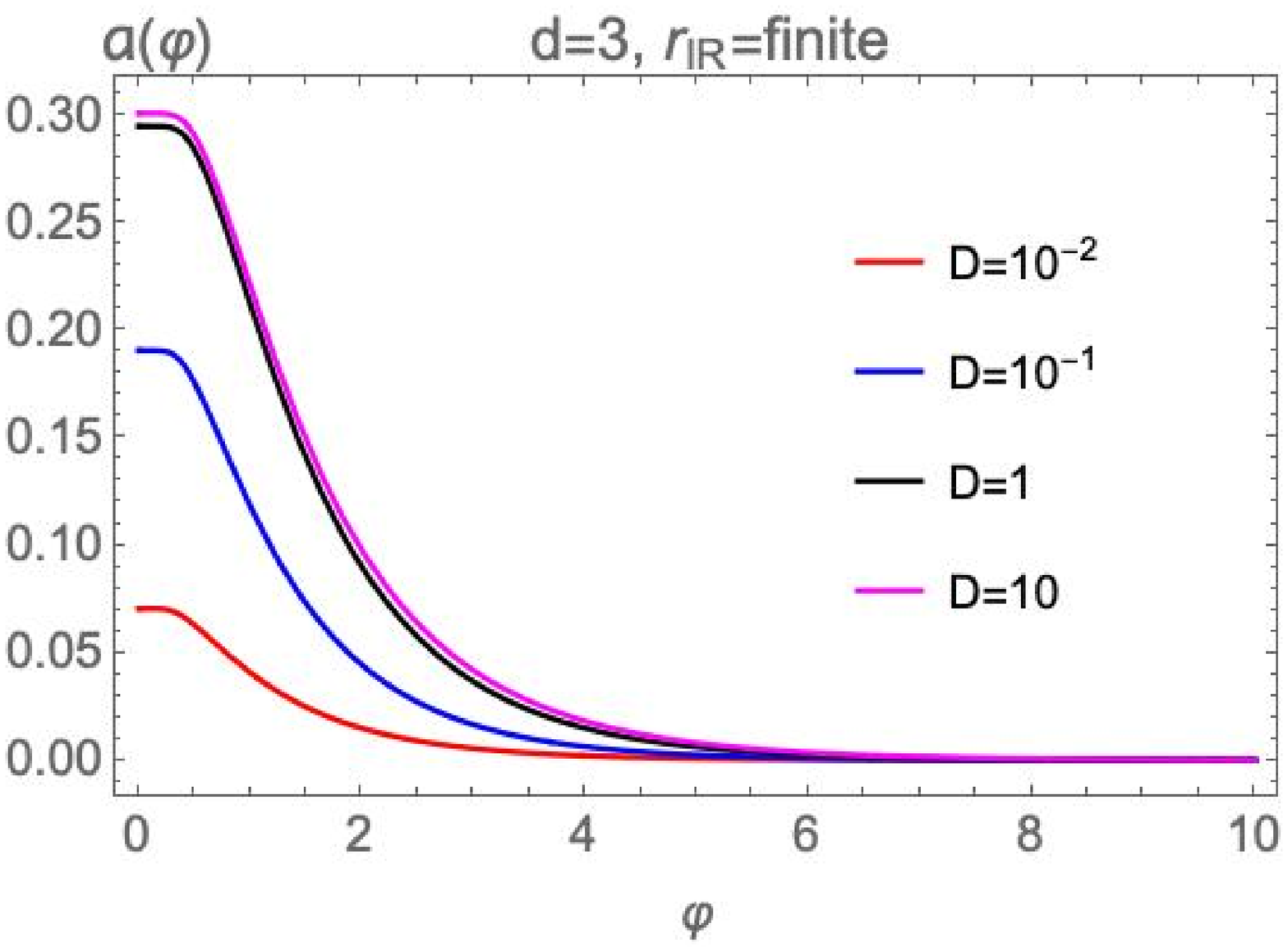}\hfil\hfil
  \includegraphics[width=.49\textwidth]{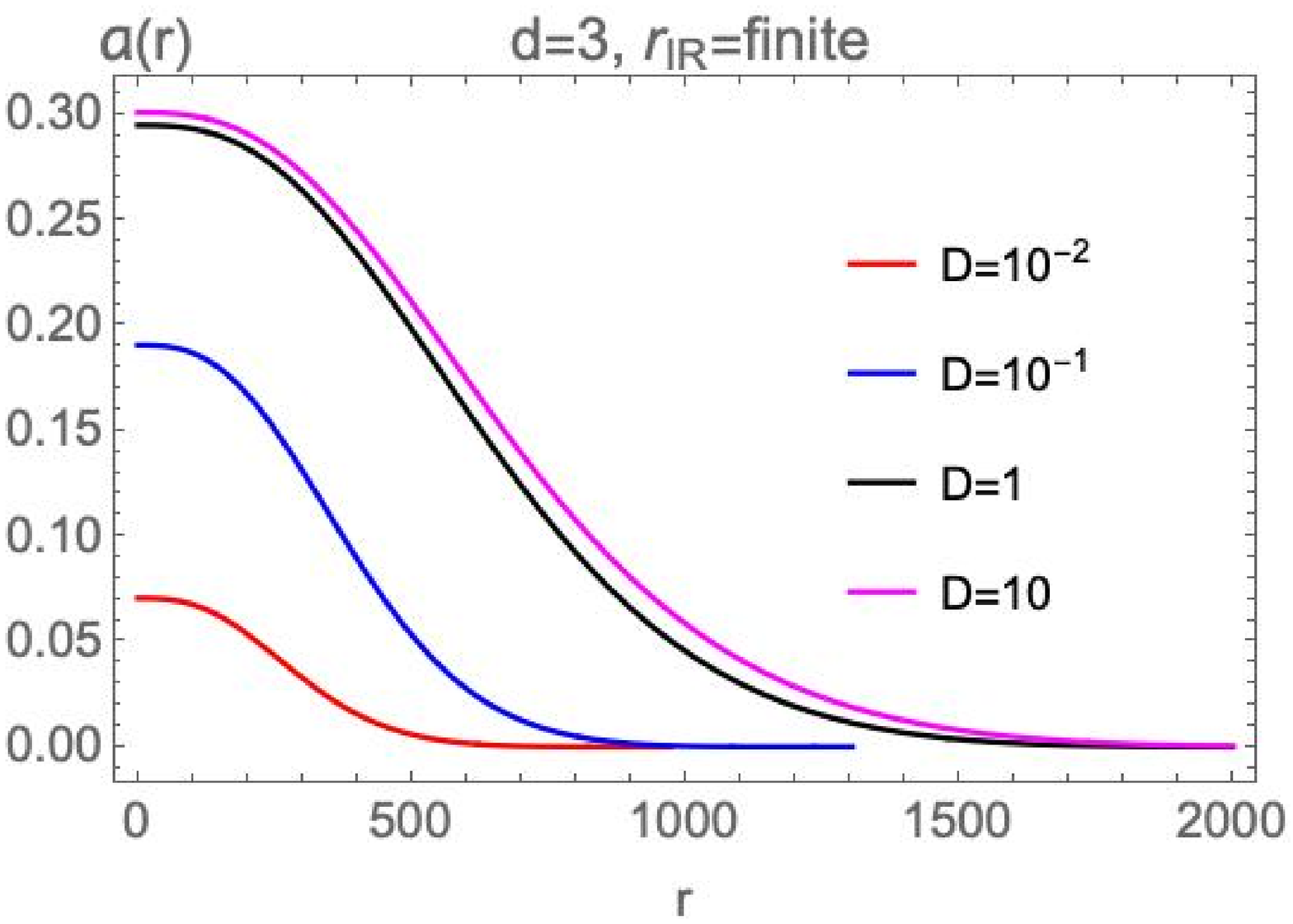}\hfil\hfil
 \end{center}
 \caption{Plots of $a(\f)$ as a function of $\f$ (\textbf{left}) and $a(r)$ as a function of $r$ (\textbf{right}) with the bulk functions \protect\eqref{Num1}. The model parameters are \protect\eqref{4d2}.
  }
  \label{fig12}
 \end{figure}

  \begin{figure}[t]
 \begin{center}
  \includegraphics[width=.49\textwidth]{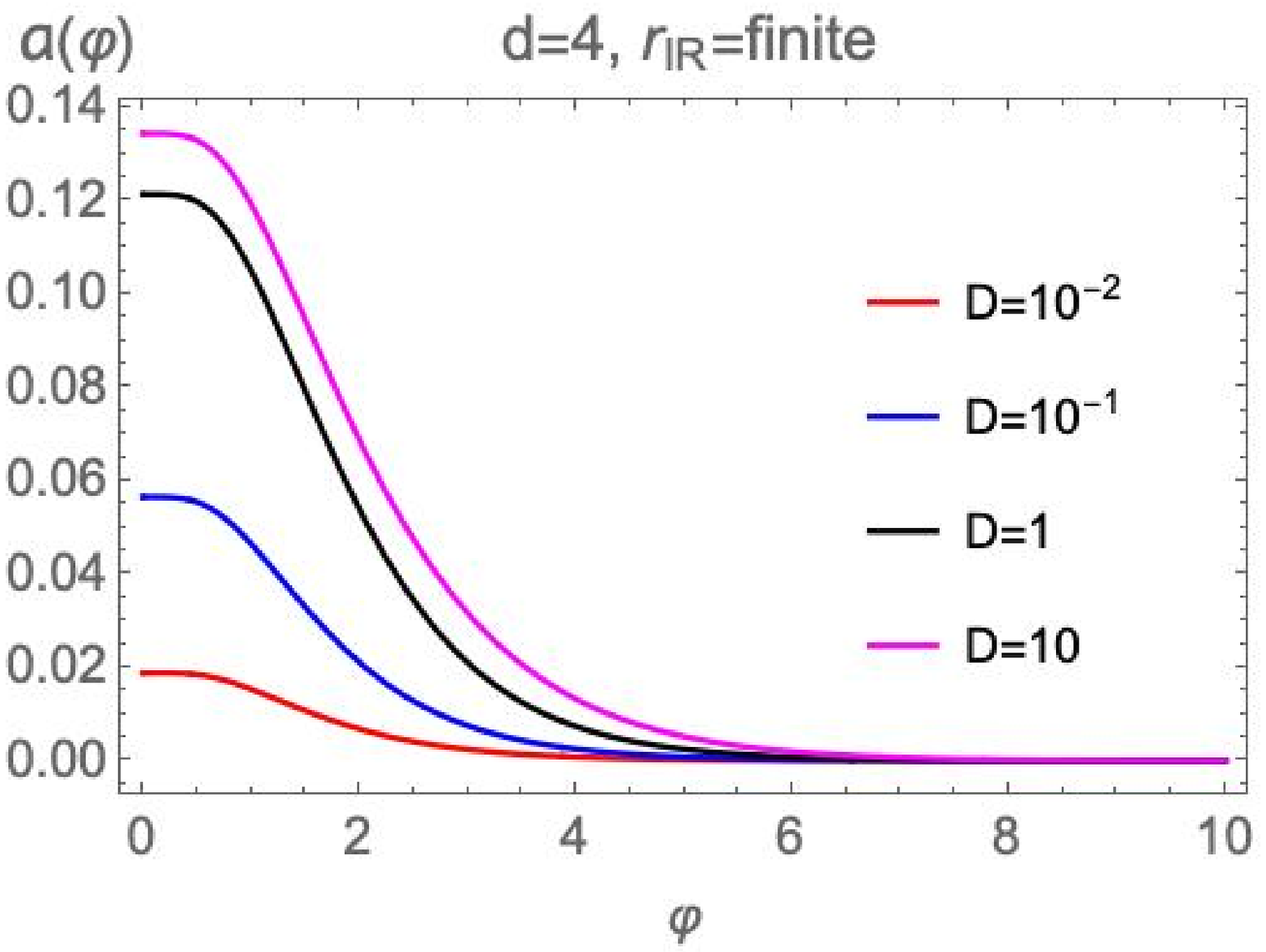}\hfil\hfil
  \includegraphics[width=.49\textwidth]{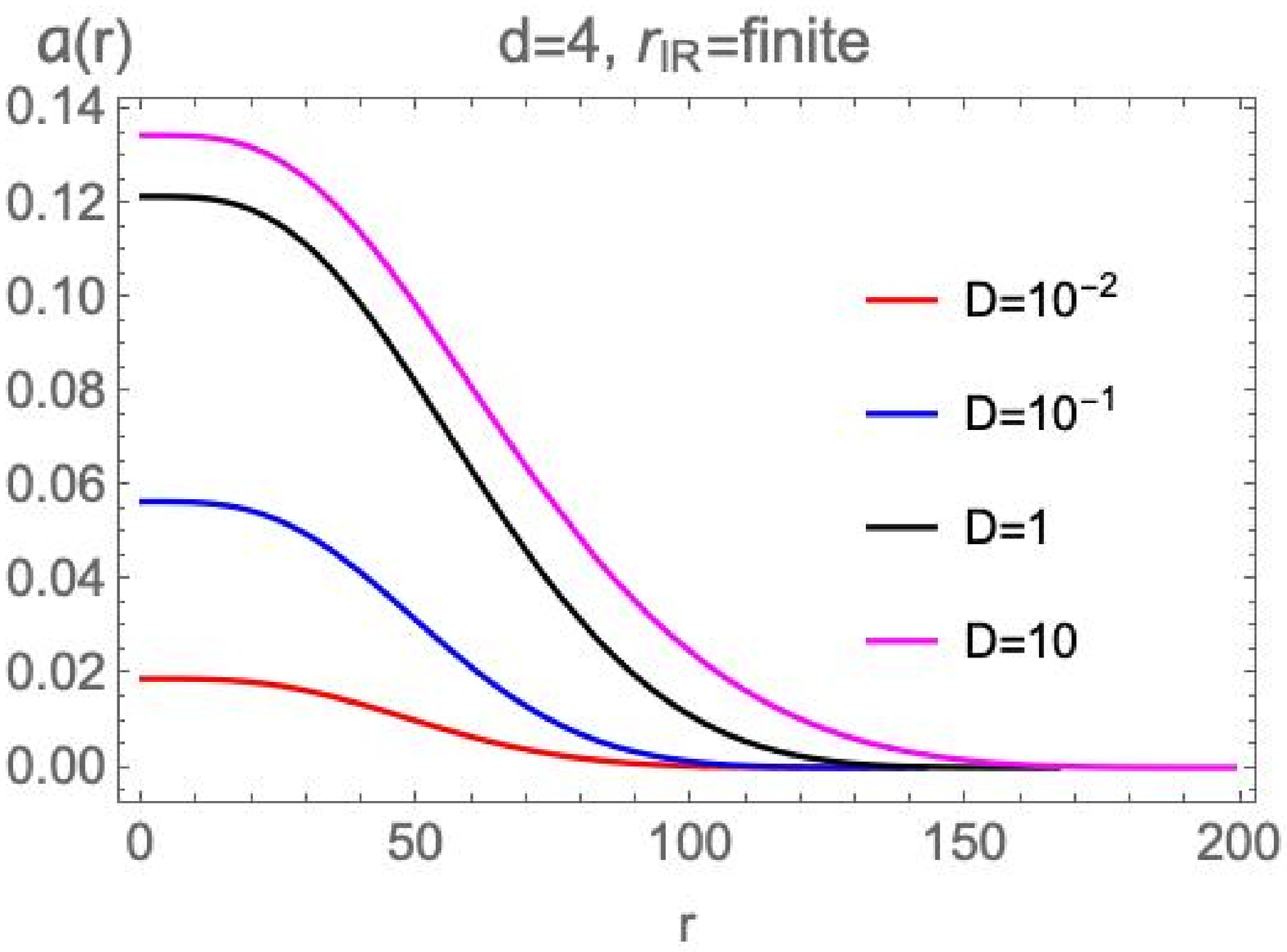}\hfil\hfil
 \end{center}
 \caption{Plots of $a(\f)$ as a function of $\f$ (\textbf{left}) and $a(r)$ as a function of $r$ (\textbf{right}) with the bulk functions \protect\eqref{Num1}. The model parameters are \protect\eqref{4d2-2}.
  }
  \label{fig12-2}
 \end{figure}

  \begin{figure}[t]
 \begin{center}
  \includegraphics[width=.49\textwidth]{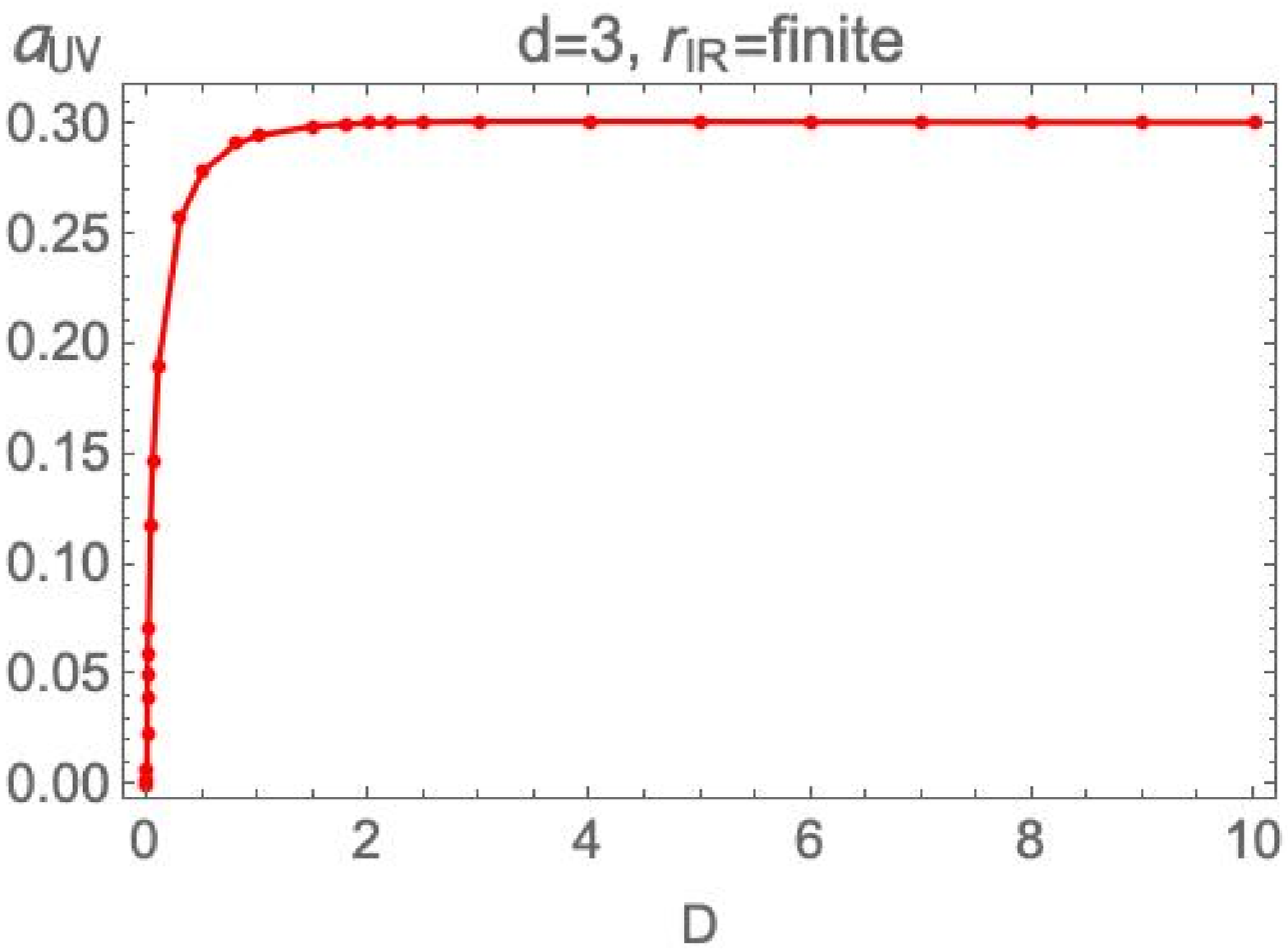}\hfil\hfil
   \includegraphics[width=.49\textwidth]{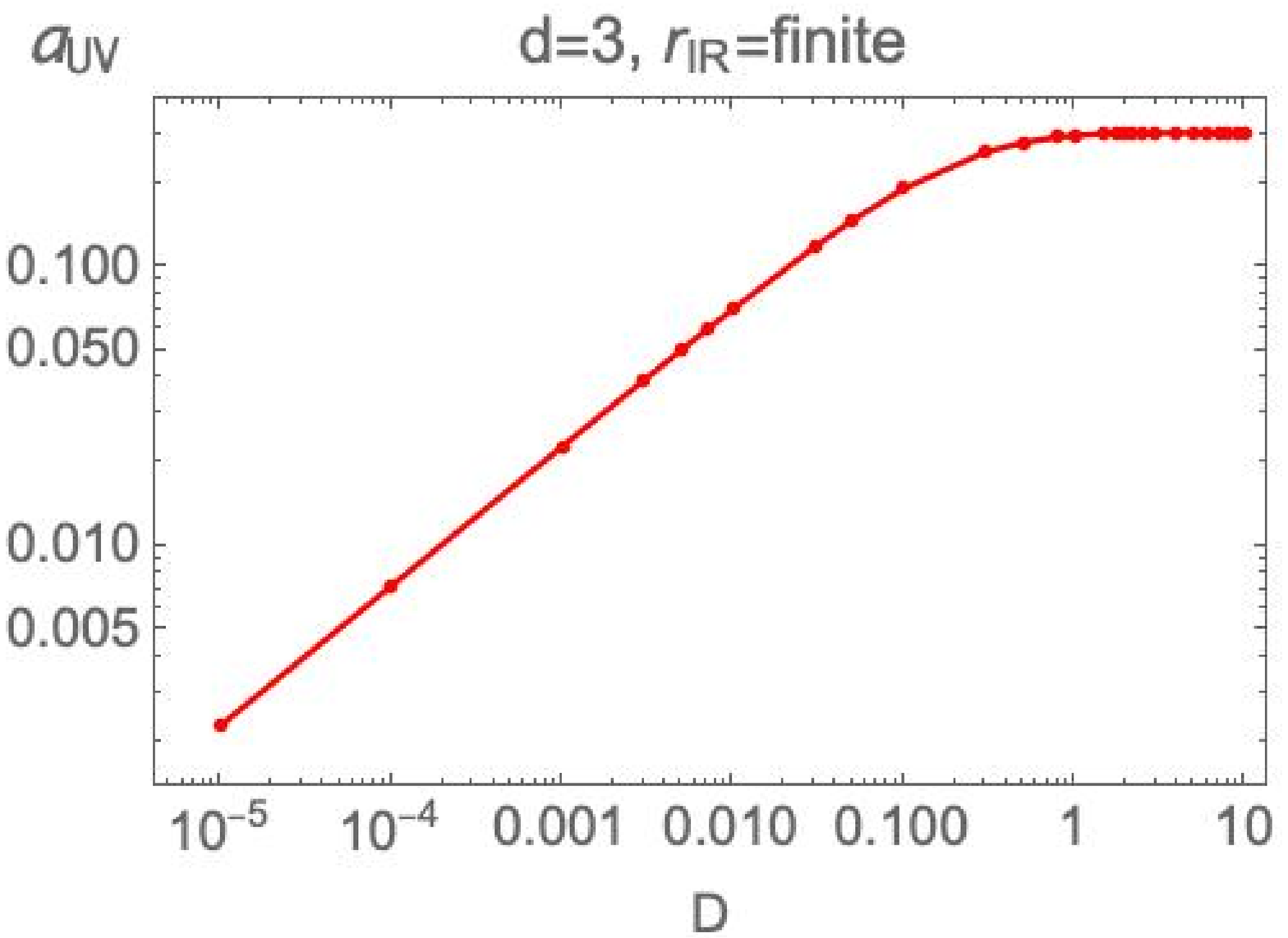}\hfil\hfil
 \end{center}
 \caption{Linear-linear (\textbf{left}) and log-log (\textbf{right}) plots of $a_{UV}$ as functions of the IR integration constant $D$,  with the bulk functions \protect\eqref{Num1}. The model parameters are \protect\eqref{4d2}. For $D\lesssim0.1$, the relation is roughly given by $a_{UV}\approx 0.7 \sqrt{D}$.
 }
  \label{fig13}
 \end{figure}

  \begin{figure}[t]
 \begin{center}
  \includegraphics[width=.49\textwidth]{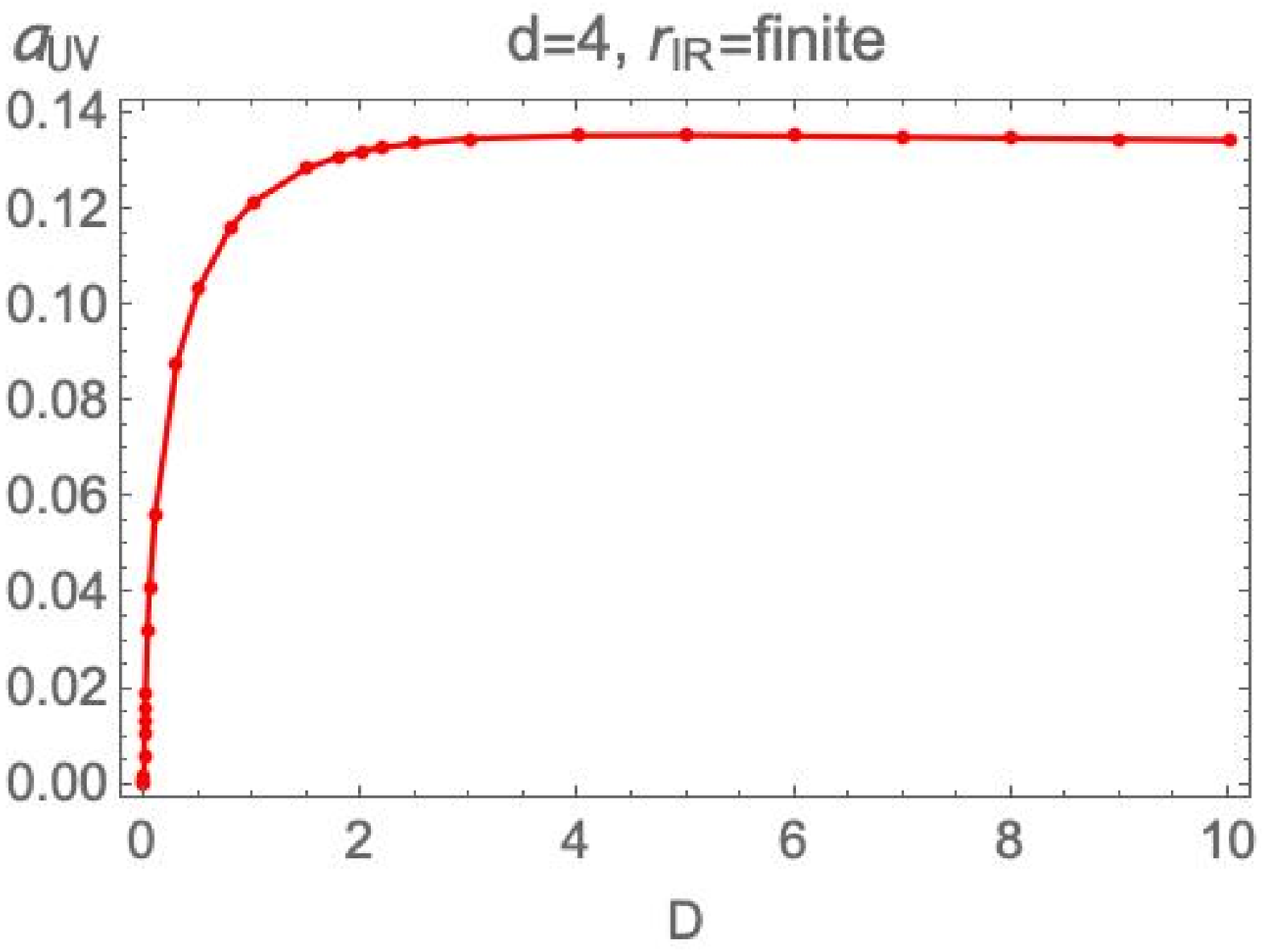}\hfil\hfil
   \includegraphics[width=.49\textwidth]{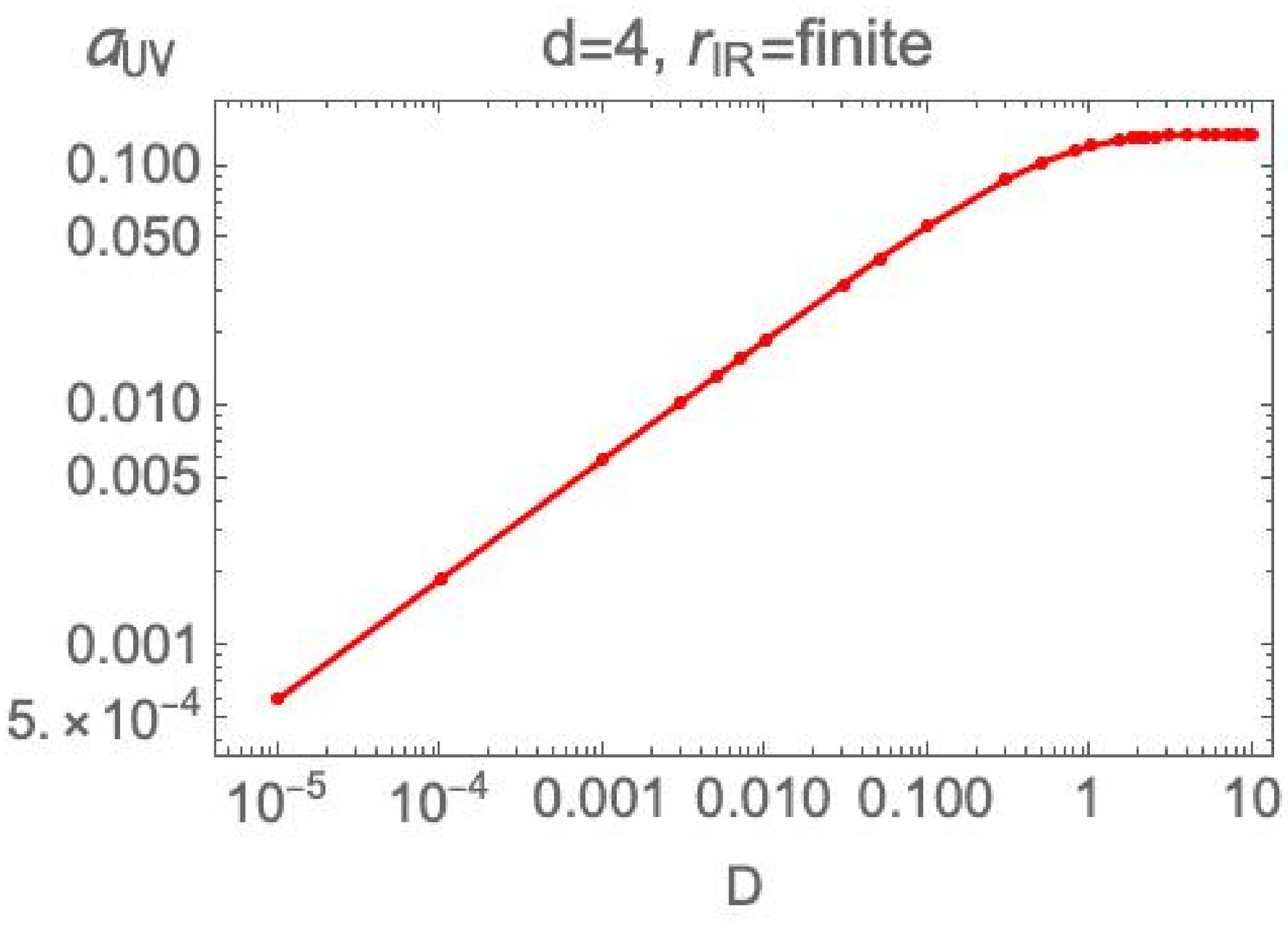}\hfil\hfil
 \end{center}
 \caption{Linear-linear (\textbf{left}) and log-log (\textbf{right}) plots of $a_{UV}$ as functions of $D$ with the bulk functions \protect\eqref{Num1}. The model parameters are \protect\eqref{4d2-2}. For $D\lesssim0.1$, the relation is roughly given by $a_{UV}\approx 0.2 \sqrt{D}$.
 }
  \label{fig13-2}
 \end{figure}

   \begin{figure}[t]
 \begin{center}
  \includegraphics[width=.49\textwidth]{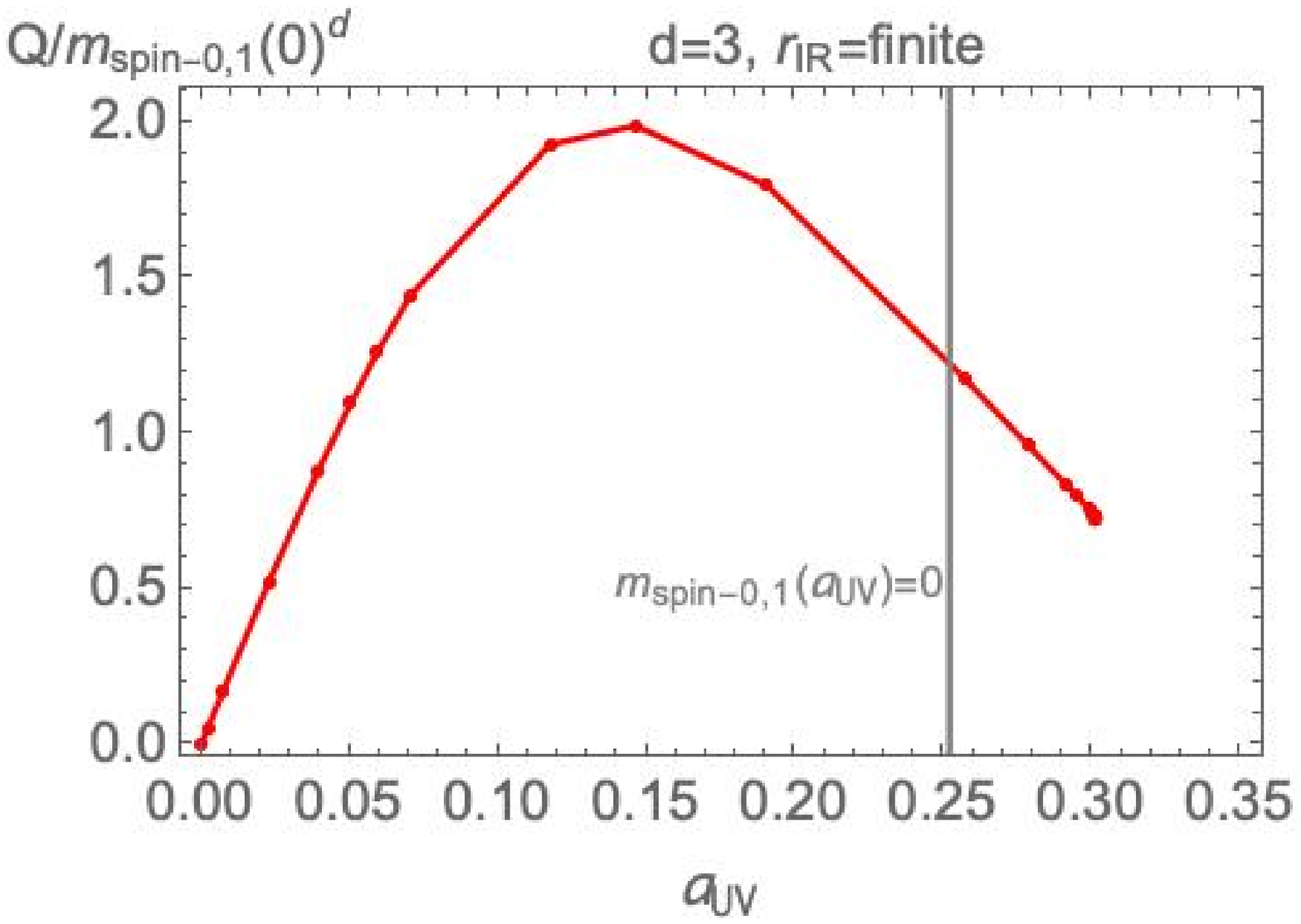}\hfil\hfil
  \includegraphics[width=.49\textwidth]{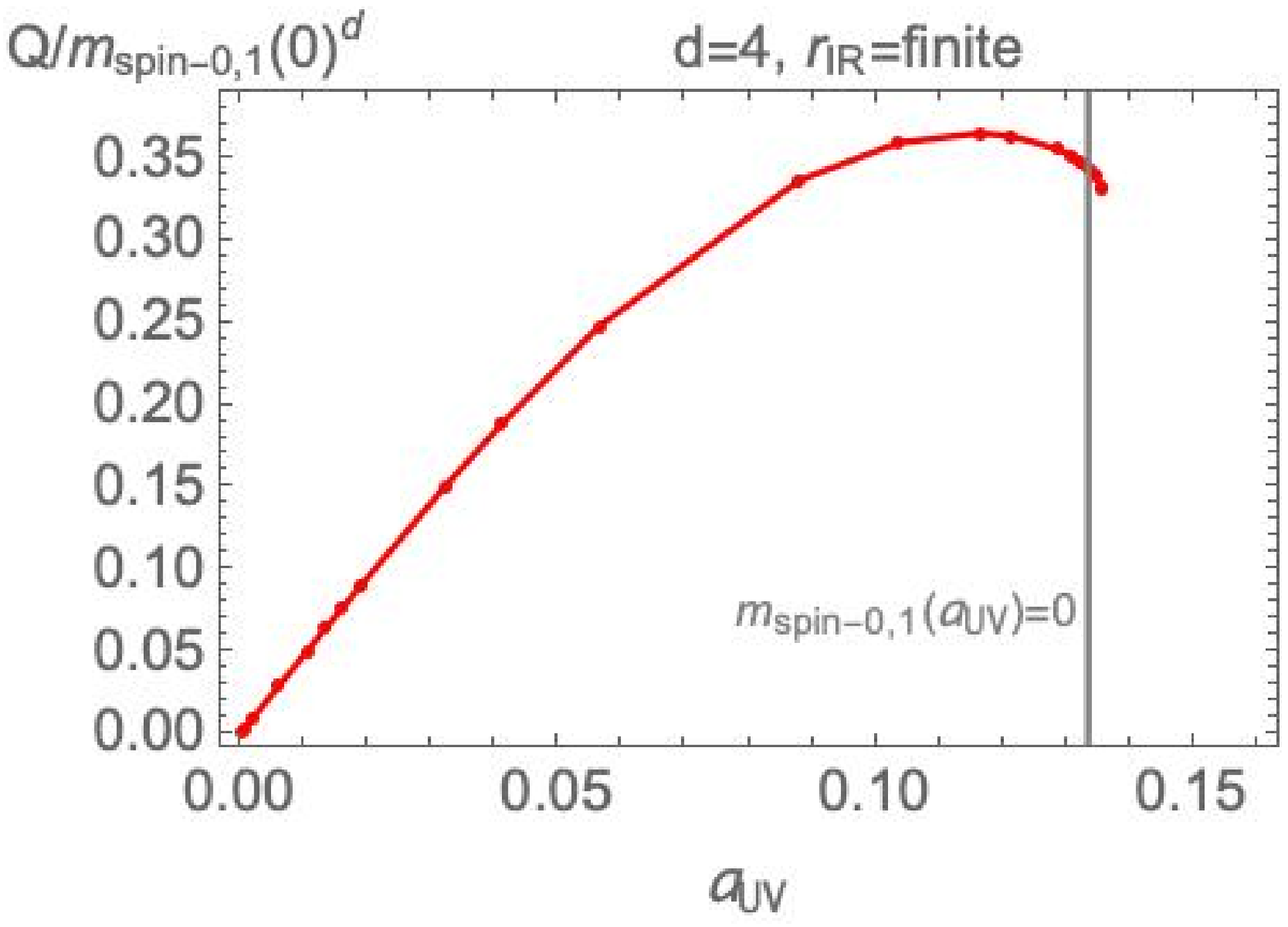}\hfil\hfil
\\
  \includegraphics[width=.49\textwidth]{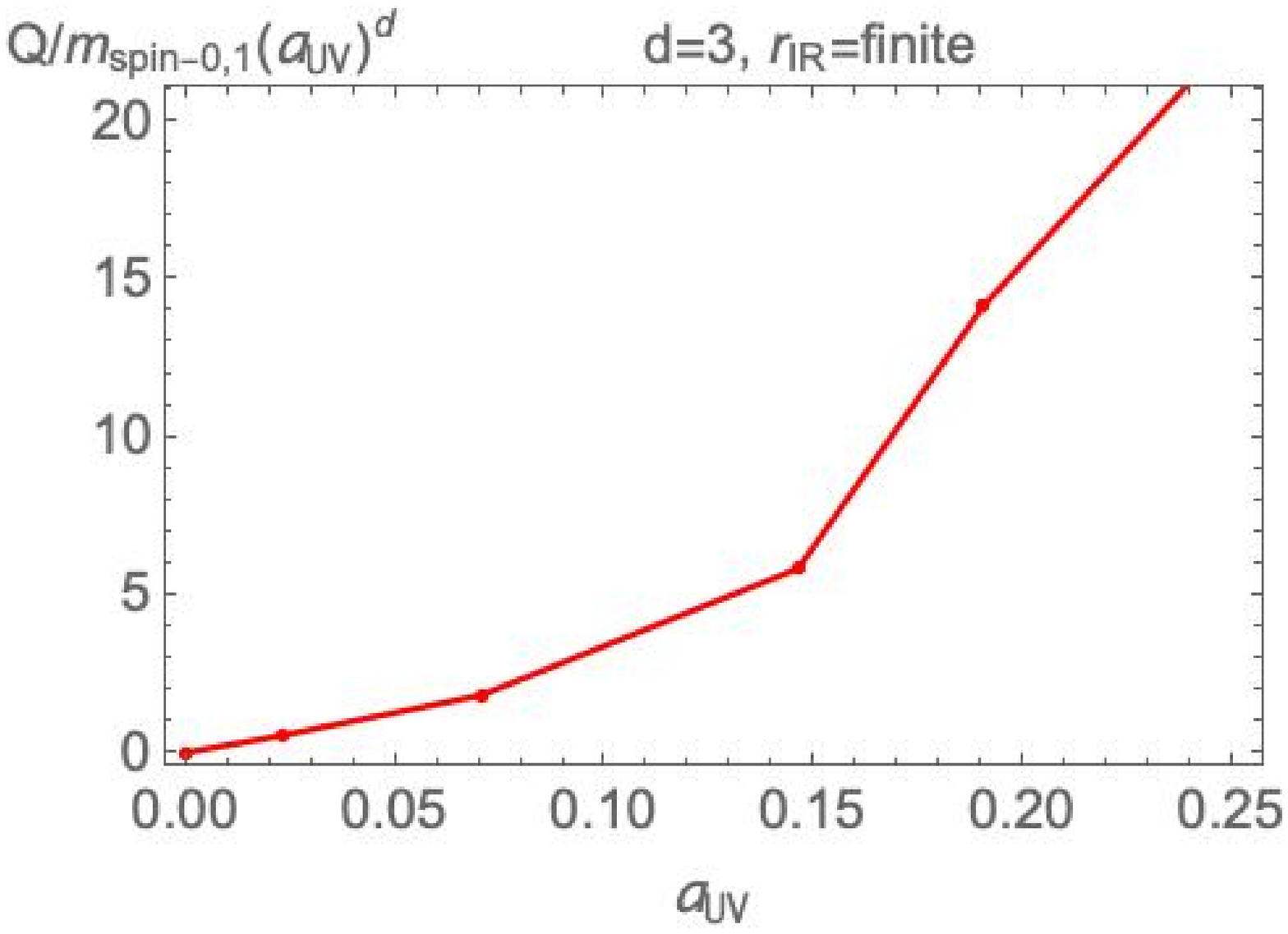}\hfil\hfil
  \includegraphics[width=.49\textwidth]{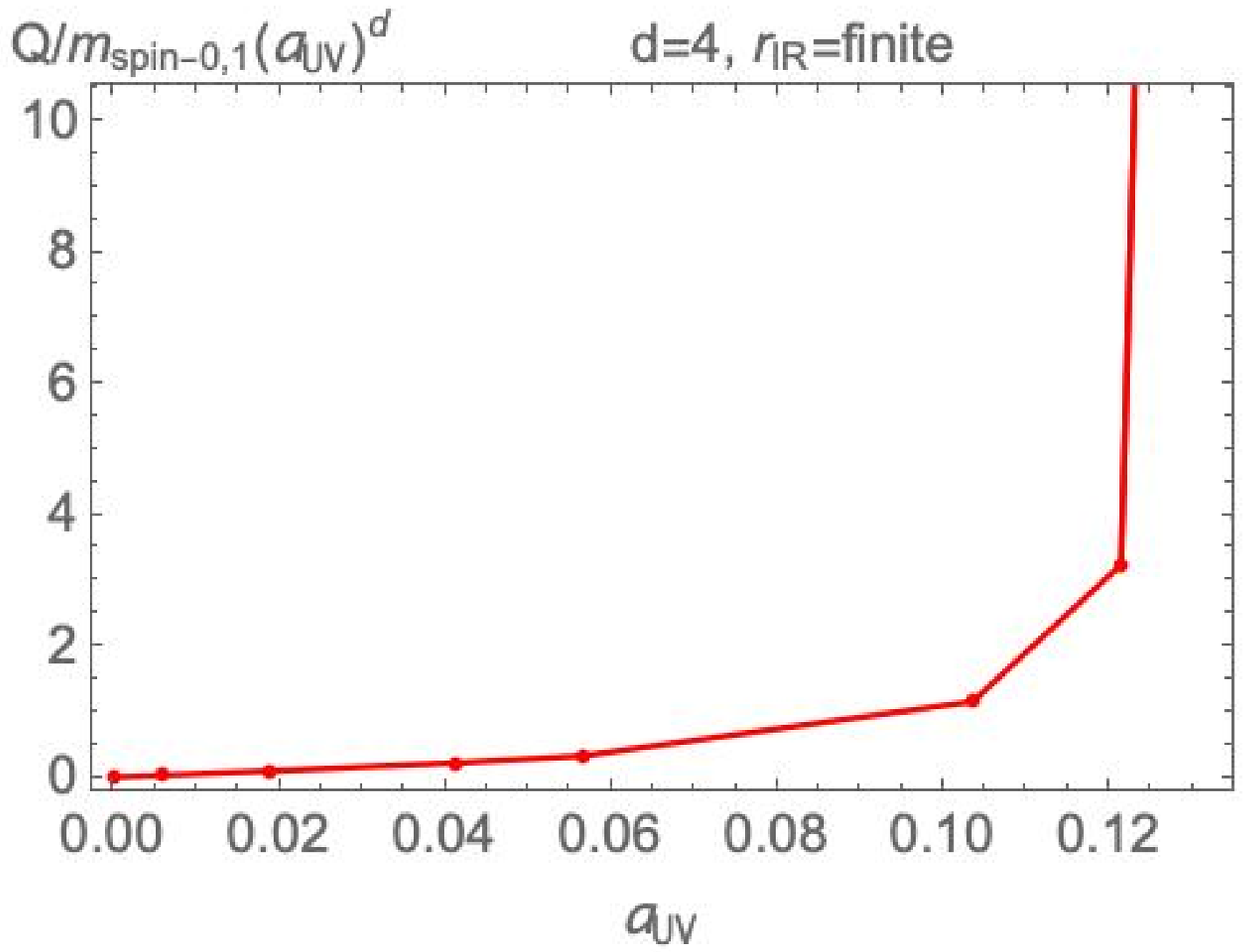}\hfil\hfil
 \end{center}
 \caption{Plots of $Q$ (that determines the vev of the instanton density) normalized by lightest glueball masses (which we shall calculate in section \protect\ref{numerical_glueball}) as functions of $a_{UV}$ with the bulk functions \protect\eqref{Num1}.
 \textbf{Upper left:} The model parameters are \protect\eqref{4d2} ($a_{UV}^{max}\approx0.3$). $Q$ is normalized by the lightest glueball mass at $a_{UV}=0$.
 \textbf{Upper right:} The model parameters are \protect\eqref{4d2-2} ($a_{UV}^{max}\approx0.14$). $Q$ is normalized by the lightest glueball mass at $a_{UV}=0$.
 \textbf{Lower left:} The model parameters are \protect\eqref{4d2}. $Q$ is normalized by the lightest glueball mass at given $a_{UV}$.
 \textbf{Lower right:} The model parameters are \protect\eqref{4d2-2}. $Q$ is normalized by the lightest glueball mass at given $a_{UV}$.}
  \label{fig3-2}
 \end{figure}

The bulk potential we use is
\be
V=
-d(d-1)-\left({1\over2}(d-\Delta_-)\Delta_--b^2V_\infty\right)\f^2-4V_\infty \sinh^2\left(b\f\over2\right),
\qquad
Y=Y_\infty  e^{\gamma\f},
\label{Num1}\ee
Here $\Delta\equiv d-\Delta_-$ is the UV dimension of the operator dual to $\f$.

We use \eqref{Num1} with
\be
d=3, \quad b=1.2, \quad \g=1.6, \quad \Delta_-=0.4, \quad \f_-=5\times10^{-2}, \quad \text{sign}(Q)=-1,\quad V_\infty=1,
\label{4d2}\ee
and
\be
d=4, \quad b=1.1, \quad \g=1.8, \quad \Delta_-=0.8, \quad \f_-=5\times10^{-2}, \quad \text{sign}(Q)=-1,\quad V_\infty=1.
\label{4d2-2}\ee
Although our primary interest is $d=4$ theories, we will analyse also $d=3$ theories as the holographic dynamics turns out to be qualitatively similar.
We have explored many other numerical solutions but we present here, typical ones.

The similarity of $d=3$ and $d=4$ Einstein-axion-dilaton theories will be useful as  in section \ref{cubic}, we shall estimate the strength of CP-violating couplings for $d=3$, where an action up to the third order in the fluctuations is known in the context of inflationary cosmology.

The scale factor $A$ and  scalar field $\f$ are plotted as functions of $r$ in figures \ref{fig42} (model parameters \eqref{4d2}, $d=3$) and \ref{fig42-2} (model parameters \eqref{4d2-2}, $d=4$). As we commented above the equation \eqref{UV1}, the scalar field $\f$ is zero at the UV, $r_{UV}=0$. As for the scale factor $A$, we determine an integration constant of $A$ by choosing $\left.A\right|_{r=1}=0$ in the numerical calculation. The constant shift of $A$ does not have a physical meaning.

In both model parameters \eqref{4d2} and \eqref{4d2-2}, we observe that $A$ is a monotonically decreasing function of $r$ while $\f$ is a monotonically increasing function.
This means that we can use $A$ or $\f$ as a coordinate instead of using $r$.
In fact, we shall use $A$ as a coordinate, when we compute the glueball spectra numerically.
Both $A$ and $\f$ diverge at finite $r$ here, which corresponds to the IR end-point.

The profiles of the axion are shown  in figures \ref{fig12} ($d=3$) and \ref{fig12-2} ($d=4$). These can be interpreted as the holographic renormalization group flow of the $\theta$-angle in the dual QFT.
 In holography, the $\theta$-angle flows to zero in the IR ($\f\to\infty$).\footnote{The IR vanishing of the $\theta$-angle has been  discussed in holography \cite{iQCD} and QFT\cite{Knizhnik:1984kn,Levine:1984pv,Latorre:1997af,Nakamura:2019ind}.}
The integration constant $D$ and the axion source $a_{UV}$ are related through the IR regularity condition \eqref{a10}. This relation is given in figure \ref{fig13} ($d=3$) and figure \ref{fig13-2} ($d=4$) numerically. As observed in \cite{axion-rg}, regularity of the axionic flow imposes an upper bound on $a_{UV}$, which we denote by $a_{UV}^{max}$.
The values of $a_{UV}^{max}$ are
\be
a_{UV}^{max} \approx
\begin{cases}
0.3 \quad \text{for \eqref{4d2}}, \\
0.14 \quad \text{for \eqref{4d2-2}}.
\end{cases}
\label{Num2}\ee
For small $D$, the axion source $a_{UV}$ is proportional to $\sqrt{D}$.

One of the important observables is the non-trivial vev of the instanton density operator, which can be holographically computed by using the right expression in \eqref{UV27}.
The values of $Q$ as functions of $a_{UV}$ are plotted in figure \ref{fig3-2}. The left and right panels correspond to $d=3$ and $d=4$, respectively.
In lower panels, $Q$ is normalized by the lightest glueball mass at given $a_{UV}$ which we denote by $m_{\text{spin-}0,1}(a_{UV})$. This shall be calculated in section \ref{numerical_glueball}.
In upper panels, $Q$ is normalized by $m_{\text{spin-}0,1}(0)$. The vertical lines in the upper panels correspond to the value of $a_{UV}$ where the lightest glueball becomes massless.

In both upper panels, we observe that $Q$ is proportional to $a_{UV}$ for small $a_{UV}$.
As $a_{UV}$ is increased, $Q$ reaches a maximum, and then decreases. The rightmost point corresponds to $a_{UV}^{max}$.
In lower panels, we observe that $Q/m_{\text{spin-}0,1}^d(a_{UV})$ diverges at some points because the lightest glueballs become massless.

To summarize, in model parameters \eqref{4d2} and \eqref{4d2-2}, we observed the qualitatively similar results for the backgrounds (scalar field, axion field, scale factor) and the vev of the instanton density.

\subsubsection{Soft potentials}
 \begin{figure}[t]
 \begin{center}
  \includegraphics[width=.49\textwidth]{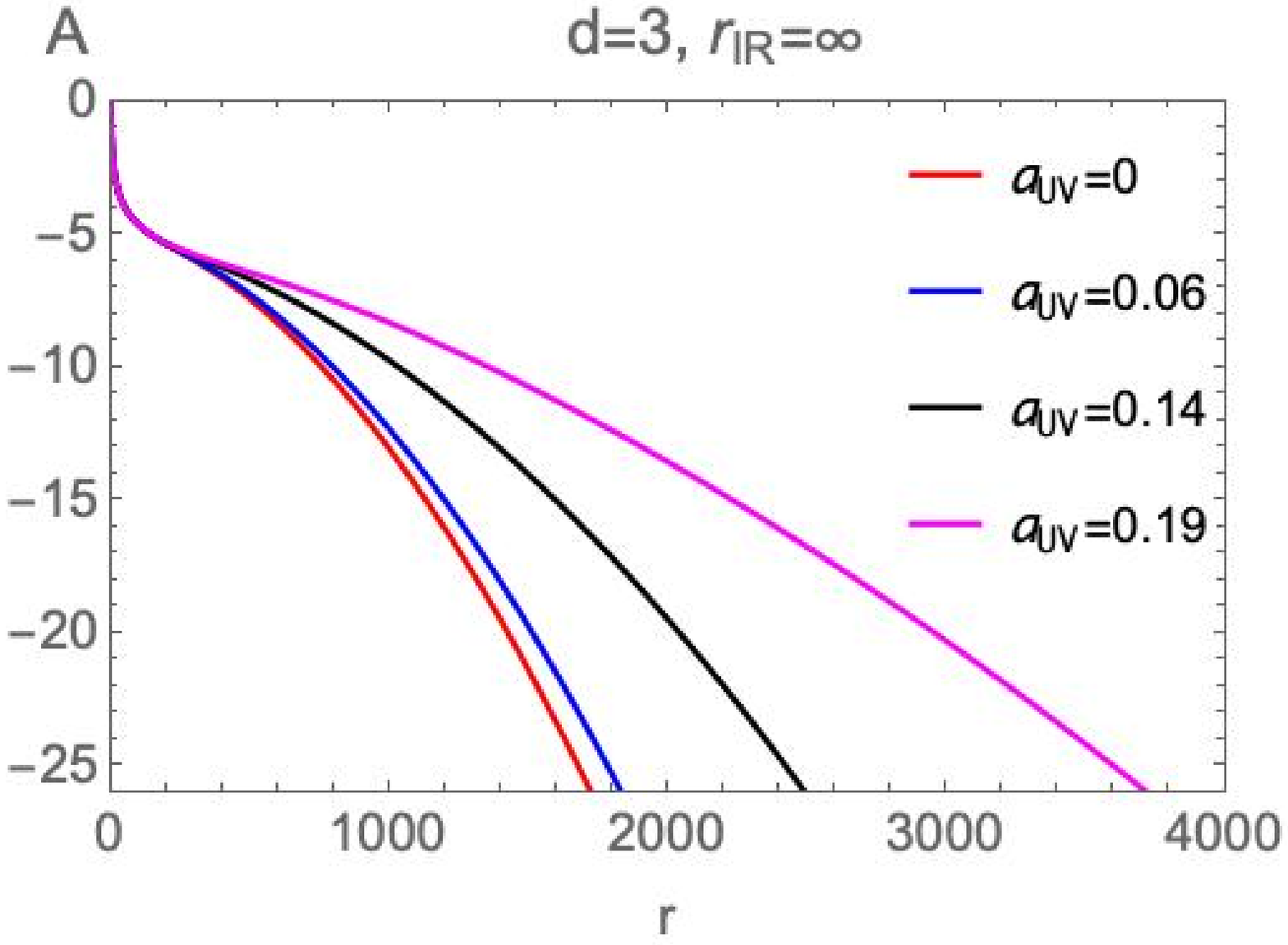}\hfil\hfil
   \includegraphics[width=.49\textwidth]{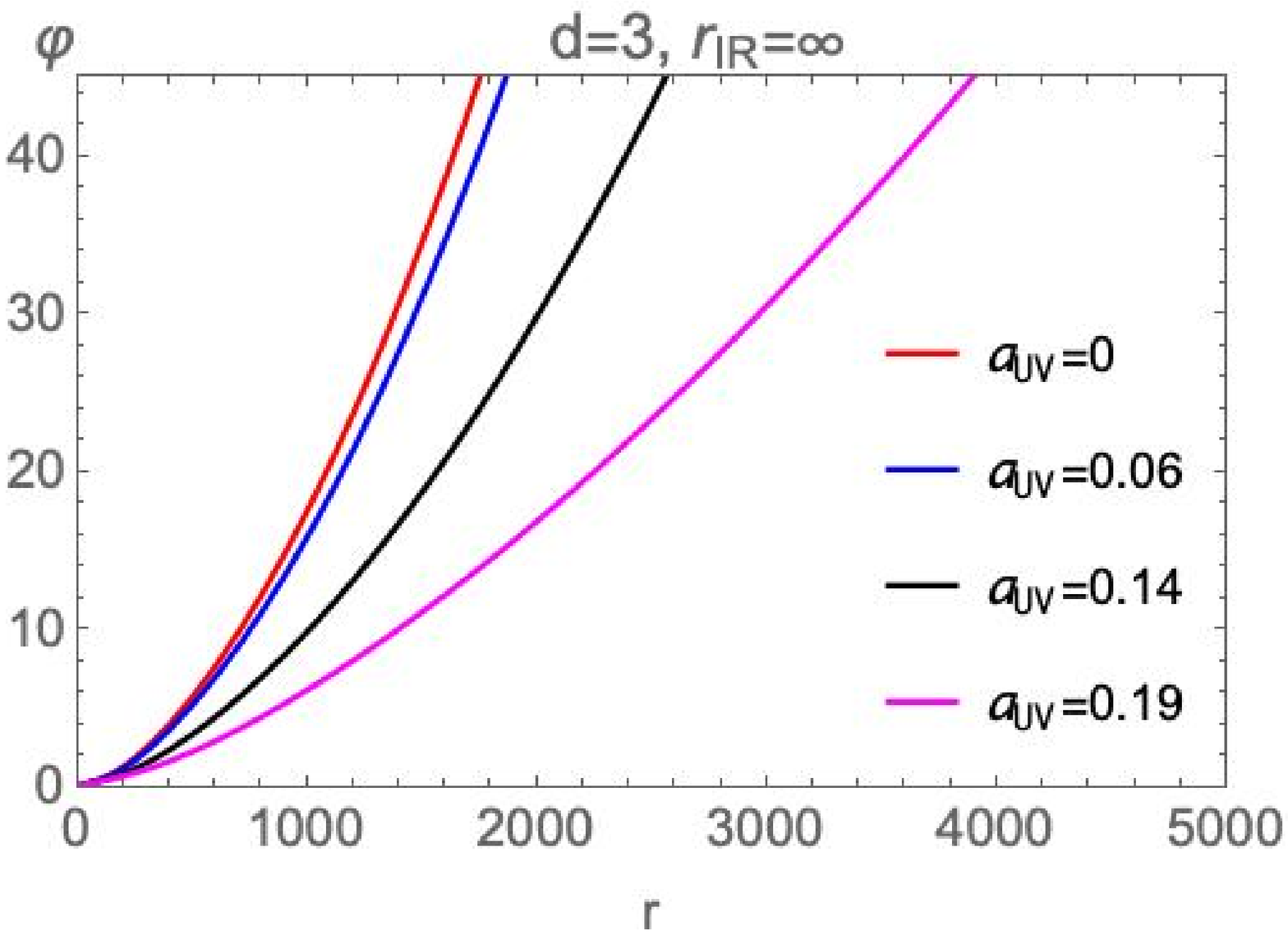}\hfil\hfil
 \end{center}
  \caption{Plots of $A$ (\textbf{left}) and $\f$ (\textbf{right}) as functions of $r$ with the bulk functions \protect\eqref{P1}. The model parameters are \protect\eqref{4d1}.
  }
  \label{fig45}
 \end{figure}

 \begin{figure}[t]
 \begin{center}
  \includegraphics[width=.49\textwidth]{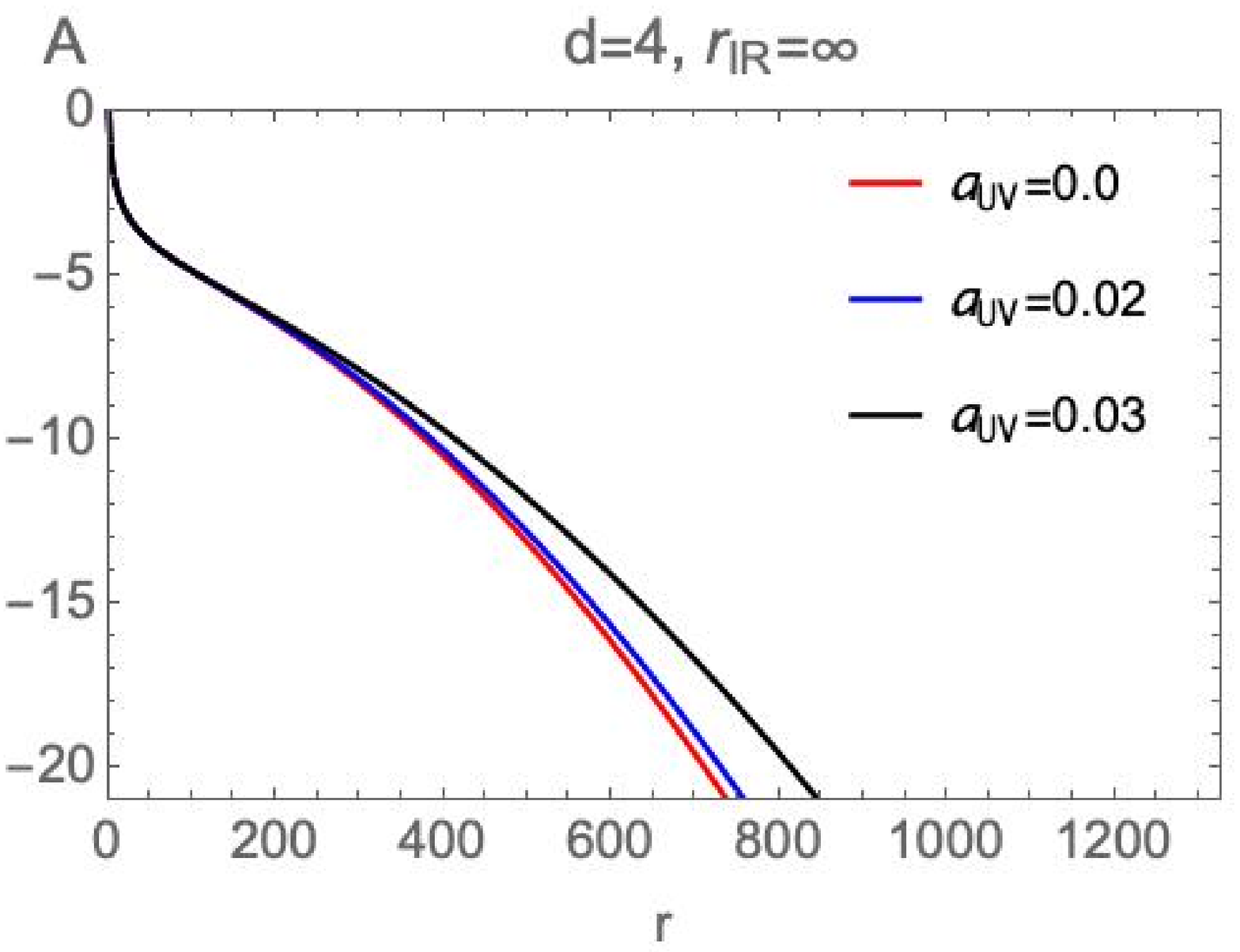}\hfil\hfil
   \includegraphics[width=.49\textwidth]{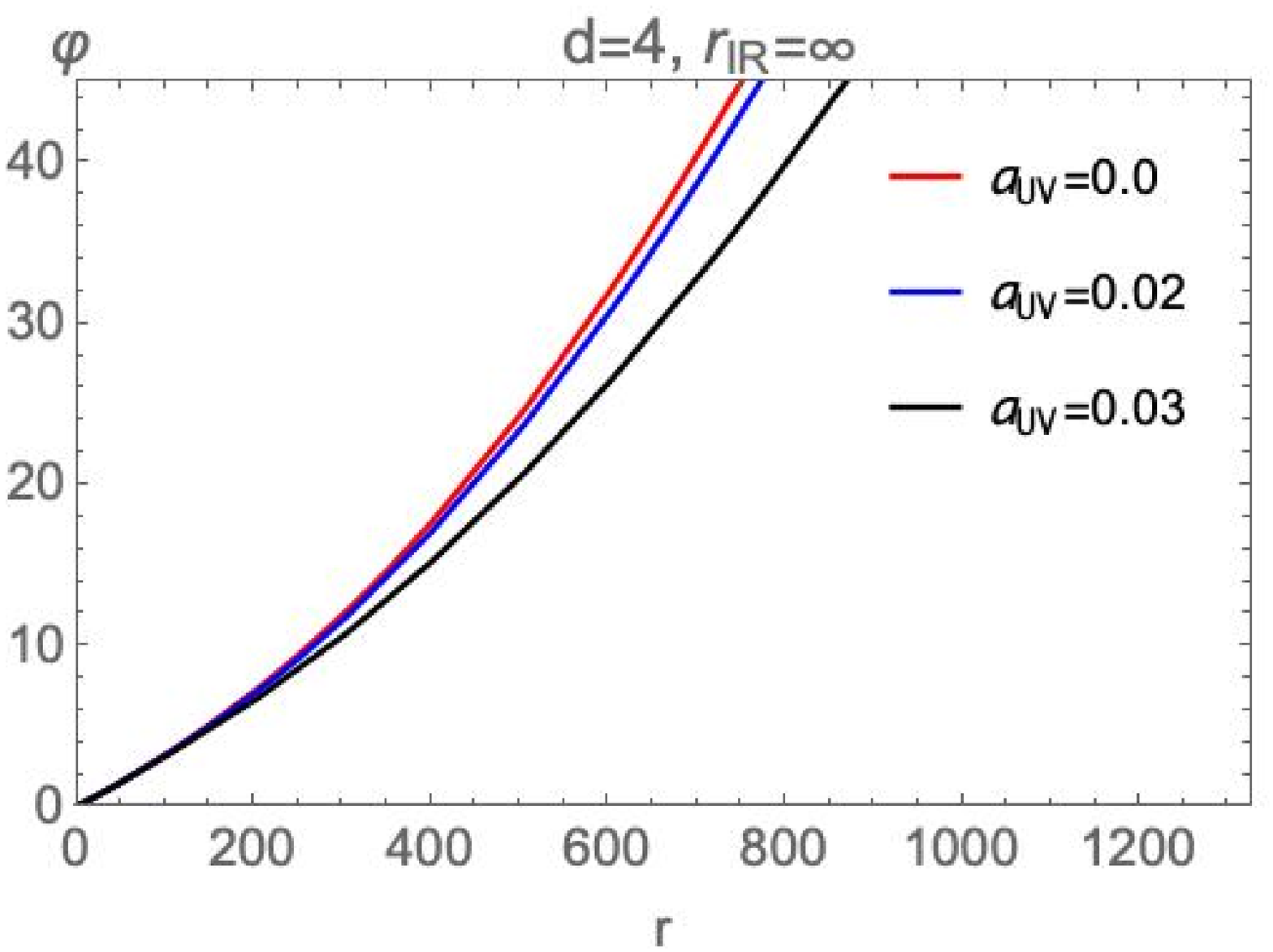}\hfil\hfil
 \end{center}
 \caption{Plots of $A$ (\textbf{left}) and $\f$ (\textbf{right}) as functions of $r$ with the bulk functions \protect\eqref{P1}. The model parameters are \protect\eqref{4d5}.
  }
  \label{fig4}
 \end{figure}

 \begin{figure}[t]
 \begin{center}
  \includegraphics[width=.49\textwidth]{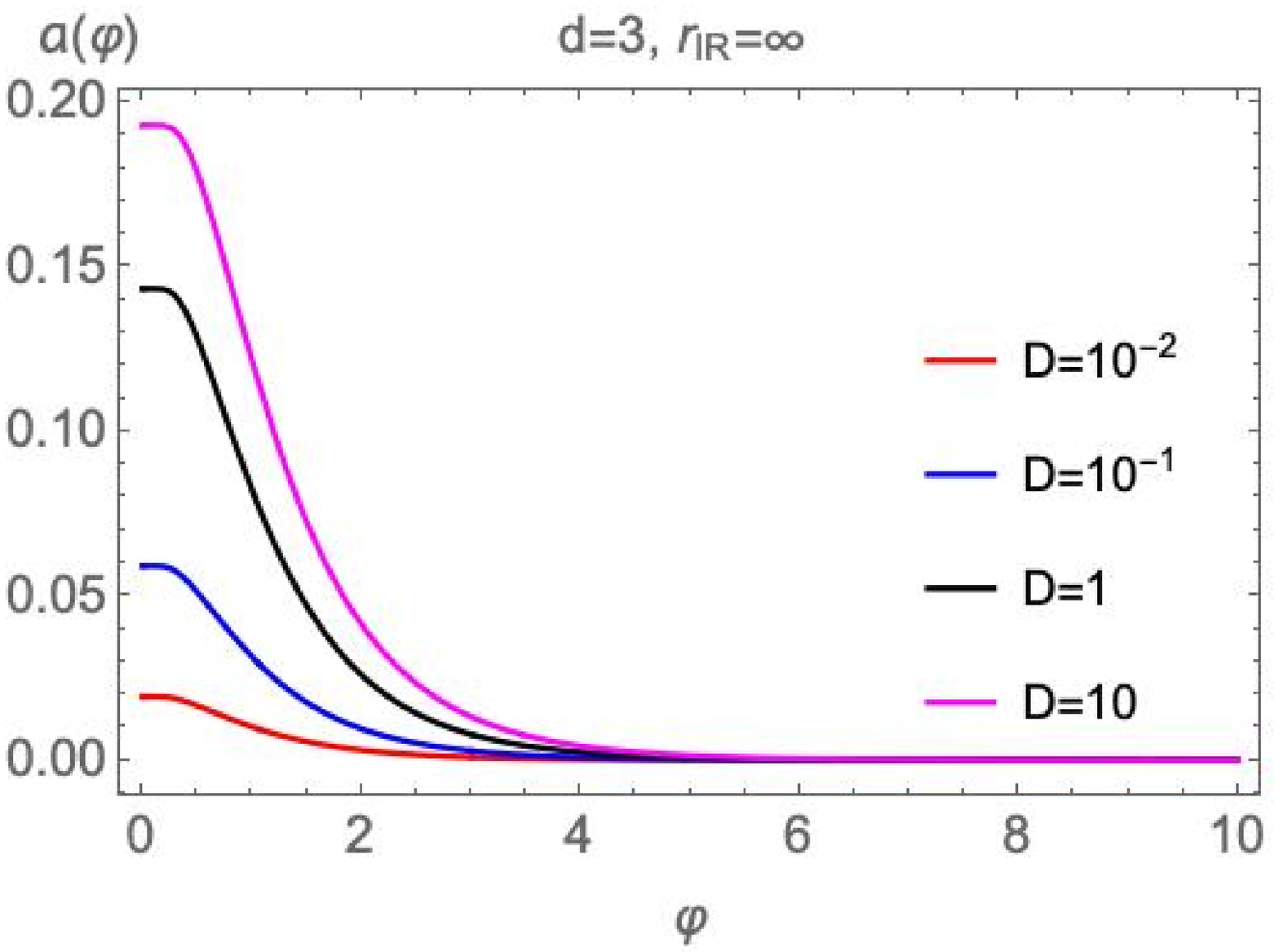}\hfil\hfil
  \includegraphics[width=.49\textwidth]{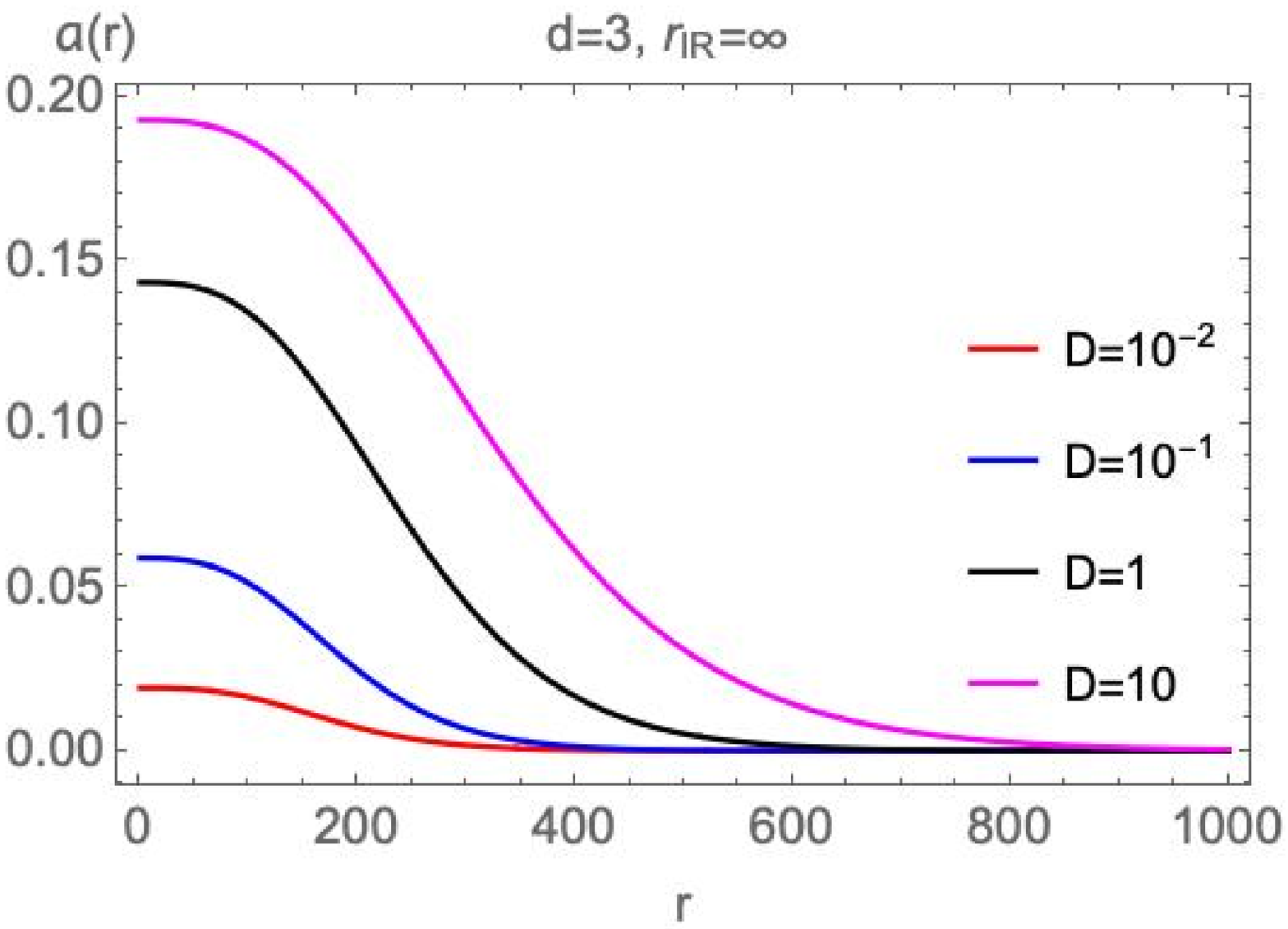}\hfil\hfil
 \end{center}
 \caption{Plots of $a(\f)$ as a function of $\f$ (\textbf{left}) and $a(r)$ as a function of $r$ (\textbf{right}) with the bulk functions \protect\eqref{P1}. The model parameters are \protect\eqref{4d1}.
  }
  \label{fig1}
 \end{figure}

 \begin{figure}[t]
 \begin{center}
  \includegraphics[width=.49\textwidth]{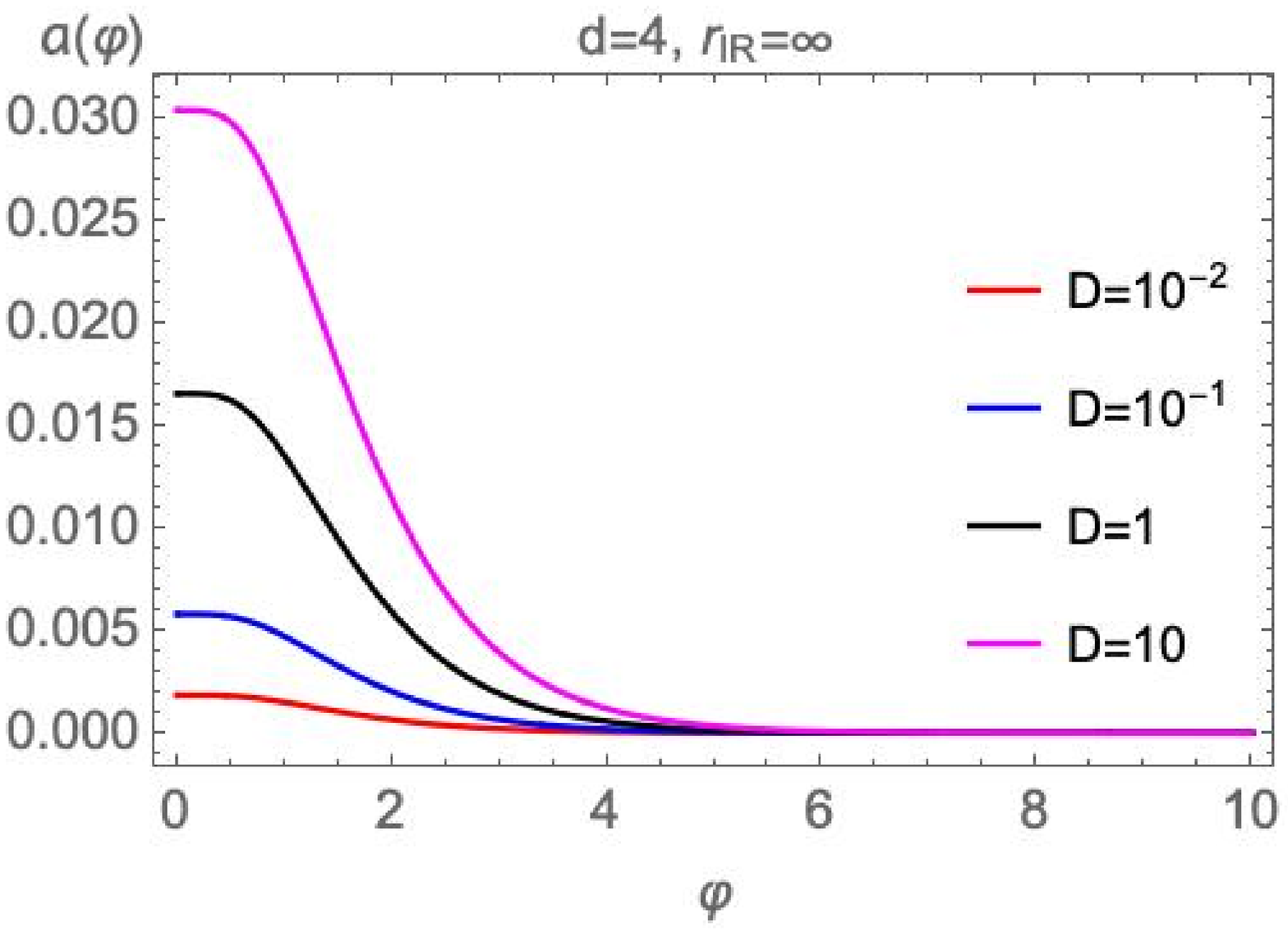}\hfil\hfil
  \includegraphics[width=.49\textwidth]{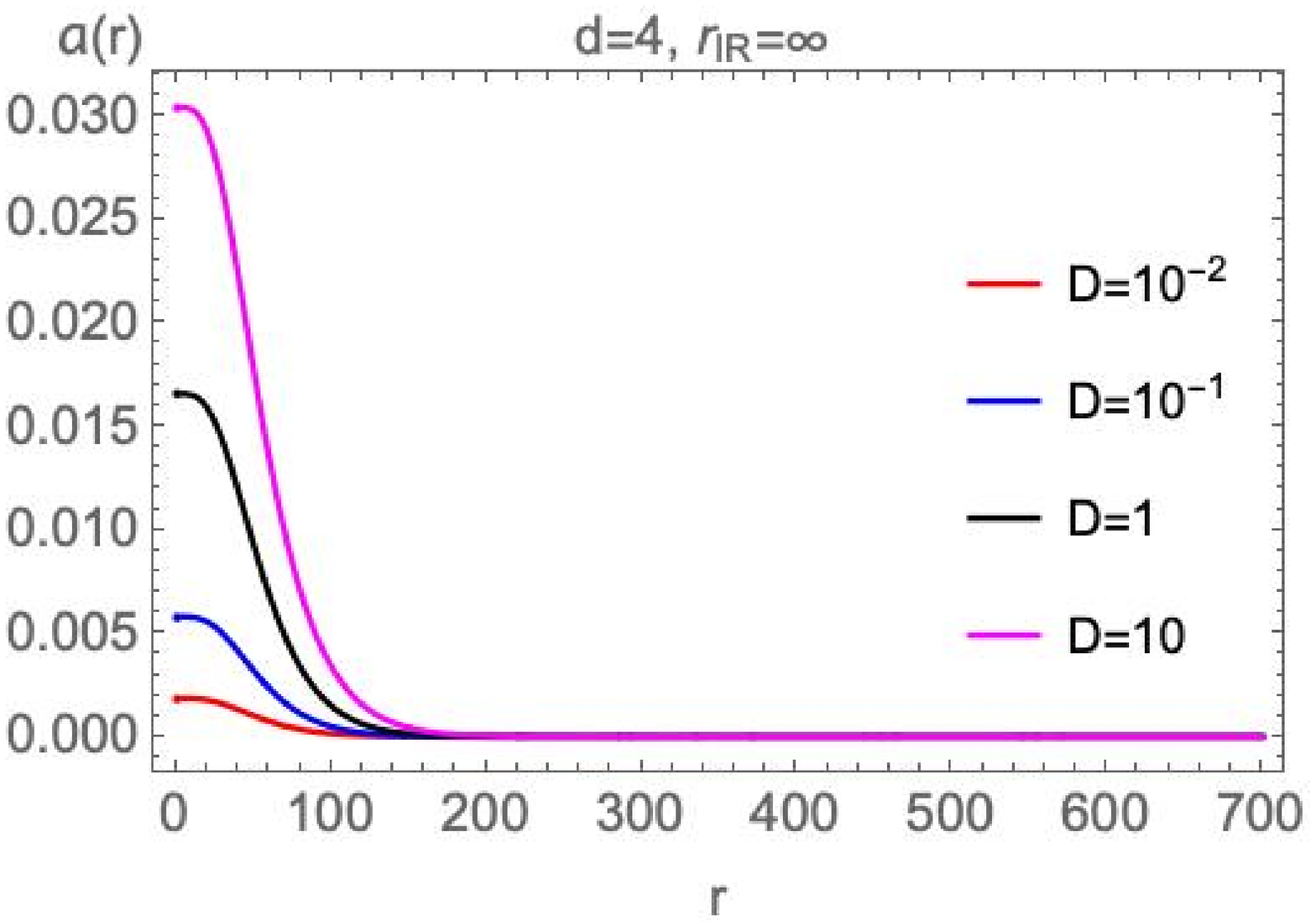}\hfil\hfil
 \end{center}
 \caption{Plots of $a(\f)$ as a function of $\f$ (\textbf{left}) and $a(r)$ as a function of $r$ (\textbf{right}) with the bulk functions \protect\eqref{P1}. The model parameters are \protect\eqref{4d5}.
  }
  \label{fig1-2}
 \end{figure}

 \begin{figure}[t]
 \begin{center}
  \includegraphics[width=.47\textwidth]{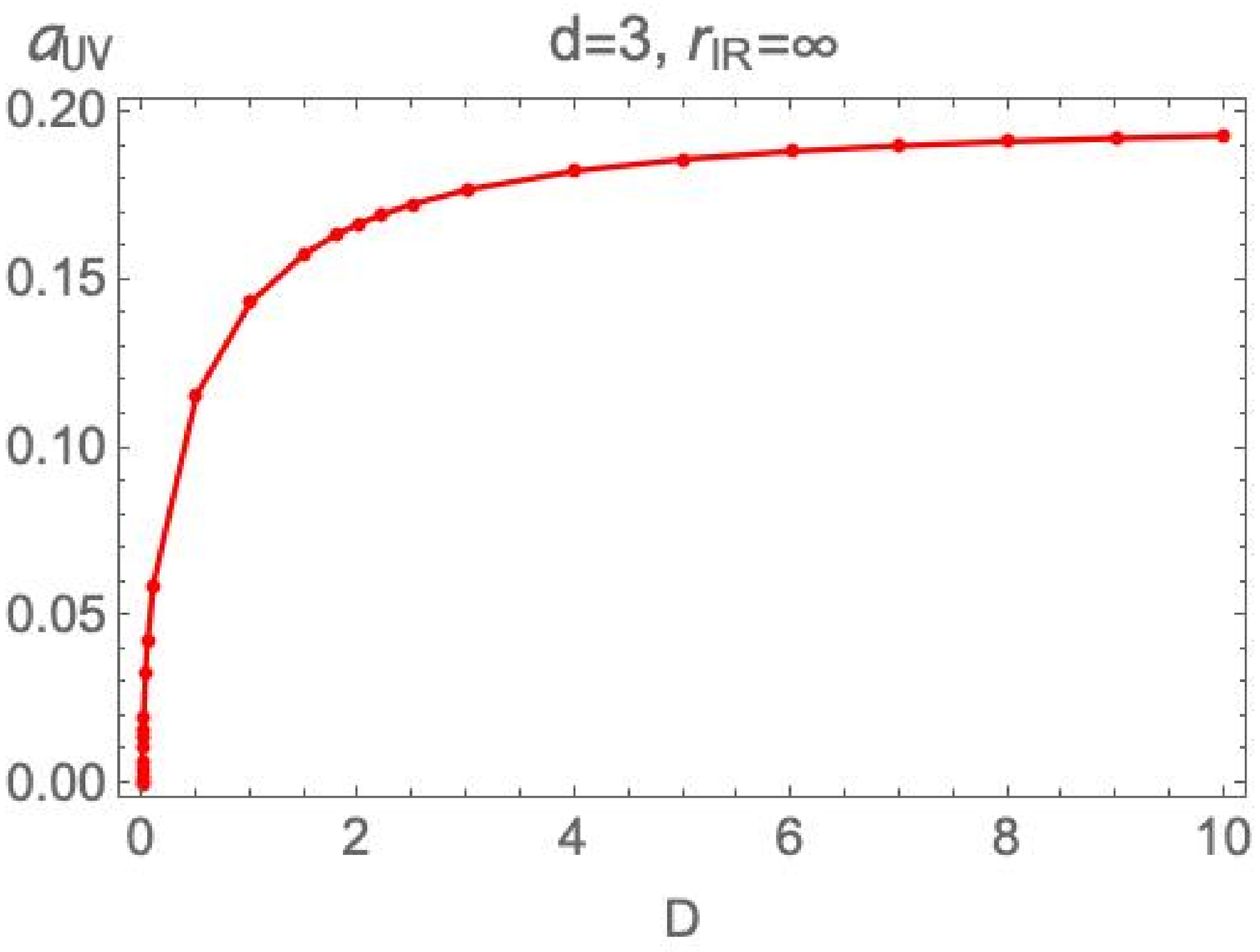}\hfil\hfil
   \includegraphics[width=.51\textwidth]{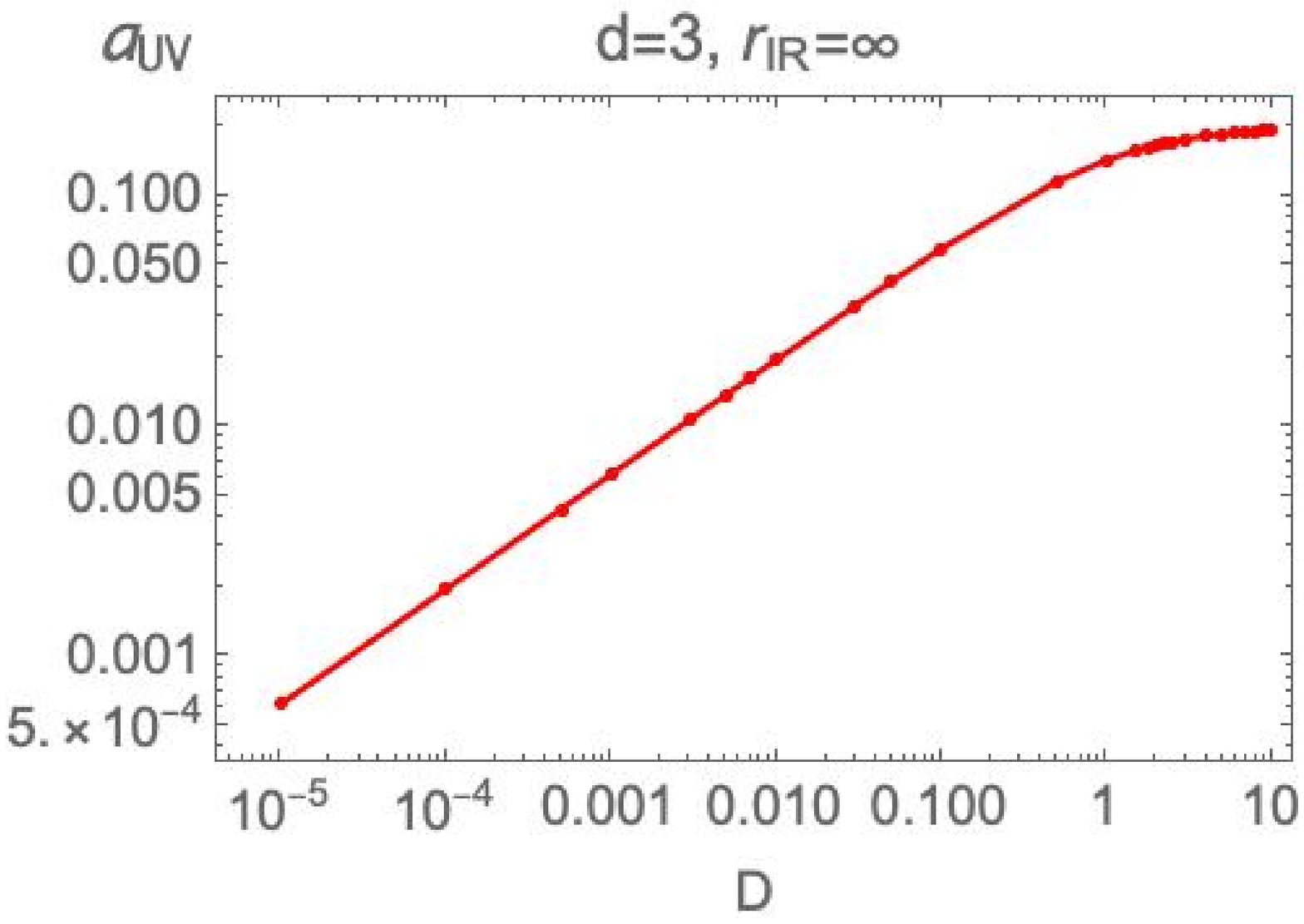}\hfil\hfil
 \end{center}
 \caption{Linear-linear (\textbf{left}) and log-log (\textbf{right}) plots of $a_{UV}$ as a function of $D$ with the bulk functions \protect\eqref{P1}. The model parameters are \protect\eqref{4d1}. For $D\lesssim0.1$, the relation is roughly given by $a_{UV}\approx 0.2 \sqrt{D}$.
 }
  \label{fig2}
 \end{figure}

 \begin{figure}[t]
 \begin{center}
  \includegraphics[width=.49\textwidth]{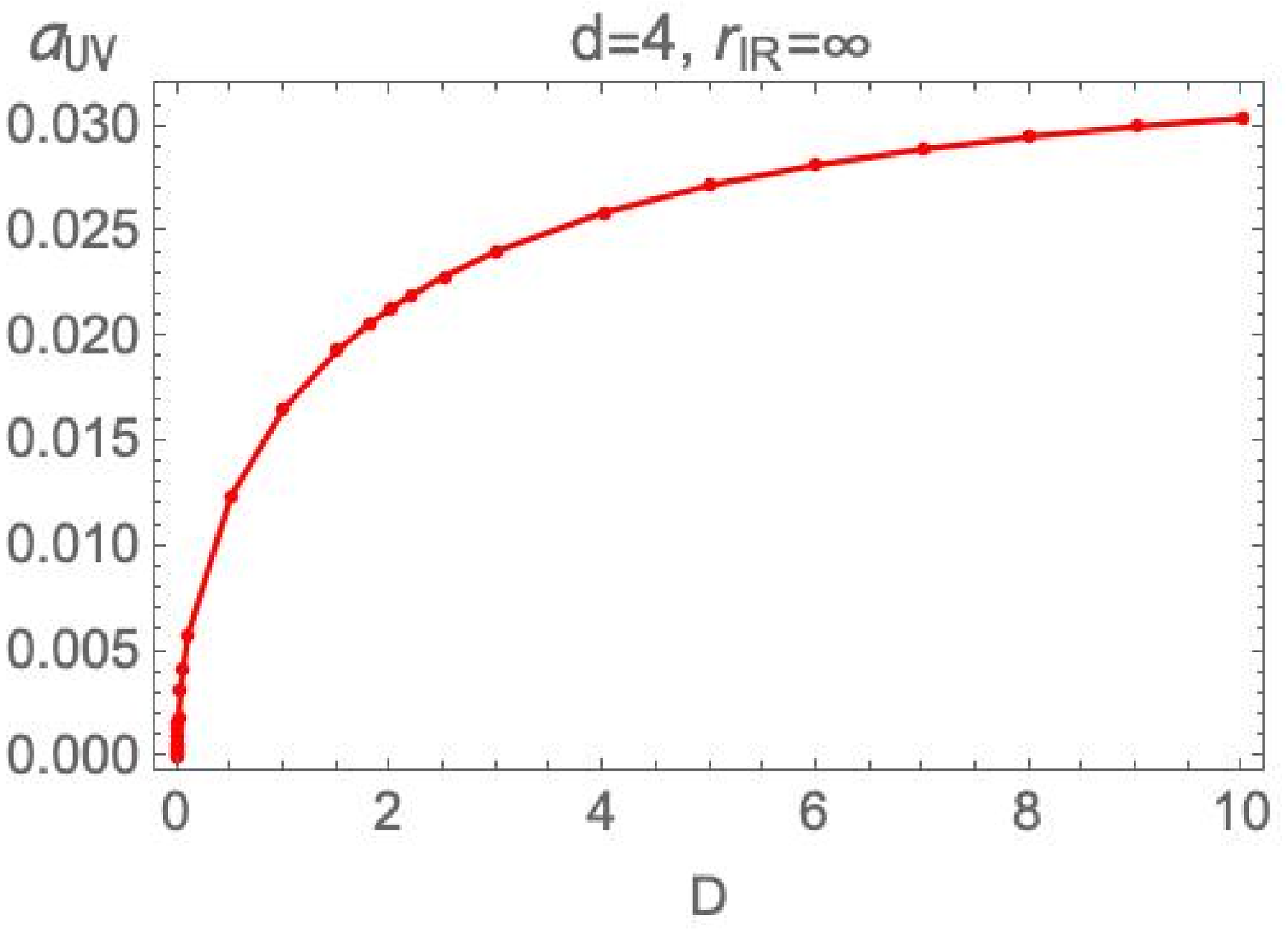}\hfil\hfil
   \includegraphics[width=.49\textwidth]{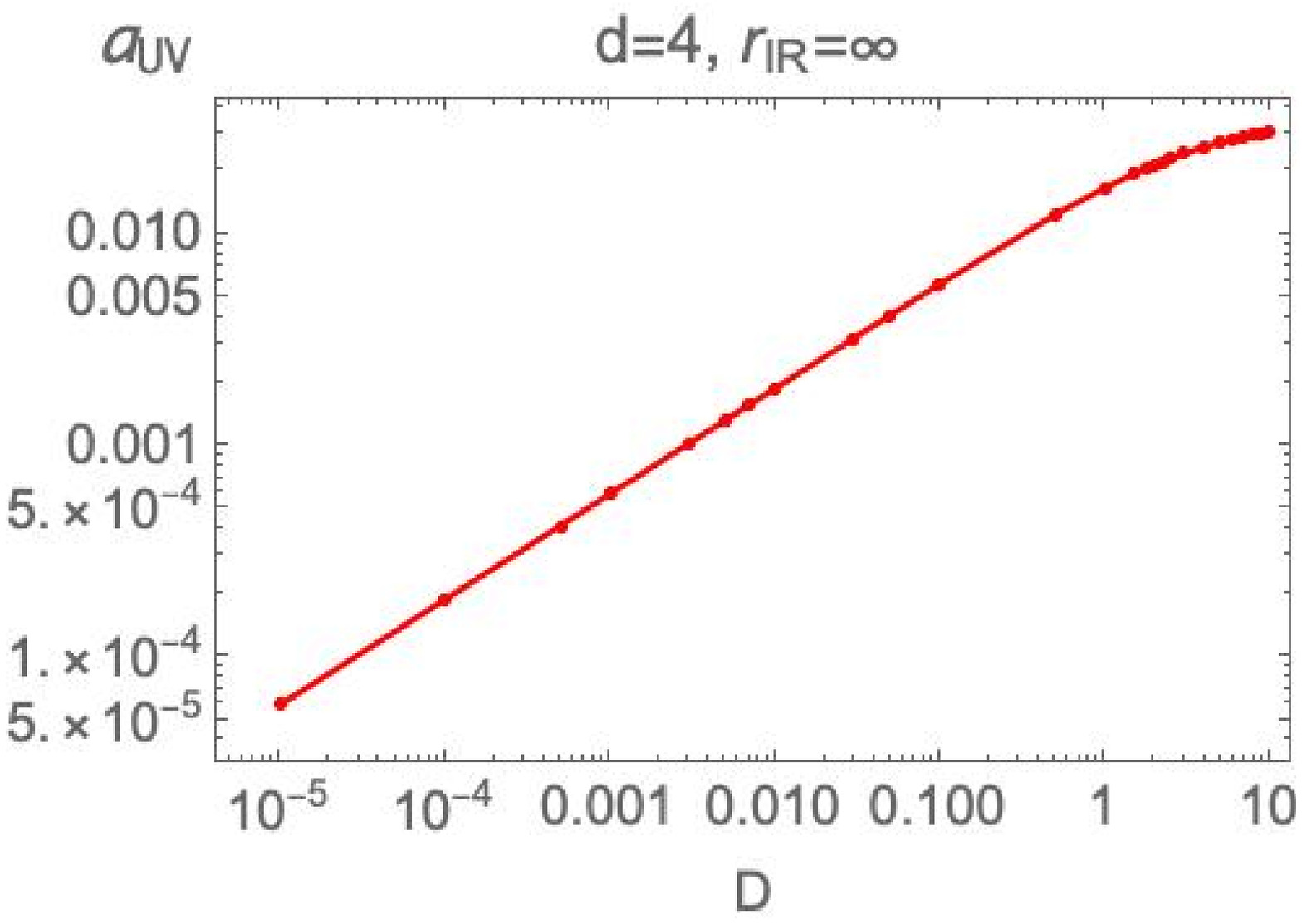}\hfil\hfil
 \end{center}
 \caption{Linear-linear (\textbf{left}) and log-log (\textbf{right}) plots of $a_{UV}$ as a function of $D$ with the bulk functions \protect\eqref{P1}. The model parameters are \protect\eqref{4d5}. For small $D\lesssim1$, the relation is roughly given by $a_{UV}\approx 0.02 \sqrt{D}$.
 }
  \label{fig2-2}
 \end{figure}

  \begin{figure}[t]
 \begin{center}
  \includegraphics[width=.47\textwidth]{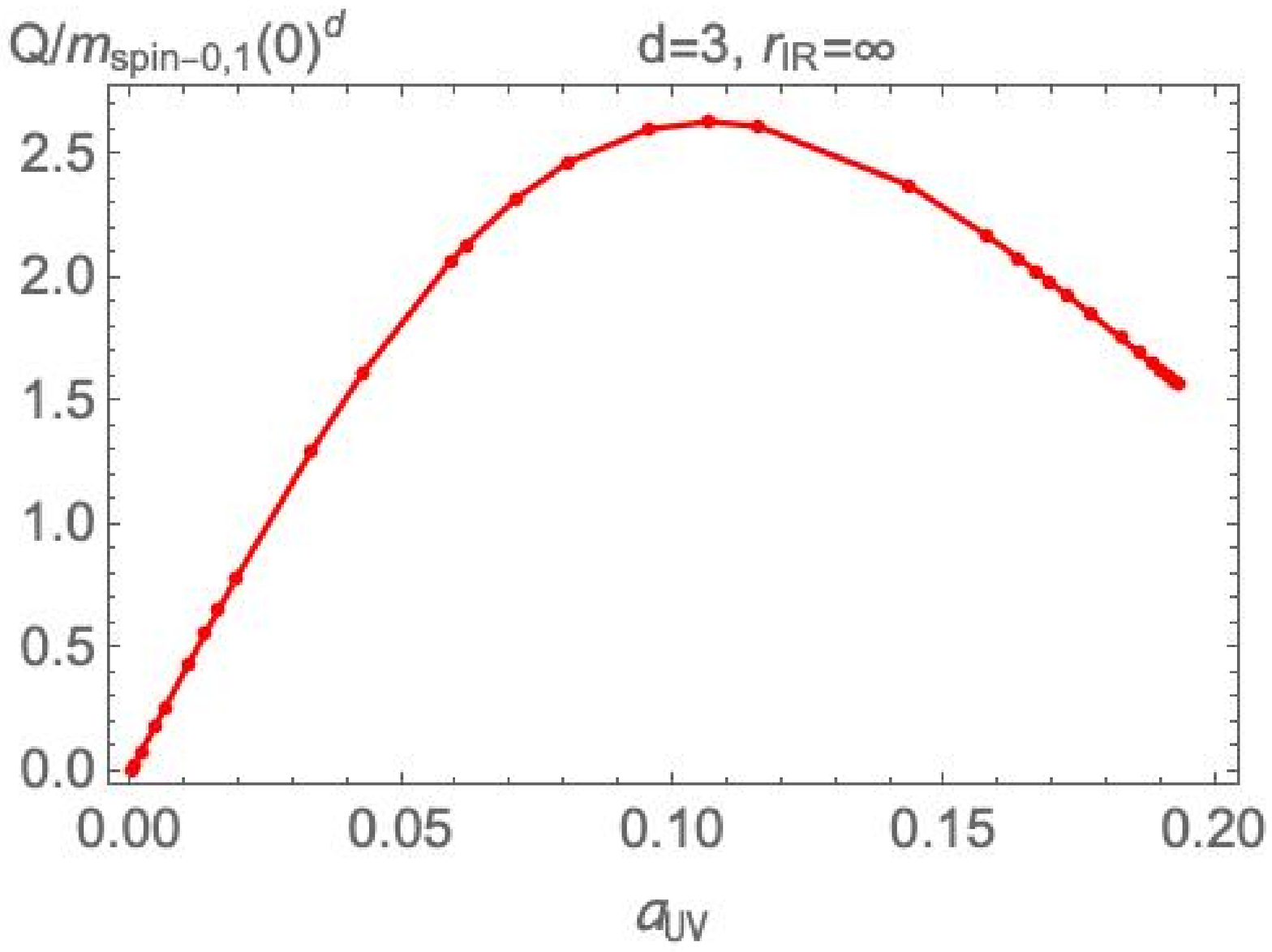}\hfil\hfil
   \includegraphics[width=.51\textwidth]{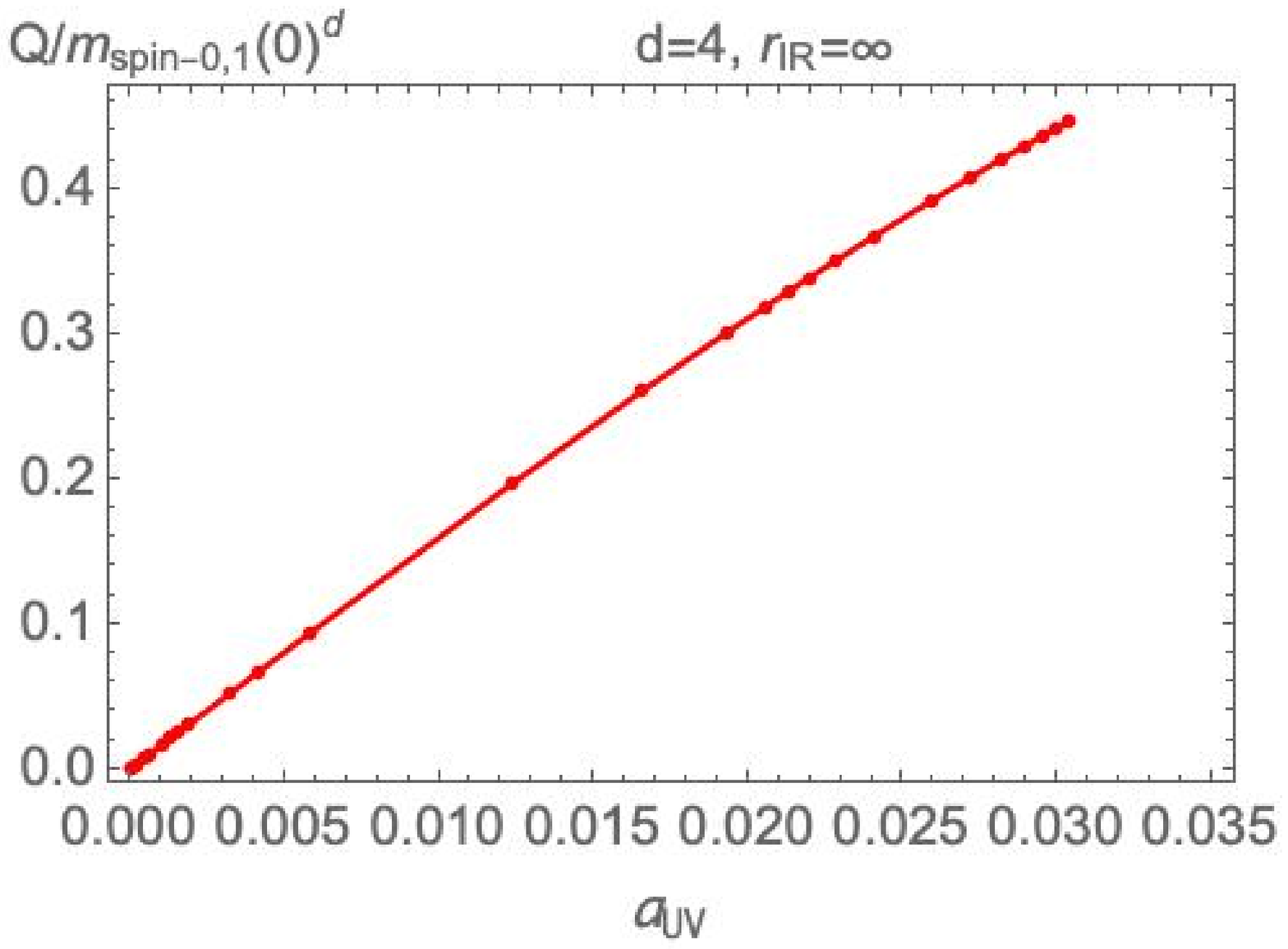}\hfil\hfil
   \\
  \includegraphics[width=.47\textwidth]{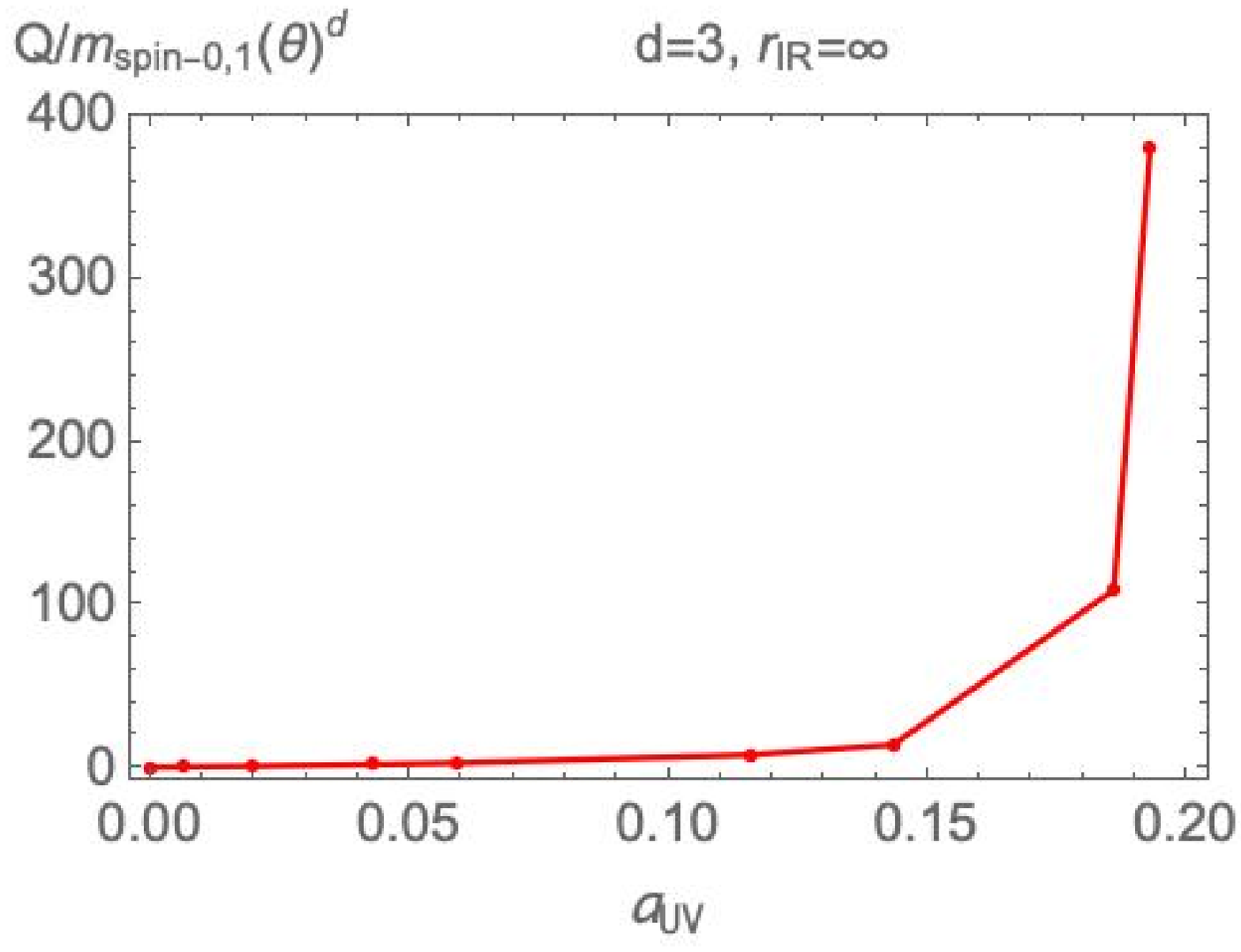}\hfil\hfil
   \includegraphics[width=.51\textwidth]{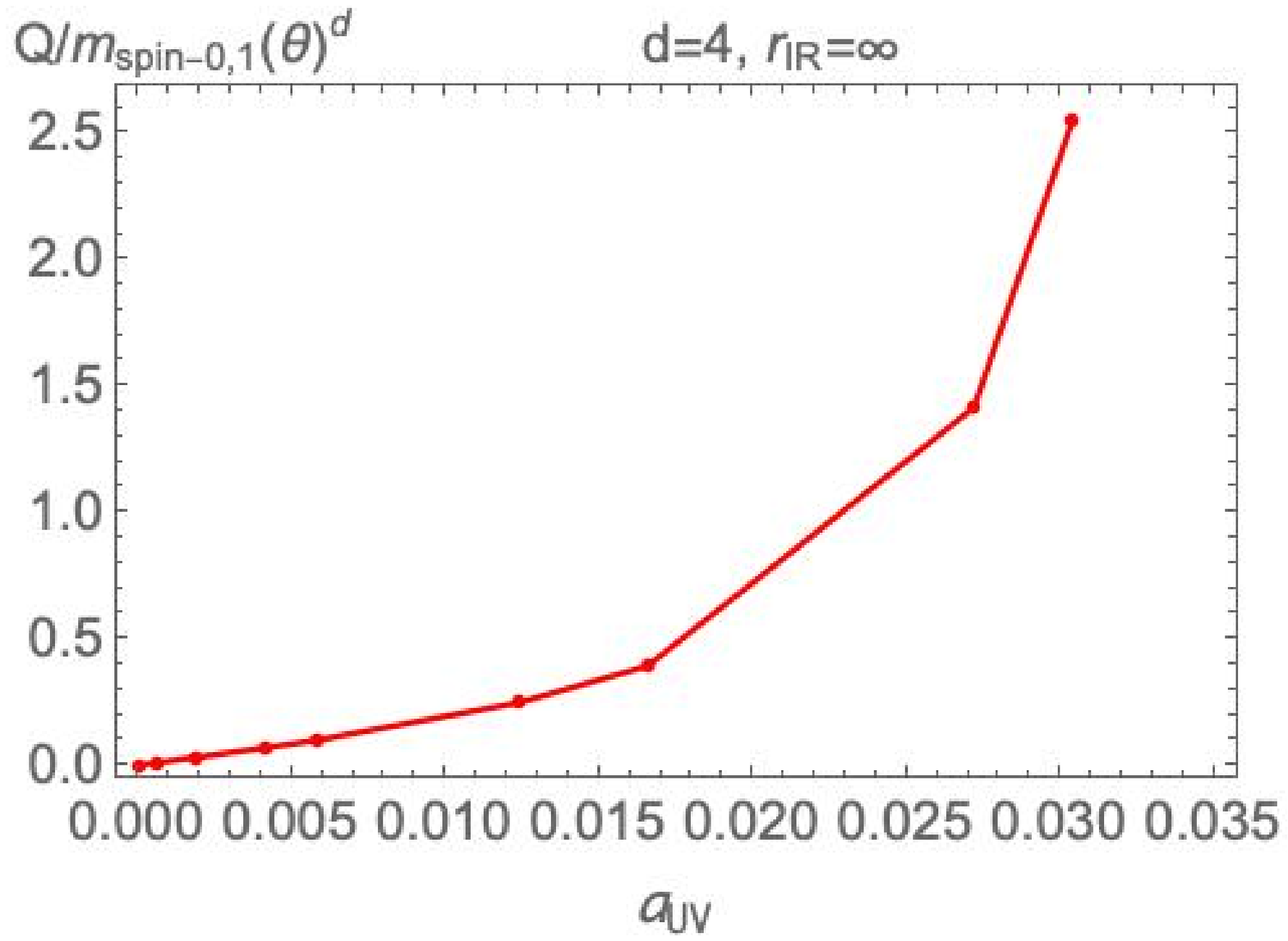}\hfil\hfil
 \end{center}
 \caption{Plot of $Q$ as a function of $a_{UV}$ with the bulk functions \protect\eqref{P1}. The model parameters are \protect\eqref{4d1} (\textbf{left}) and \protect\eqref{4d5} (\textbf{right}).
  Plots of $Q$ normalized by lightest glueball masses as functions of $a_{UV}$ with the bulk functions \protect\eqref{P1}.
 \textbf{Upper left:} The model parameters are \protect\eqref{4d1} ($a_{UV}^{max}$ is $0.19$). $Q$ is normalized by the lightest glueball mass at $a_{UV}=0$.
 \textbf{Upper right:} The model parameters are \protect\eqref{4d5} ($a_{UV}^{max}$ is $0.03$). $Q$ is normalized by the lightest glueball mass at $a_{UV}=0$.
 \textbf{Lower left:} The model parameters are \protect\eqref{4d1}. $Q$ is normalized by the lightest glueball mass at given $a_{UV}$.
 \textbf{Lower right:} The model parameters are \protect\eqref{4d5}. $Q$ is normalized by the lightest glueball mass at given $a_{UV}$.
 }
  \label{fig3}
 \end{figure}

The bulk potentials are
\be
V=
-d(d-1)-\left({1\over2}(d-\Delta_-)\Delta_--\frac{2V_\infty}{d-1} \right)\f^2-4V_\infty \left(1+\f^2\right)^{\frac{P}{2}}\sinh^2\left(\frac{\f}{\sqrt{2(d-1)}}\right),
\label{P1}\ee
$$
Y = Y_c + Y_\infty e^{\sqrt{2(d-1)}\f},
$$
with $\frac{1}{d+1}< P <1$ (see \eqref{s29-5} for a lower bound).
We use \eqref{P1} with
\be
d=3, \quad P=0.5, \quad \Delta_-=0.4, \quad Y_c = 1, \quad \f_-=5\times10^{-2}, \quad \text{sign}(Q)=-1,
\label{4d1}\ee
and
\be
d=4, \quad P=0.5, \quad \Delta_-=0.8, \quad Y_c = 1, \quad \f_-=5\times10^{-2}, \quad \text{sign}(Q)=-1.
\label{4d5}\ee
As in the previous background, we shall observe qualitative similarities in these two model parameters, at least in the region where $a_{UV}$ is small.

The scale factor and scalar field are plotted in figures \ref{fig45} and \ref{fig4} as functions of $r$. They are monotonic functions of $r$. As $a_{UV}$ is increased, $\f$ as a function of $r$ grows slower. Similarly, $A$ as a function of $r$ decreases slower as $a_{UV}$ is increased.
Holographic renormalization group flows of the $\theta$-angle are plotted in figures \ref{fig1} ($d=3$) and \ref{fig1-2} ($d=4$). The $\theta$-angle goes to zero in the IR, as in the previous background.
Figures \ref{fig2} ($d=3$) and \ref{fig2-2} ($d=4$) show the relation between $a_{UV}$ and $D$. For small $D$, $a_{UV}$ is proportional to $\sqrt{D}$ while $a_{UV}$ saturates to a maximum value, $a_{UV}^{max}$, for large $D$.
The values of $a_{UV}^{max}$ are
\be
a_{UV}^{max}\approx
\begin{cases}
0.19 \quad \text{for \eqref{4d1}}, \\
0.03 \quad \text{for \eqref{4d5}}.
\end{cases}
\label{Num3}\ee
The values of $Q/m_{\text{spin-}0,1}^d(0)$ (upper) and $Q/m_{\text{spin-}0,1}^d(a_{UV})$ (lower) are shown in figure \ref{fig3}, where the left and right panels correspond to $d=3$ and $d=4$.
This quantity is related to the vev of the instanton density through \eqref{UV27}.
In upper panels, $Q$ is proportional to $a_{UV}$ for small $a_{UV}$. In the upper right panel, it seems that $Q\propto a_{UV}$ holds in the whole region. On the other hand, in the upper left panel, $Q$ is no longer proportional to $a_{UV}$ for larger values of $a_{UV}\gtrsim0.05$.
In lower panels, both functions are monotonically increasing. This corresponds to the fact that glueball masses decrease as functions of $a_{UV}$, as we shall observe in section \ref{numerical_glueball}.

To summarize, in model parameters \eqref{4d1} and \eqref{4d5}, we observed the qualitatively similar results for the backgrounds (scalar field, axion field, scale factor). The $a_{UV}$ dependence of the vev of the instanton density is different for larger values of $a_{UV}$, but is the same for smaller values of $a_{UV}$.

\section{Linear perturbations and glueball spectra}\label{Quadratic}

In this section, we discuss the spectra and wave-functions of linear perturbations around the vacuum  in the  in the presence of $\theta$-angle. Normalizable fluctuations correspond to  glueballs, which in the linearized approximations are non-interacting single particle states. We will include lowest-order non-linearities in the next section.
Note that the $\theta$-dependence of glueball spectra was discussed in top-down holographic QCD \cite{Dubovsky:2011tu,BC1} and lattice QCD \cite{DelDebbio:2006yuf}. 

We shall study the gauge invariant  linear perturbations of $(d+1)$-dimensional Einstein-axion-dilaton theory.
The perturbations of the $(d+1)$-dimensional metric are parametrized as
\be
ds^2=e^{2A(r)} \left[ (1+2\phi) dr^2 + 2 B_\mu dx^\mu dr + \left( \eta_{\mu\nu} + h_{\mu\nu} \right)  dx^\mu dx^\nu \right]\equiv e^{2A} \tilde{g}_{ab} dx^a dx^b.
\label{pt2-2}\ee
As for the scalar fields, we use $\f$ and $a$ for the background value, and use $\delta \f$ and $\delta a$ for the fluctuations.
We further decompose $B_\mu$ and $h_{\mu\nu}$  as
\be
B_\mu = \de_\mu W + B_\mu^T,
\quad \de^\mu B_\mu^T=0,
\label{pt6-1}\ee
\be
h_{\mu\nu}=2\eta_{\mu\nu}\psi + 2\partial_\mu \partial_\nu E + 2 \partial_{(\mu}V_{\nu)}^T + h_{\mu\nu}^{TT},
\quad
h=2d \psi + 2\de^\mu\de_\mu E,
\label{pt6}\ee
where
$$\partial^\mu V_\mu^T=\partial^\mu  h_{\mu\nu}^{TT}= \eta^{\mu\nu} h_{\mu\nu}^{TT}=0\;.$$

After the calculation described in appendix \ref{quadratic}, it turns out that the equation for the transverse-traceless tensor mode is
\be
{h^{TT}_{\mu\nu}}''
+(d-1)A' {h^{TT}_{\mu\nu}}'
+\p^2 {h^{TT}_{\mu\nu}}
=0,
\label{pt22}\ee
where a prime stands for an $r$-derivative.
The equations for the scalar perturbations are
\be
\left[
-\de^\mu \de_\mu
- \frac{d^2}{dr^2}
+ \begin{pmatrix}
V_{\zeta} & 2 A' \eta_\perp \left(B_\zeta' + \frac{(A' \eta_\perp)'}{A' \eta_\perp} + \frac{d}{dr}\right) \\
2 A' \eta_\perp \left(B_\zeta' - \frac{d}{dr}\right) & V_{{\cal S}}
\end{pmatrix}
\right]
\begin{pmatrix}
\psi_\zeta \\
\psi_{\cal S}
\end{pmatrix}
=0.
\label{re6}\ee
where we have defined
\be
\frac{\psi_\zeta}{e^{\frac{d-1}{2}A}}=
\frac{\sigma'}{A'} \psi -  \frac{\f'\delta \f + Y a' \delta a}{\sigma'},
\quad
\frac{\psi_{\cal S}}{e^{\frac{d-1}{2}A}\sqrt{Y}} =
\frac{-a'\delta \f + \f' \delta a}{\sigma'},
\label{re1}\ee
and
\be
\sigma'=\sqrt{\f'^2+Y a'^2},
\quad
\eta_\perp=
\sqrt{Y} a' \frac{e^{2A} \de_\f V}{A'\sigma'^2},
\label{re7}\ee
$$
V_\zeta= B_\zeta'' + \left( B_\zeta' \right)^2,
\quad
V_{\cal S} = \frac{d-1}{2}A'' + \left(\frac{d-1}{2}A' \right)^2 + e^{2A} m_s^2,
$$
$$
B_\zeta \equiv \frac{d-1}{2}A+\frac{1}{2}\log\frac{\sigma'^2}{A'^2},
\quad
m_s^2\equiv V_{;ss} + H^2 \eta_\perp^2  - (d-1) \epsilon H^2 R_{\text{fs}}.
$$
Here $H, \epsilon, V_{;ss}$ and $R_{\text{fs}}$ are defined in \eqref{a-thi13}, \eqref{a-thi14} \eqref{a-thi15}, and \eqref{a-thi6}, respectively.

The first derivative terms in the equations (\ref{re6}) can be eliminated by the redefinition,
\be
\begin{pmatrix}
\psi_\zeta \\
\psi_{\cal S}
\end{pmatrix}
=
{\cal R}
\begin{pmatrix}
\tilde{\psi}_\zeta \\
\tilde{\psi}_{\cal S}
\end{pmatrix},
\quad
{\cal R}\equiv
\begin{pmatrix}
\cos\Phi & \sin\Phi \\
-\sin\Phi & \cos\Phi
\end{pmatrix}
,\quad
\Phi' =
A' \eta_\perp,
\label{re8}\ee
where the above equations define $\Phi$ up to a constant, which amounts to an irrelevant constant rotation.  The differential equation \eqref{re6} becomes
\be
\left[
-\de^\mu \de_\mu
- \frac{d^2}{dr^2}
+ {\cal R}^{-1}
\begin{pmatrix}
V_\zeta - \left(A' \eta_\perp \right)^2 & 2 A' \eta_\perp \left(B_\zeta' + \frac{(A' \eta_\perp)'}{2A' \eta_\perp}\right) \\
2 A' \eta_\perp \left(B_\zeta' + \frac{(A' \eta_\perp)'}{2A' \eta_\perp}\right) & V_{\cal S} - \left(A' \eta_\perp \right)^2
\end{pmatrix}
{\cal R}
\right]
\begin{pmatrix}
\tilde{\psi}_\zeta \\
\tilde{\psi}_{\cal S}
\end{pmatrix}
=0,
\label{re9}\ee
where \eqref{re8} is used.
The integration constant of $\Phi$ corresponds to the freedom to rotate the basis with a $r$-independent matrix.

\subsection{Glueball spectra}
In this subsection, we will convert the metric and scalar equations into coupled Schr\"odinger-like problems in order to calculate the spectra of spin-0 and spin-2 glueballs.

\subsubsection{Spin-$0$ glueballs}
\noindent \textbf{No axionic flow}

When the axion field is identically zero,  equation \eqref{re9} is simplified.
The two equations in \eqref{re9} for $\psi_\zeta$ and $\psi_{\cal S}$ are decoupled:
\be
\de^\mu\de_\mu \psi_\zeta + \psi_\zeta''- V_\zeta \psi_\zeta=0,
\quad
B_\zeta = \frac{d-1}{2}A + \frac{1}{2} \log\left(\frac{\f'}{A'}\right)^2,
\label{pt31-3}\ee
\be
\de^\mu\de_\mu \psi_{\cal S} + \psi_{\cal S}'' - V_{\cal S} \psi_{\cal S}=0,
\quad
V_{\cal S} = B_{\cal S}'' + B_{\cal S}'^2,
\quad
B_{\cal S} = \frac{d-1}{2}A + \frac{1}{2} \log Y.
\label{pt31-4}\ee
In this case, CP is conserved, and $\delta a$ will generate the tower of $0^{+-}$ glueballs while $\delta\f$ will generate the tower of $0^{++}$ glueballs.

We decompose the fields in eigenmodes of the radial Hamiltoinian, i.e. we write
\be
\psi_\zeta(r,x) = \sum_{i=1}^\infty \psi_{\zeta, i}^{(d)}(x) \psi_{\zeta, i}^{(r)}(r) ,
\quad
\psi_{\cal S}(r,x) = \sum_{\alpha=1}^\infty \psi_{{\cal S},\alpha}^{(d)}(x) \psi_{{\cal S},\alpha}^{(r)}(r),
\label{pt32}\ee
where the radial wave-functions solve the eigenvalue problem,
\be
-{\psi_{\zeta,i}^{(r)}}'' + V_\zeta \psi_{\zeta,i}^{(r)} = m_{\zeta,i}^2 \psi_{\zeta,i}^{(r)},
\quad
- {\psi_{{\cal S},\alpha}^{(r)}}'' + V_{\cal S} \psi_{{\cal S},\alpha}^{(r)} = m_{{\cal S},\alpha}^2 \psi_{{\cal S},\alpha}^{(r)}.
\label{pt34}\ee
The corresponding  eigenvalues $m_i^2$ and  $m_\alpha^2$ are the squared masses of the scalar and pseudoscalar glueballs, and the fluctuation  equations (\ref{pt31-3}-\ref{pt31-4}) reduce to $d$-dimensional Klein-Gordon  equations with masses  $m_i^2$ and  $m_\alpha^2$  for the space-time fields $ \psi_{\zeta, i}^{(d)}(x)$ and $\psi_{{\cal S},\alpha}^{(d)}(x)$. 

The orthonormality condition of the wave-functions is:
\be
\int {\psi_{\zeta,j}^{(r)*}} \psi_{\zeta,i}^{(r)} =\delta_{ij},
\quad
\int {\psi_{{\cal S},\beta}^{(r)*}} \psi_{{\cal S},\alpha}^{(r)} =\delta_{\alpha\beta}.
\label{pt34-2}\ee
This is read-off from the scalar product which makes the radial Hamiltonian Hermitianm, i.e.  is the standard $L^2$ norm on $\psi_{\zeta,i}^{(r)}$ and   $\psi_{{\cal S},\alpha}^{(r)}$ on $[0, r_{IR}]$.

\noindent \textbf{Non-trivial axionic flow}

In the presence of the non-trivial axion source $a_{UV}\neq0$, CP is no-longer a symmetry and equations \eqref{re9} can not be diagonalized. Consequently, there is no invariant distinction between the scalar and pseudo-scalar glueballs. We need to solve the eigenvalue problem of the coupled equation. As in \cite{VQCD,disc}, for numerical purposes, it turns out to be convenient to use  $A$ as a radial coordinate, rather than $r$.
Moreover, by an appropriate rotation, the first derivative terms in the differential equation can be eliminated. The resultant equation for the scalar fluctuations is (the derivation is presented in appendix \ref{transformation})
\be
\left[
-\frac{\de^\mu \de_\mu}{A'^2}
- \frac{d^2}{dA^2}
+ {\cal R}^{-1}
\begin{pmatrix}
\frac{d^2 B_{\zeta A}}{dA^2} + \left( \frac{dB_{\zeta A}}{dA} \right)^2 - \eta_\perp^2 & 2\eta_\perp \frac{dB_{\zeta A}}{dA}+\frac{d\eta_\perp}{dA} \\
2\eta_\perp \frac{dB_{\zeta A}}{dA}+\frac{d\eta_\perp}{dA} & \frac{d^2 B_{ {\cal S} A}}{dA^2} + \left( \frac{dB_{ {\cal S} A}}{dA} \right)^2  + m_{{\cal S}A}^2
\end{pmatrix}
{\cal R}
\right]
\begin{pmatrix}
\psi_{\zeta A} \\
\psi_{{\cal S}A}
\end{pmatrix}
=0.
\label{tr20}\ee
The quantity $\eta_\perp$ and ${\cal R}$ are defined in \eqref{re7} and \eqref{re8}. We have also defined
\be
B_{\zeta A}=\frac{d}{2}A+\frac{1}{2}\log\frac{S^2+\frac{T}{Y}}{W},
\quad
B_{ {\cal S} A}=\frac{d}{2}A+\frac{1}{2}\log W,
\quad
m_{{\cal S}A}^2= \frac{e^{2A}V_{;ss}}{A'^2}  - (d-1) \epsilon R_{\text{fs}},
\label{tr16}\ee
where $\epsilon, V_{;ss}$ and $R_{\text{fs}}$ are defined in \eqref{a-thi14} \eqref{a-thi15}, and \eqref{a-thi6}, respectively.
The integration constant of $\Phi$ corresponds to the freedom to rotate the basis with a $r$-independent matrix, which does not affect the physics.

By looking for the solution of \eqref{tr20} with the plane wave ansatz $\psi_{\zeta A}, \psi_{{\cal S} A}\propto e^{iq_\mu x^\mu}$ where $q$ and $x$ are the $d$-dimensional momentum and coordinate\footnote{This is the same as assuming that the $d$-dimensional part of the wave-function solves Klein-Gordon's equation, an alternative, and equivalent  way  as the decomposition (\ref{pt32})  which leads to the radial Hamiltonian eigenvalue problem. Only, this time the Hamiltonian is not diagonal in  the $\psi_\zeta,
\psi_{\cal S}$ basis.}, the glueball mass is identified as $m^2=-q^2$. 
Then, the glueball mass $m_{\text{spin-}0,i}^2$ corresponding to $i$-th mode is calculated by solving the Schr\"odinger equation
\be
\left[
- \frac{m_{\text{spin-}0,i}^2}{A'^2}
- \frac{d^2}{dA^2}
+ {\cal R}^{-1}
\begin{pmatrix}
\frac{d^2 B_{\zeta A}}{dA^2} + \left( \frac{dB_{\zeta A}}{dA} \right)^2 - \eta_\perp^2 & 2\eta_\perp \frac{dB_{\zeta A}}{dA}+\frac{d\eta_\perp}{dA} \\
2\eta_\perp \frac{dB_{\zeta A}}{dA}+\frac{d\eta_\perp}{dA} & \frac{d^2 B_{ {\cal S} A}}{dA^2} + \left( \frac{dB_{ {\cal S} A}}{dA} \right)^2  + m_{{\cal S}A}^2
\end{pmatrix}
{\cal R}
\right]
\begin{pmatrix}
\psi_{\zeta A, i}^{(A)} \\
\psi_{{\cal S}A, i}^{(A)}
\end{pmatrix}
=0.
\label{tr21}\ee
The normalization of $\psi_{\zeta A, i}^{(A)}, \psi_{{\cal S} A, i}^{(A)}$ is
\be
\int \frac{dA}{A'^2} \left( \left|\psi_{\zeta A, i}^{(A)}\right|^2 + \left|\psi_{{\cal S} A, i}^{(A)}\right|^2 \right) = 1.
\label{cu3}\ee
This is the norm on a doublet of $L^2[0,r_{IR}]$ eigenfunctions with respect to the conformal coordinate, upon a ghange of variables  to an integral over the scale factor $A$.

\subsubsection{Spin-2 glueballs}
Contrary to the spin-0 glueballs, there is no mixing in the equation of the spin-2 fluctuation \eqref{pt22}.
The equation \eqref{pt22} is controlled by the bulk Laplacian and is equivalent to
\be
\left( \frac{d^2}{dr^2} +\p^2 - \frac{d-1}{2}\left( A''+\frac{d-1}{2}A'^2 \right) \right) \psi_t = 0,
\quad \psi_t \equiv e^{\frac{d-1}{2}A}h^{TT}_{\mu\nu}.
\label{t1}\ee
As in the previous subsections, the spin-2 glueball masses, $m_{\text{spin-}2,i}^2$, are determined by solving the Schr\"odinger equation
\be
-{\psi_{t,i}^{(r)}}'' +V_t \psi_{t,i}^{(r)} = m_{\text{spin-}2,i}^2 \psi_{t,i}^{(r)},
\quad
V_t = \frac{d-1}{2}\left( A''+\frac{d-1}{2}A'^2 \right)
\label{t3}\ee
where we normalize $\psi_{t,i}^{(r)}$ as
\be
\int {\psi_{t,j}^{(r)*}} \psi_{t,i}^{(r)} =\delta_{ij}.
\label{t4}\ee

\subsection{Numerical results for the glueball spectra}\label{numerical_glueball}

 \begin{figure}[t]
 \begin{center}
  \includegraphics[width=.49\textwidth]{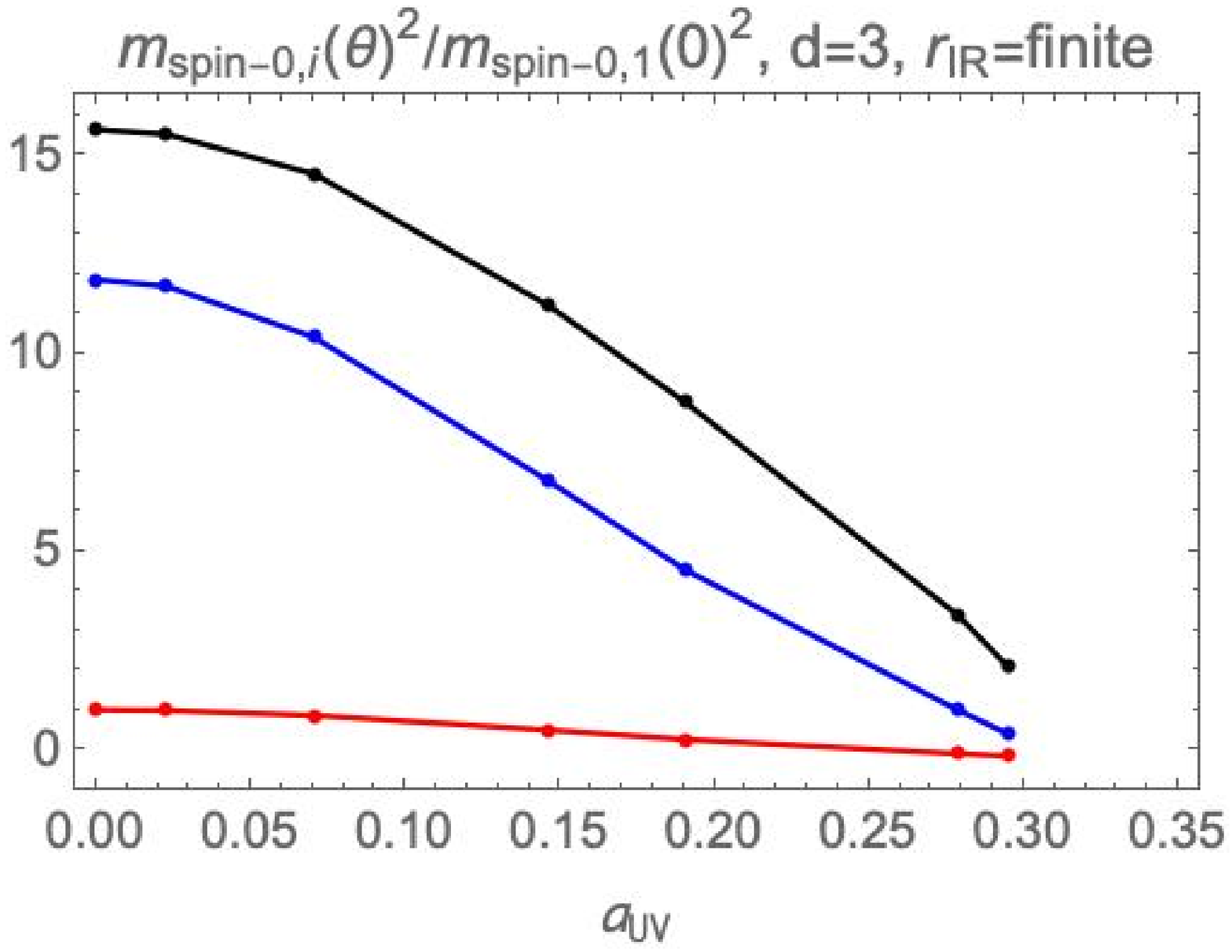}\hfil\hfil
  \includegraphics[width=.49\textwidth]{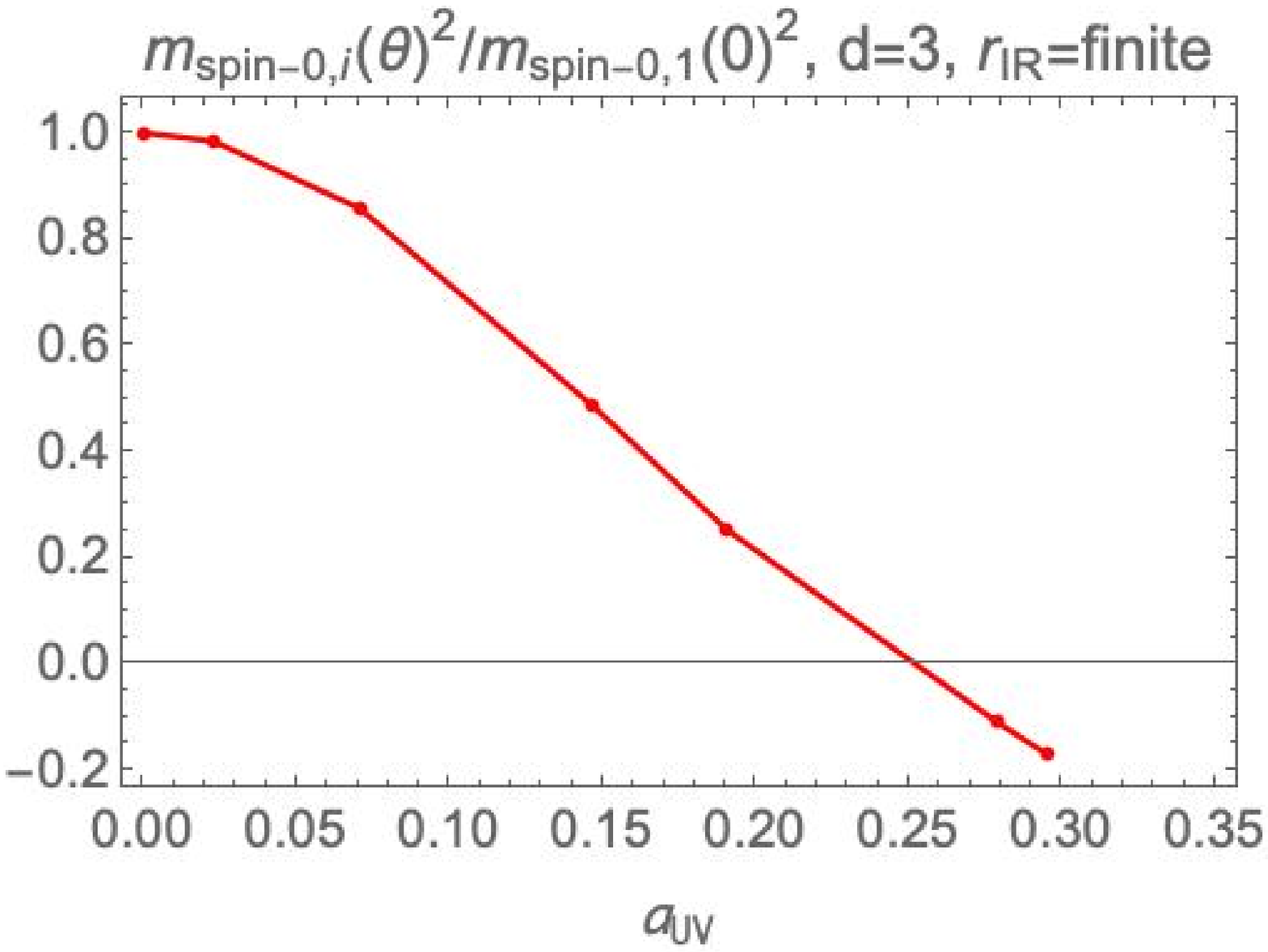}\hfil\hfil
 \end{center}
 \caption{Plots of the lowest three (\textbf{left}) and the lightest (\textbf{right}) spin-0 glueball mass squared normalized by that of the lightest glueball at $a_{UV}=0$, with the bulk functions \protect\eqref{Num1}. The model parameters are \protect\eqref{4d2}. The lightest glueball becomes tachyonic for $0.25\lesssim a_{UV}\leq a_{UV}^{max}\approx0.3$.
 }
  \label{fig20}
 \end{figure}

  \begin{figure}[t]
 \begin{center}
  \includegraphics[width=.49\textwidth]{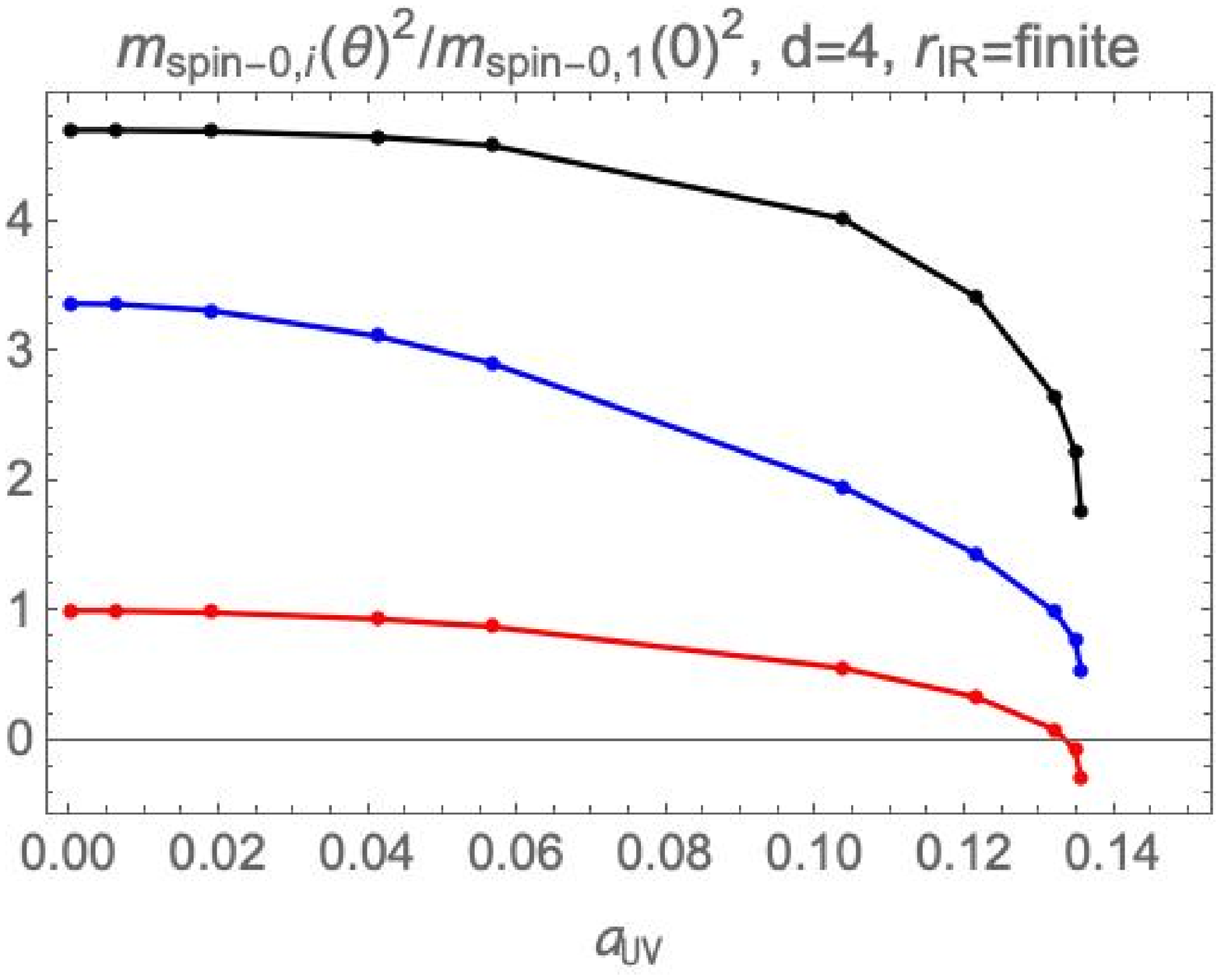}\hfil\hfil
  \includegraphics[width=.49\textwidth]{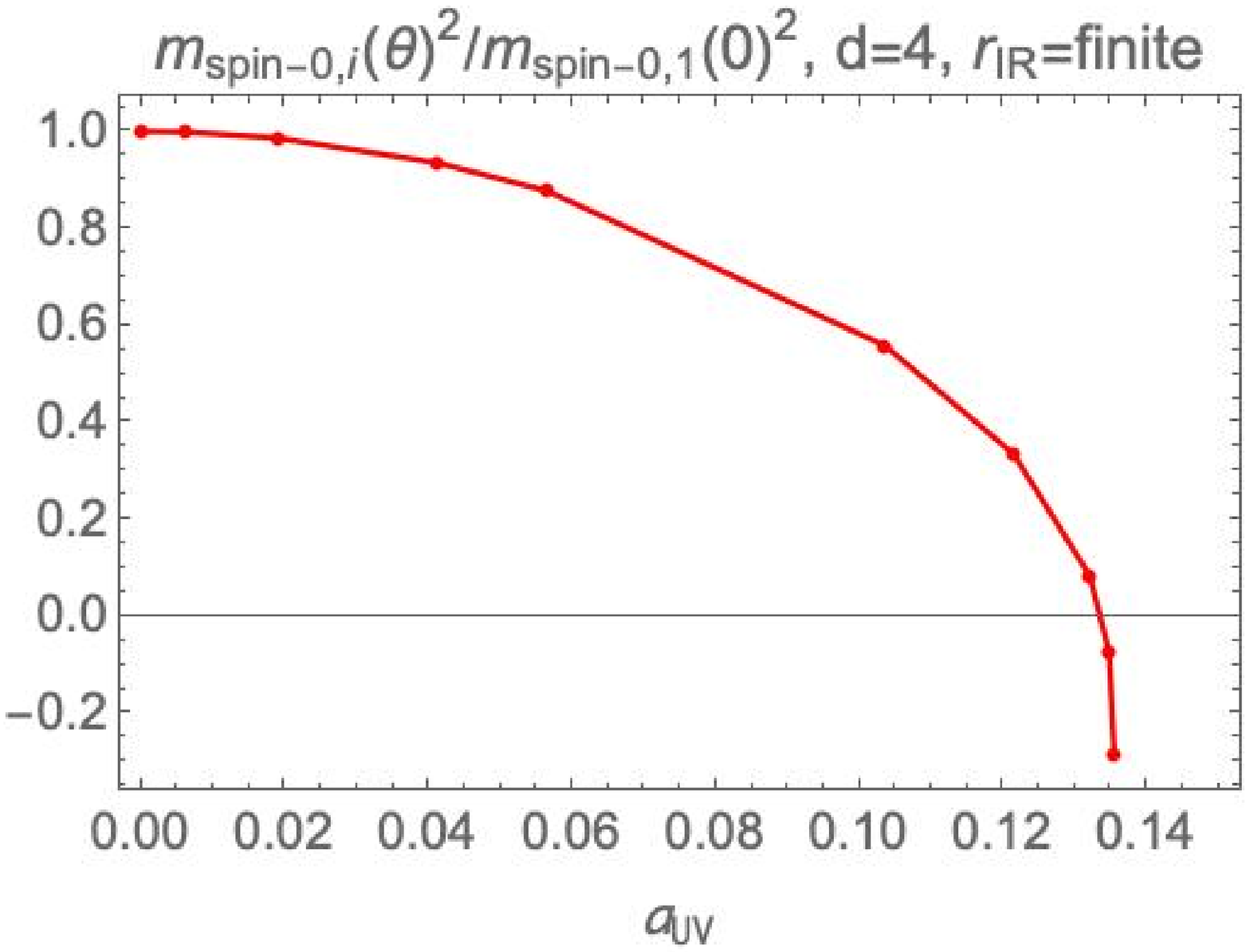}\hfil\hfil
 \end{center}
 \caption{Plots of the lowest three (\textbf{left}) and the lightest (\textbf{right}) spin-0 glueball mass squared normalized by that of the lightest glueball at $a_{UV}=0$, with the bulk functions \protect\eqref{Num1}. The model parameters are \protect\eqref{4d2-2}. The lightest glueball becomes tachyonic for $0.13\lesssim a_{UV}\leq a_{UV}^{max}\approx0.135$.
 }
  \label{fig20-2}
 \end{figure}

   \begin{figure}[t]
 \begin{center}
  \includegraphics[width=.49\textwidth]{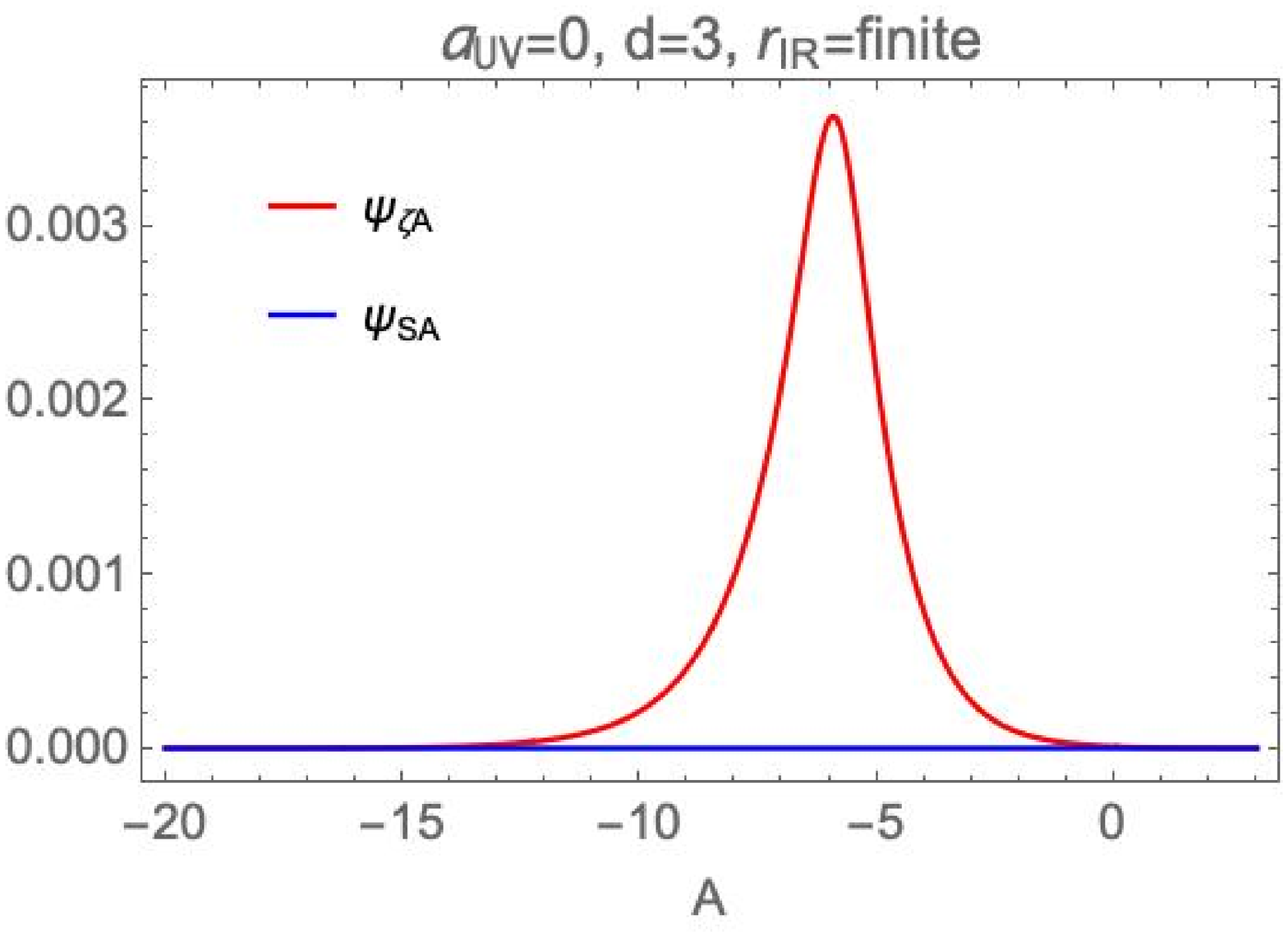}\hfil\hfil
   \includegraphics[width=.49\textwidth]{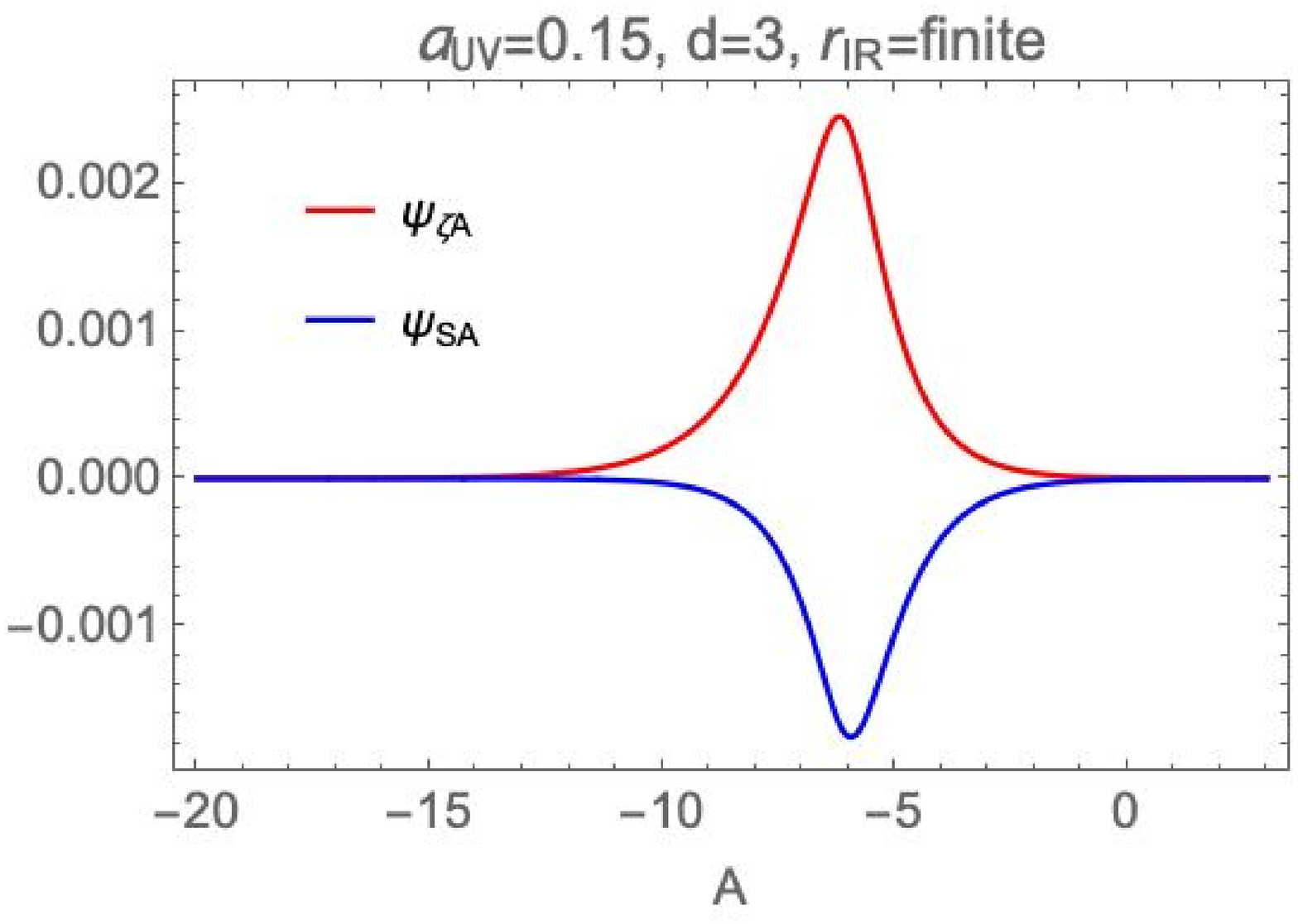}\hfil\hfil
   \\
  \includegraphics[width=.49\textwidth]{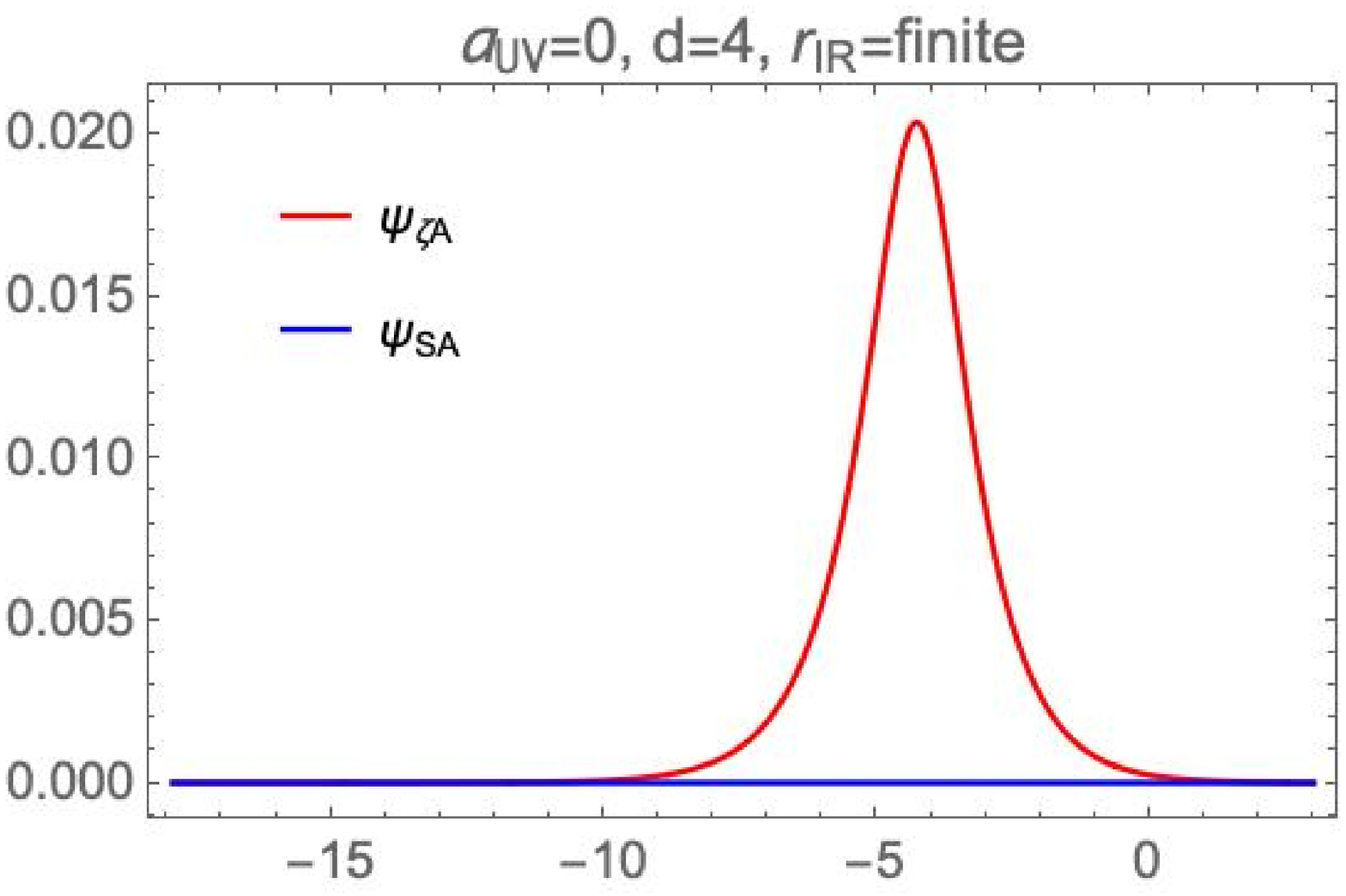}\hfil\hfil
   \includegraphics[width=.49\textwidth]{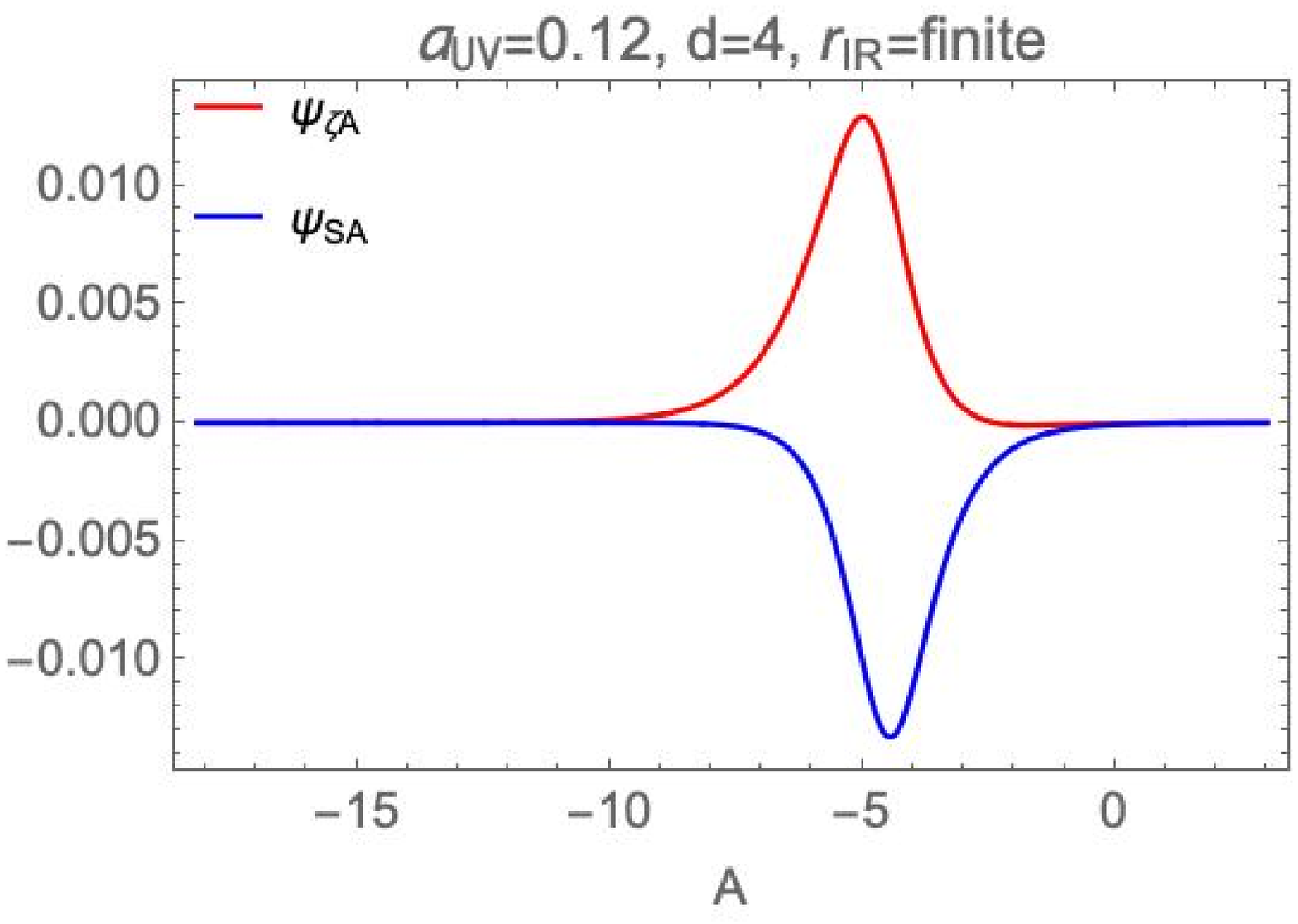}\hfil\hfil
 \end{center}
 \caption{Plots of wave-functions $\psi_{\zeta A}$ and $\psi_{{\cal S} A}$ corresponding to the lightest glueball as functions of $A$ with the bulk functions \protect\eqref{Num1}.
 \textbf{Upper left:} The model parameters are \protect\eqref{4d2} with $a_{UV}=0$.
 \textbf{Upper right:} The model parameters are \protect\eqref{4d2} with $a_{UV}=0.15$.
 \textbf{Lower left:} The model parameters are \protect\eqref{4d2-2} with $a_{UV}=0$.
 \textbf{Lower right:} The model parameters are \protect\eqref{4d2-2} with $a_{UV}=0.12$.
 }
  \label{fig8}
 \end{figure}

 \begin{figure}[t]
 \begin{center}
  \includegraphics[width=.49\textwidth]{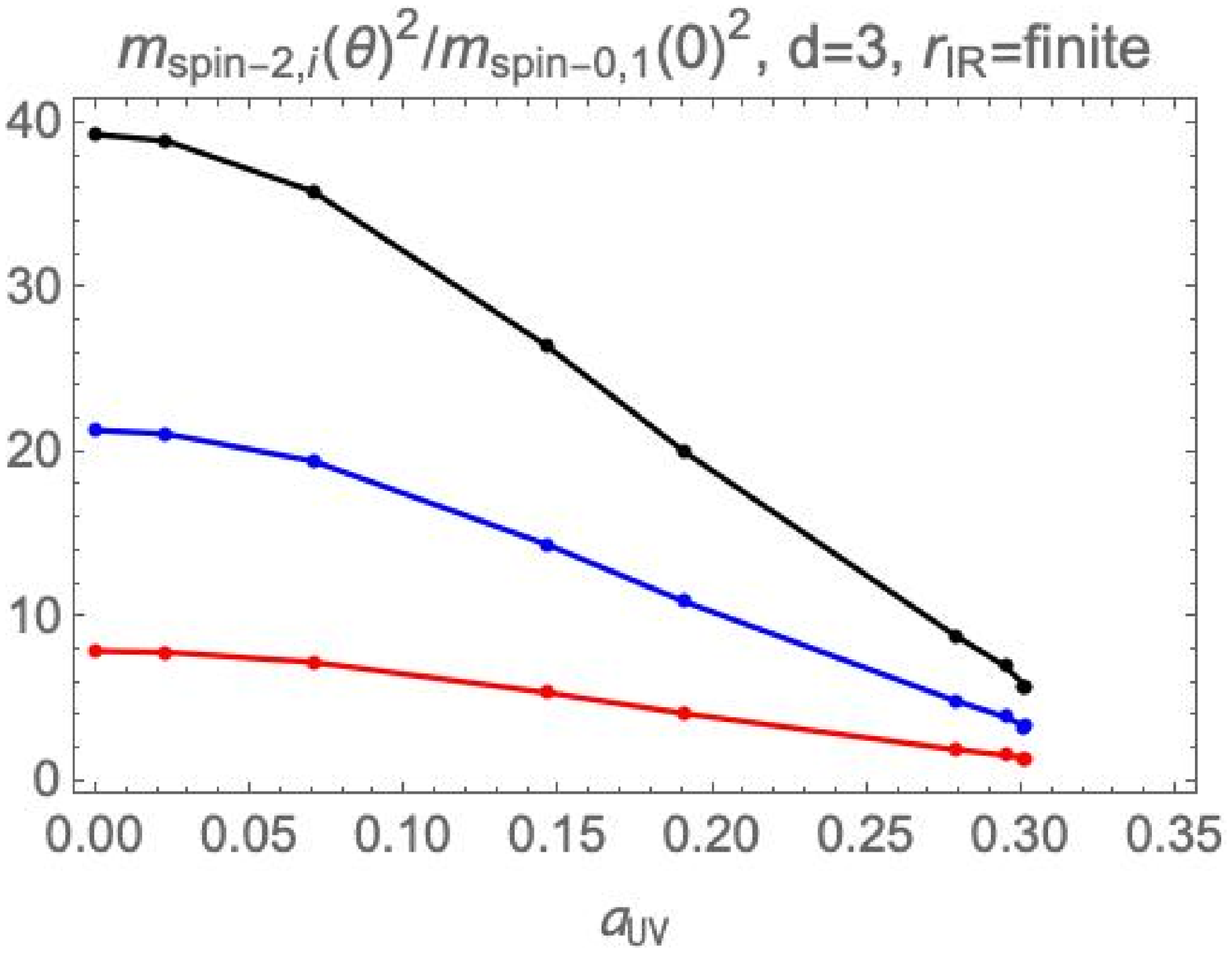}\hfil\hfil
  \includegraphics[width=.49\textwidth]{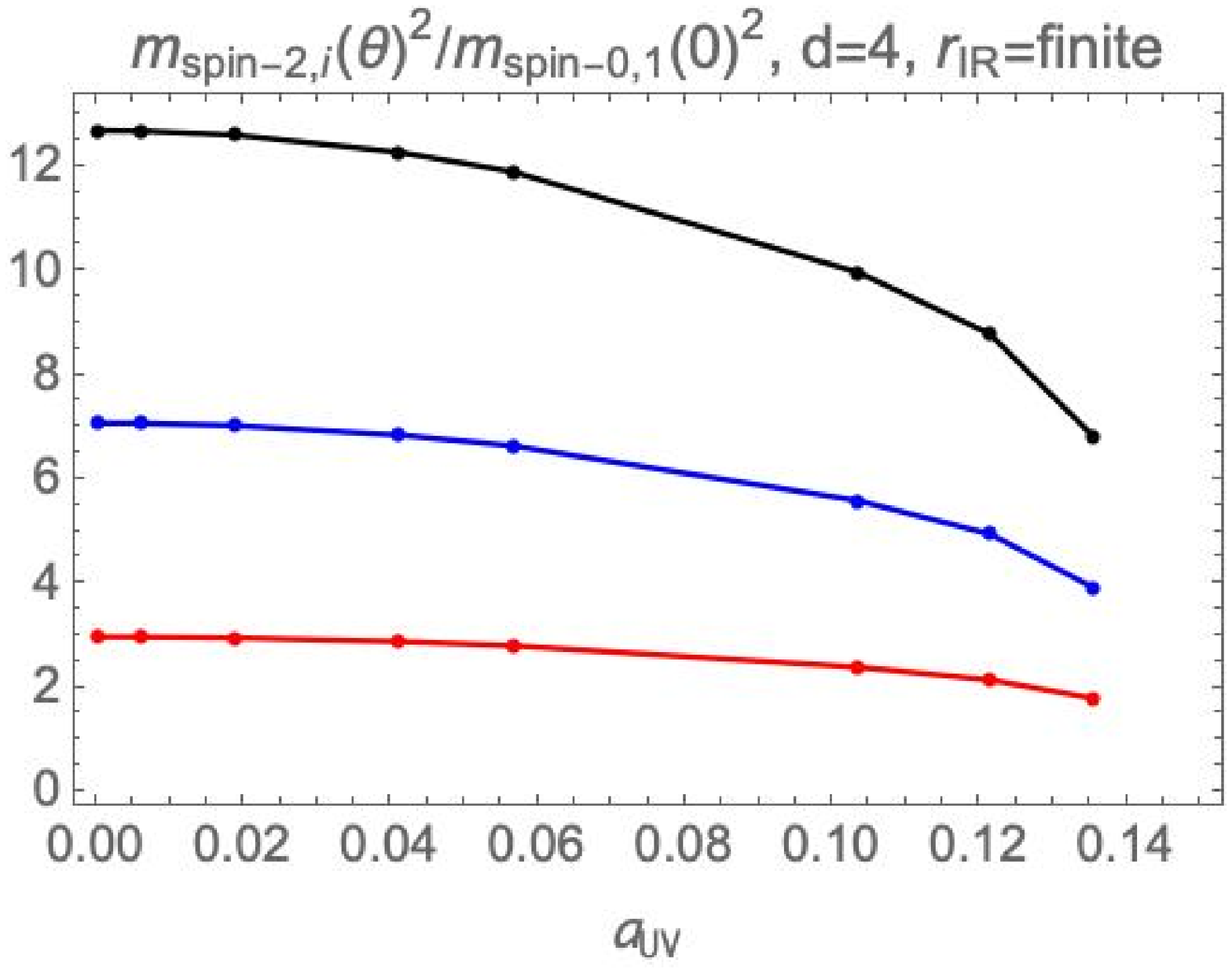}\hfil\hfil
 \end{center}
 \caption{Plots of the lowest three spin-2 glueball mass squared normalized by that of the lightest glueball at $a_{UV}=0$, with the bulk functions \protect\eqref{Num1}. The model parameters are \protect\eqref{4d2} (\textbf{left}) and \protect\eqref{4d2-2} (\textbf{right}). We do not observe the tachyonic instability of the spin-2 glueballs.
 }
  \label{fig20-3}
 \end{figure}

   \begin{figure}[t]
 \begin{center}
  \includegraphics[width=.49\textwidth]{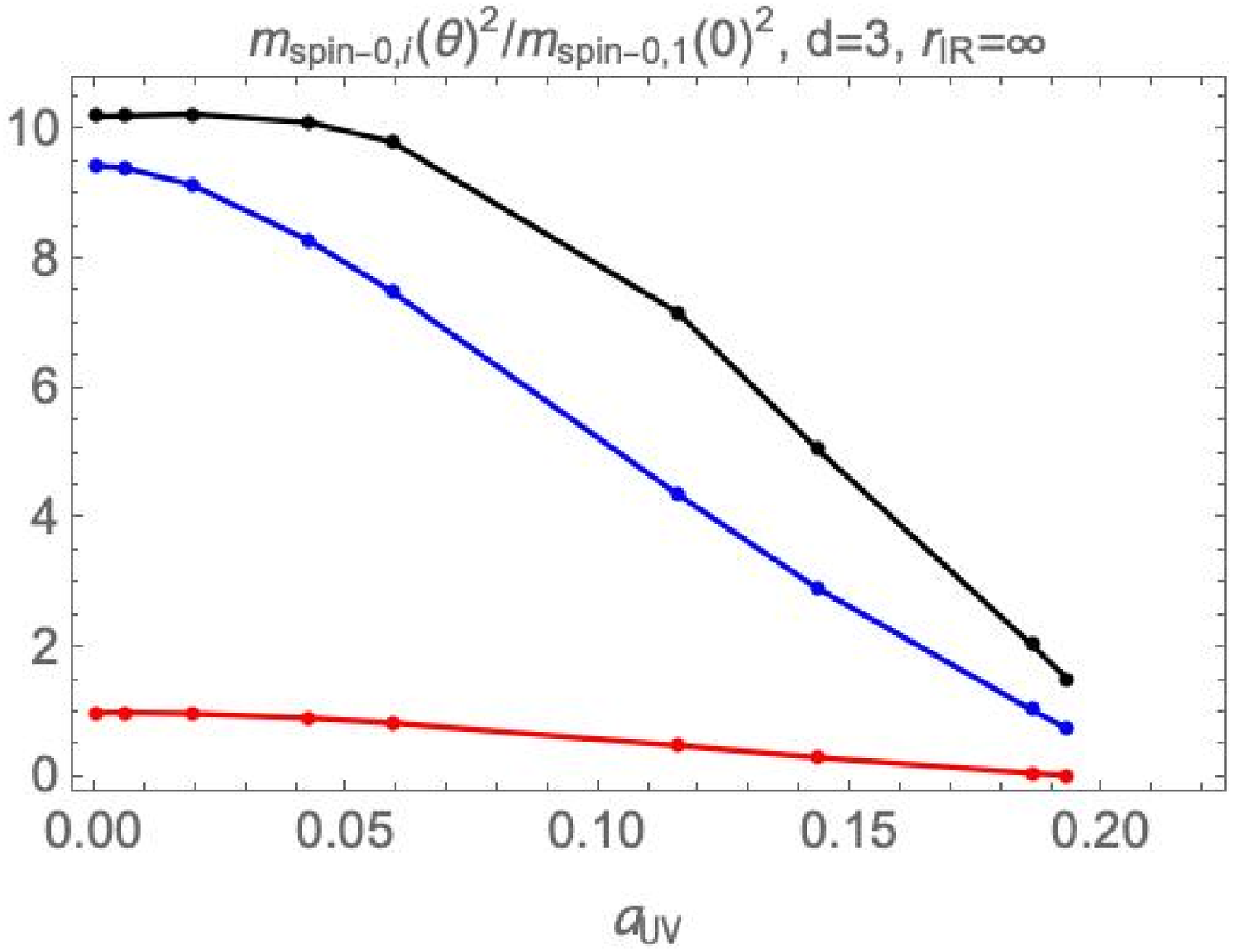}\hfil\hfil
  \includegraphics[width=.49\textwidth]{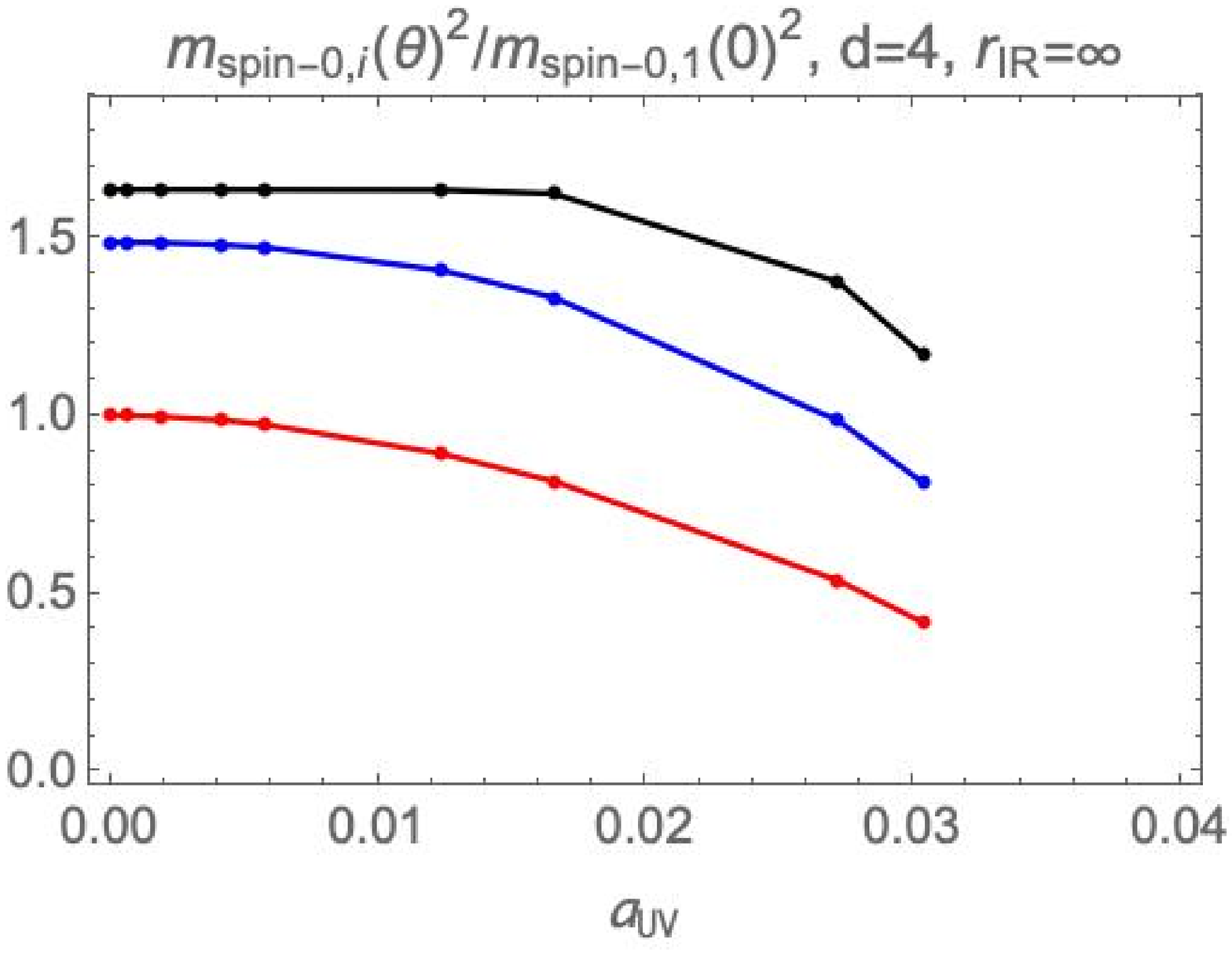}\hfil\hfil
 \end{center}
 \caption{Plots of the lowest three spin-0 glueball mass squared with the bulk functions \protect\eqref{P1}. The model parameters are \protect\eqref{4d1} (\textbf{left})  and \protect\eqref{4d5} (\textbf{right}). The masses are normalized by mass squared of the lightest glueball mass at $a_{UV}=0$. Note that there is no invariant distinction between the scalar and pseudo-scalar glueballs for $a_{UV}\neq0$.
 }
  \label{fig10}
 \end{figure}

   \begin{figure}[t]
 \begin{center}
  \includegraphics[width=.49\textwidth]{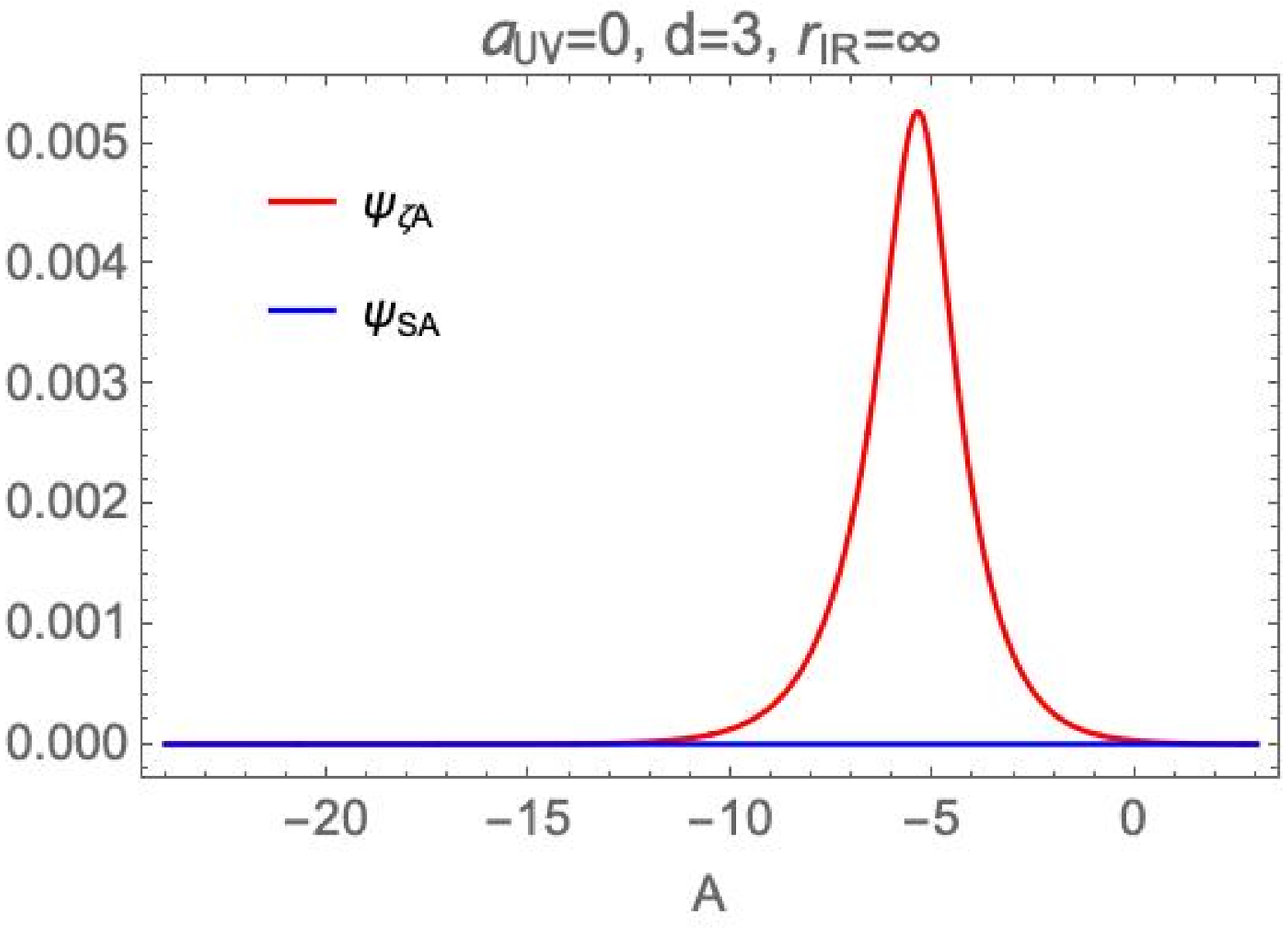}\hfil\hfil
   \includegraphics[width=.49\textwidth]{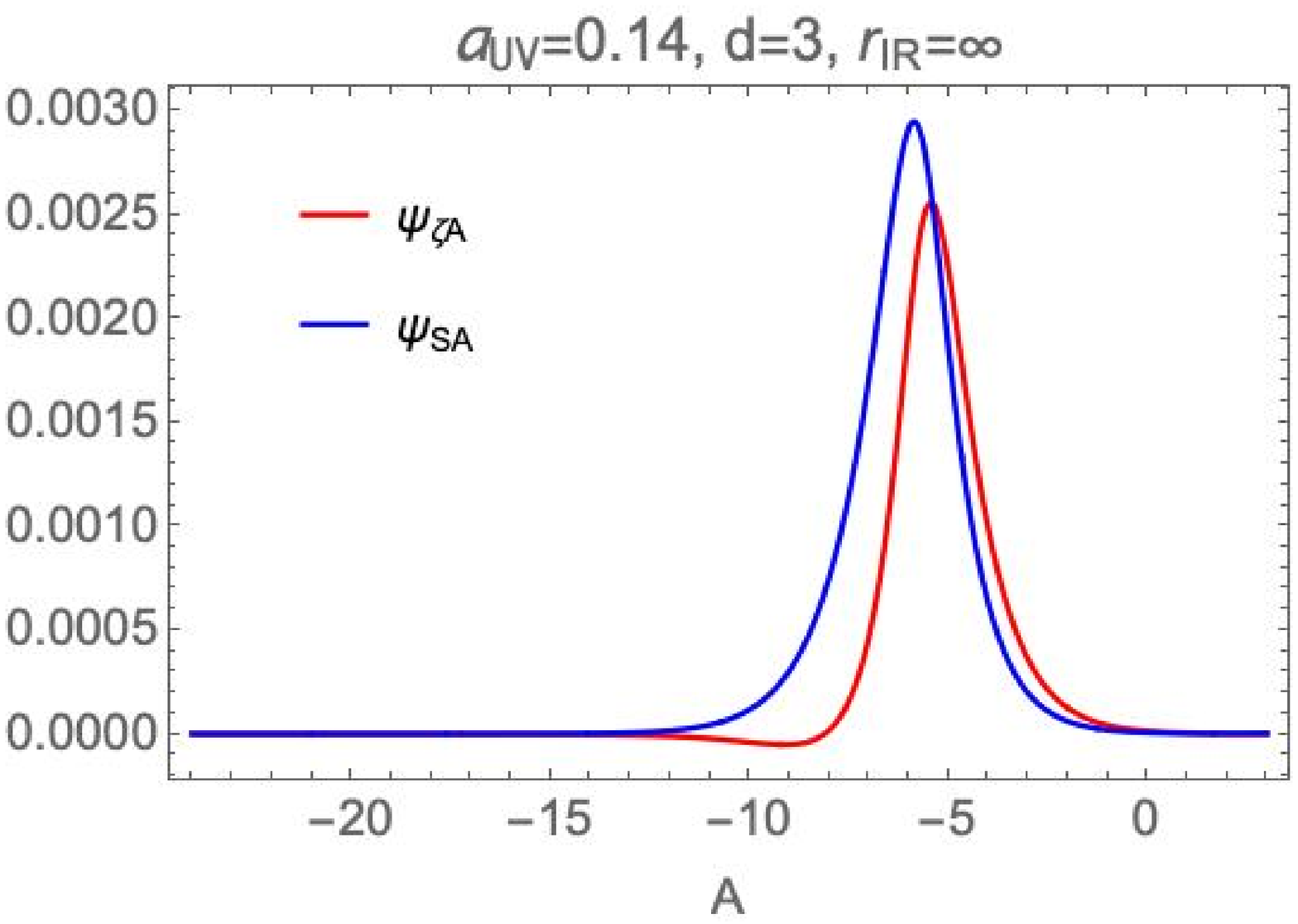}\hfil\hfil
   \\
  \includegraphics[width=.49\textwidth]{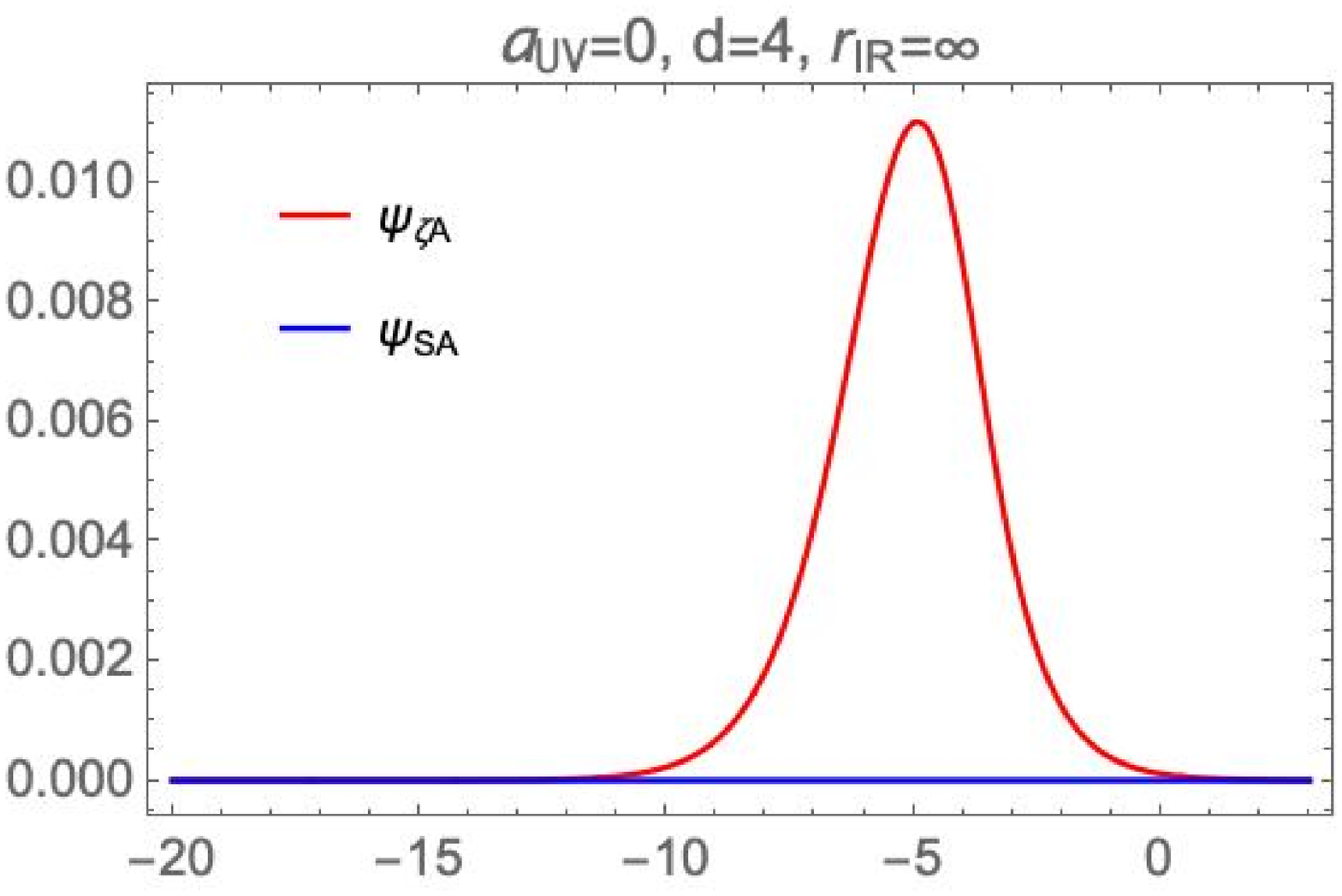}\hfil\hfil
   \includegraphics[width=.49\textwidth]{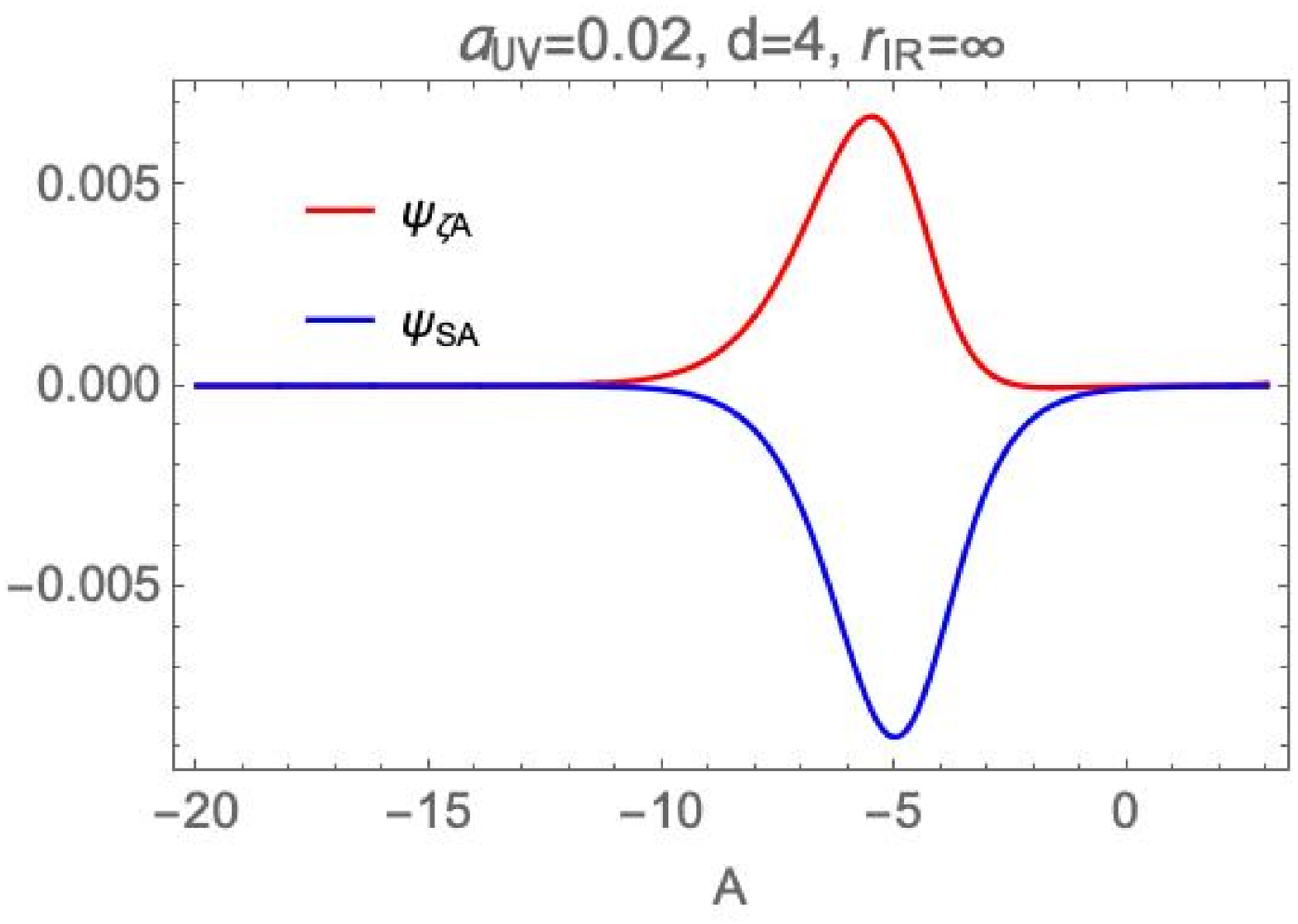}\hfil\hfil
 \end{center}
 \caption{Plots of wave-functions $\psi_{\zeta A}$ and $\psi_{{\cal S} A}$ corresponding to the lightest glueball as functions of $A$ with the bulk functions \protect\eqref{P1}.
 \textbf{Upper left:} The model parameters are \protect\eqref{4d1} with $a_{UV}=0$.
 \textbf{Upper right:} The model parameters are \protect\eqref{4d1} with $a_{UV}=0.14$.
 \textbf{Lower left:} The model parameters are \protect\eqref{4d5} with $a_{UV}=0$.
 \textbf{Lower right:} The model parameters are \protect\eqref{4d5} with $a_{UV}=0.02$.
 }
  \label{fig8-2}
 \end{figure}

   \begin{figure}[t]
 \begin{center}
  \includegraphics[width=.49\textwidth]{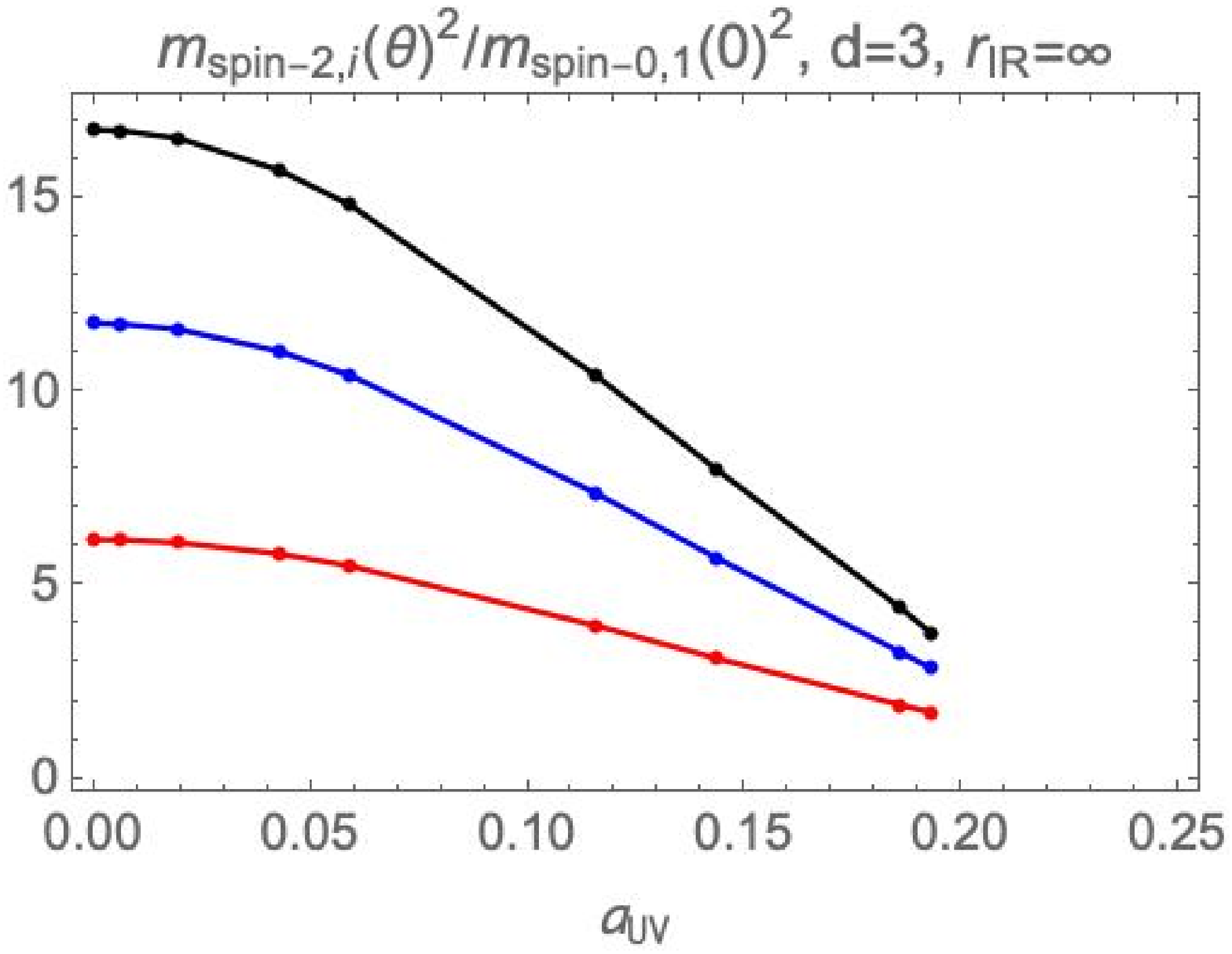}\hfil\hfil
  \includegraphics[width=.49\textwidth]{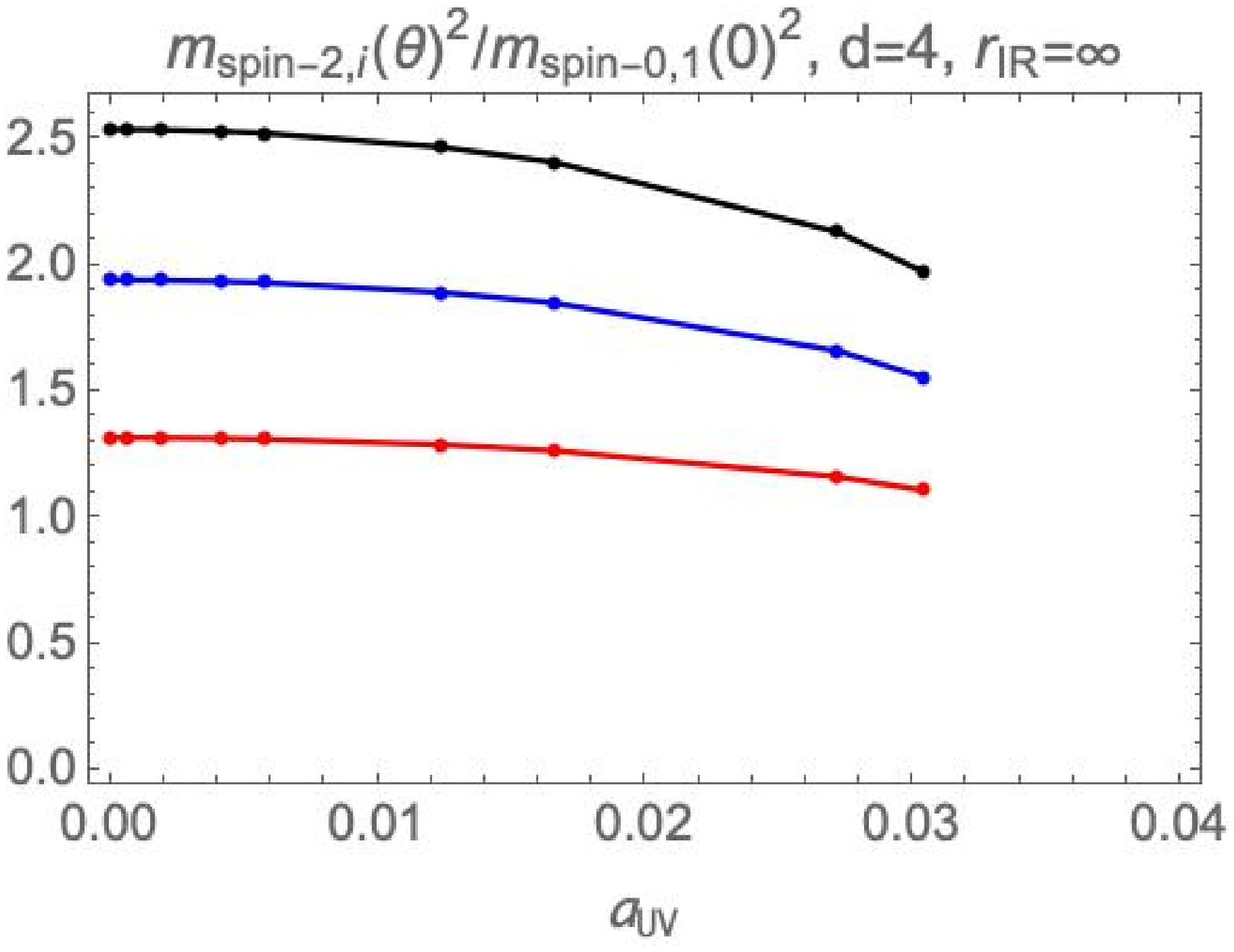}\hfil\hfil
 \end{center}
 \caption{Plots of the lowest three spin-2 (\textbf{right}) glueball mass squared with the bulk functions \protect\eqref{P1}. The model parameters are \protect\eqref{4d1} (\textbf{left})  and \protect\eqref{4d5} (\textbf{right}). The masses are normalized by mass squared of the lightest glueball mass at $a_{UV}=0$.
 }
  \label{fig10-2}
 \end{figure}

As we discussed so far, we solve the Schr\"odinger equations \eqref{tr21} to obtain the mass and wave-function of the scalar and pseudo-scalar glueballs. Note that, in the presence of the non-trivial axion source $a_{UV}\neq0$, there is no distinction between the scalar and pseudo-scalar glueballs.
We use the method described in appendix H in \cite{disc}. For bulk functions, we consider \eqref{Num1} and \eqref{P1}. These correspond to steep and soft potentials, respectively.

As in section \ref{background_numerical}, we use model parameters with $d=3$ and $d=4$. We shall observe that the $\theta$-dependence of glueball masses and wave-functions is qualitatively similar in $d=3$ and $d=4$.

\subsubsection{Steep potentials}

We use the bulk functions \eqref{Num1} with the model parameters \eqref{4d2} and \eqref{4d2-2}.
In the absence of the axion source, we obtain
\be
m_\zeta^2 ~\f_-^{-2/\Delta_-}  \approx
\begin{cases}
20, \quad 250, \quad 570, \quad \ldots, \quad \quad \text{for \eqref{4d2}}, \\
1.2, \quad 6.2, \quad 13, \quad \ldots, \quad \quad \text{for \eqref{4d2-2}},
\end{cases}
\label{4d3}\ee
\be
m_{\cal S}^2 ~\f_-^{-2/\Delta_-} \approx
\begin{cases}
180, \quad 420, \quad 820, \quad \ldots, \quad \quad \text{for \eqref{4d2}}, \\
4.1, \quad 11, \quad 19, \quad \ldots, \quad \quad \text{for \eqref{4d2-2}}.
\end{cases}
\label{4d4}\ee

In the presence of $a_{UV}$, the lowest three scalar and pseudoscalar glueball masses as functions of $a_{UV}$ are shown in figures \ref{fig20} and \ref{fig20-2}. The right panels are enlarged views of the lightest glueball mass.
The masses are normalized by the lightest glueball mass at $a_{UV}=0$.
We observe that all glueball masses are decreasing functions of $a_{UV}$.
Moreover, we find that the lightest glueballs become tachyonic for $0.25\lesssim a_{UV}\leq a_{UV}^{max}\approx0.3$ in figure \ref{fig20}, and for $0.13\lesssim a_{UV}\leq a_{UV}^{max}\approx0.135$ in figure \ref{fig20-2}.

Wave-functions of the lightest glueball are shown in figure \ref{fig8}. The upper and lower figures correspond to the model parameter \eqref{4d2} and \eqref{4d2-2}, respectively. The upper left and right panels use $a_{UV}=0$ and $a_{UV}=0.15$, while the lower left and right panel use $a_{UV}=0$ and $a_{UV}=0.12$. We obtain $\psi_{\zeta A}\neq0$ and $\psi_{{\cal S} A}=0$ for $a_{UV}=0$, as it should be.
For a larger value of $a_{UV}=0.12 \text{ or } 0.15$, an amplitude of $\psi_{{\cal S} A}$ is comparable with that of $\psi_{\zeta A}$.

The decrease of all glueball masses, as $\theta$ is increased from zero, found here, is similar to what was observed in \cite{Dubovsky:2011tu,BC1}  using   Witten's black D$_4$ holographic model. The tachyon instability however does not appear there. In our case (steep potentials), the dilaton potential in the IR is steeper and this may be at the origin of the instability.

In lattice QCD, the leading $\theta^2$ correction to the glueball mass, also turns out to be negative, as reported in \cite{DelDebbio:2006yuf}. This is consistent with our holographic computation.

Masses of the spin-2 glueballs are plotted in figure \ref{fig20-3}. The masses are normalized by the lightest spin-0 glueball mass at $a_{UV}=0$. The left and right panels correspond to the model parameters \eqref{4d2} and \eqref{4d2-2}, respectively. As in the spin-0 glueballs, the all spin-2 glueball masses are monotonically decreasing functions of $a_{UV}$. We do not observe the tachyonic instability of the spin-2 glueballs.

\subsubsection{Soft potentials}

We use the bulk functions \eqref{P1} with the model parameters \eqref{4d1} and \eqref{4d5}.
In the absence of the axion source, the masses of the scalar and pseudo-scalar glueballs are
\be
m_\zeta^2 ~\f_-^{-2/\Delta_-} \approx
\begin{cases}
35, \quad 320, \quad 510, \quad \ldots, \quad \quad \text{for \eqref{4d1}}, \\
0.95, \quad 1.6, \quad  2.1, \quad \ldots, \quad \quad \text{for \eqref{4d5}},
\end{cases}
\label{u1}\ee
\be
m_{\cal S}^2 ~\f_-^{-2/\Delta_-} \approx
\begin{cases}
350, \quad 530, \quad 700, \quad \ldots, \quad \quad \text{for \eqref{4d1}}, \\
1.4, \quad 2.1, \quad 2.7, \quad \ldots, \quad \quad \text{for \eqref{4d5}}.
\end{cases}
\label{u2}\ee
The masses are modified in the presence of the non-trivial axionic flow.
The lowest three glueball masses squared are plotted in figure \ref{fig10} as functions of $a_{UV}$.
As in the previous figures, the masses are normalized by the lightest glueball mass at $a_{UV}=0$.
We again observe that all glueball masses are decreasing function of $a_{UV}$.
However, contrary to the case of steep potentials,  in this background we do not observe  tachyonic instabilities of the glueballs for both parameter choices \eqref{4d1} and \eqref{4d5}.
In figure \ref{fig8-2}, we plot the wave-functions corresponding to the lightest three glueballs.
We again observe that an amplitude of $\psi_{{\cal S} A}$ is comparable with that of $\psi_{\zeta A}$ in the presence of the axion source.

Finally, spin-2 glueball masses are plotted in figure \ref{fig10-2}, where the left and right panels correspond to \eqref{4d1} and \eqref{4d5}, respectively. The spin-2 glueball masses are monotonically decreasing functions of $a_{UV}$, and do not exhibit the tachyonic instability.

\section{The cubic interaction terms and dynamical CP-violation}\label{cubic}

It was long suspected that renormalization effects drive the effective $\theta$-angle to zero in the IR, therefore softening the strong CP-problem, \cite{Polyakov}. In holographic theories, we have the first explicit example of this phenomenon, by interpreting the axion solution in the standard holographic fashion.  
It is therefore a nice laboratory to study the potential softening of CP-violating effects. We will investigate this question here, in the effective theory of scalar glueballs.

Therefore, in this section, we shall  compute the cubic coupling among the spin-0 glueballs. To this end, we need to know the action up to the third order in the fluctuations. In this paper, instead of computing the cubic action directly in our setup, a highly non-trivial task, we borrow the result \cite{Garcia-Saenz:2019njm} of the calculation of non-gaussianities in the context of the inflationary cosmology.
By performing the proper analytic continuation \cite{McFadden:2009fg,McFadden:2010vh}, we obtain the desired action.
Since the calculation of cosmology is performed in $d=3$, we focus on the four-dimensional bulk space-time, which would be dual to a three-dimensional quantum field theory.

Using the wave-functions obtained by solving the Schr\"odinger equations, we evaluate the strength of CP-violating couplings.
Since the $\theta$-angle flows to zero in the IR, a naive expectation is that CP-violating couplings are suppressed by the effect of the running of the bulk axion.

Our goal is to check whether this naive expectation is correct by a holographic computation.
Although we work in $d=3$, we believe that qualitatively similar results hold for $d=4$,  because we observed similar behavior in $d=3$ and $d=4$ in sections \ref{background_numerical} and \ref{numerical_glueball}.\footnote{In the background corresponding to $r_{IR}=$finite, we observed the similarity between $d=3$ and $d=4$ theories in the whole region of $a_{UV}$. In the background corresponding to $r_{IR}=\infty$, we observed the similarity for small $a_{UV}$.}

The cubic action of \cite{Garcia-Saenz:2019njm} and the procedure of the analytic continuation \cite{McFadden:2009fg,McFadden:2010vh} are summarized in appendix \ref{continuation}.
After the computation, it turns out that there are four types of cubic couplings depending on the structure of the momenta. Here we concentrate on cubic couplings without $x^\mu$-derivatives.
The action at the cubic order in the scalar fluctuations of our action without $x^\mu$-derivatives, $S^{(3)}_{\Psi^3}$,  is\footnote{The full cubic order action including derivative couplings is given in (\ref{a-thi19-2}, \ref{a-thi19-3}).}
\be
S^{(3)}_{\Psi^3}= \int^{r_{IR}}_{r_{UV}} dr \int d^3x \,e^A {\cal L}^{(3)}_{\Psi^3}
= -\int^{\infty}_{-\infty} dA \int d^3x \, \frac{e^A}{A'} {\cal L}^{(3)}_{\Psi^3},
\label{thi19-2}\ee
\be
\frac{{\cal L}_{\Psi^3}^{(3)}}{M_p^2}
=-2 e^{A}
\epsilon (\epsilon - \eta) \zeta'^2 \zeta
+e^{3A} \bigg[
\frac{1}{2} m_s^2 (\epsilon + \mu_s) \zeta {\cal F}^2
-(2\epsilon -\eta -2 \lambda_\perp) \frac{\sigma' \eta_\perp}{e^{2A}}  \zeta \zeta' {\cal F}+
\label{thi19-3}\ee
$$
+ \frac{\sigma' \eta_\perp}{e^{3A} H}  {\cal F} \zeta'^2
+ \frac{H \eta_\perp^2 - 2 \epsilon H R_{\text{fs}}}{e^A} \zeta' {\cal F}^2
- \frac{V_{;sss} + 2e^{-A} \sigma' H \eta_\perp R_{\text{fs}} -2 \epsilon H^2 R_{\text{fs},s}}{6} {\cal F}^3
- 2 e^{-2A}\epsilon \, \zeta {\cal F}'^2
\bigg].
$$
Here $\epsilon, \eta, m_s^2, \mu_s, \lambda_\perp, \eta_\perp, V_{;sss}, R_\text{fs}, R_{\text{fs},s}$ are defined in \eqref{a-thi14}, \eqref{a-thi24}, \eqref{re7}, \eqref{a-thi23}, \eqref{a-thi23-2}, \eqref{a-thi16}, \eqref{a-thi6}, and \eqref{a-thi23-3}, respectively.
As we explained in section \ref{background_numerical}, we can use $A$ as a coordinate instead of using $r$. The fields $\zeta, {\cal F}$ are related with the glueball wave-functions $(\psi_{\zeta A}, \psi_{{\cal S}A})$ as
\be
\begin{pmatrix}
\zeta \\
{\cal F}
\end{pmatrix}
=\frac{e^{-\frac{3}{2}A} }{\sqrt{W}}
\begin{pmatrix}
1 & 0 \\
0 & -1
\end{pmatrix}
\begin{pmatrix}
\cos\Phi & \sin\Phi \\
-\sin\Phi & \cos\Phi
\end{pmatrix}
\begin{pmatrix}
\psi_{\zeta A} \\
\psi_{{\cal S}A}
\end{pmatrix}.
\label{cu1}\ee
The effective cubic interaction term without $x^\mu$-derivative, $D_{\Psi^3, ijk}$, is
\be
S=\int d^3x
\,D_{\Psi^3, ijk} \psi^{(3)}_i\psi^{(3)}_j\psi^{(3)}_k,
\label{thi29-7}\ee
where $\psi^{(3)}_i$ is the three-dimensional part of the KK decomposition \eqref{pt32}.
Note that $i=1,2,\ldots$ correspond to the lightest spin-0 glueball, second lightest spin-0 glueball, and so on. For example, $D_{\Psi^3, 111}$ is the cubic interactions among the lightest spin-0 glueballs.
The cubic term $D_{\Psi^3, ijk}$ is computed by
\be
D_{\Psi^3, ijk} = \int^{A_{UV}}_{A_{IR}} dA\, C_{\Psi^3, ijk},
\quad
\frac{{\cal L}_{\Psi^3}^{(3)}}{M_p^2}\equiv -\frac{A'}{e^A} C_{\Psi^3, ijk} \psi^{(3)}_i\psi^{(3)}_j\psi^{(3)}_k.
\label{thi29-8}\ee

 \begin{figure}[t]
 \begin{center}
 \includegraphics[width=.49\textwidth]{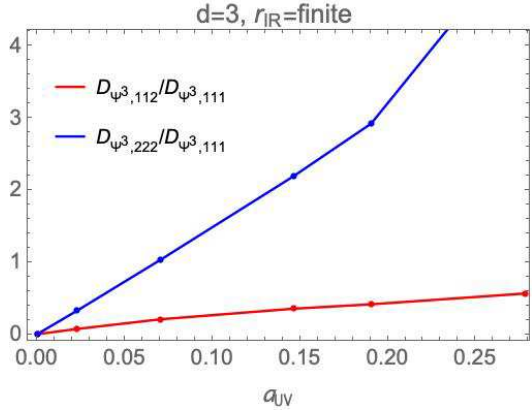}\hfil\hfil
 \end{center}
 \caption{Plot of ratios $D_{{\Psi^3},112}/D_{{\Psi^3},111}$ and $D_{{\Psi^3},222}/D_{{\Psi^3},111}$ as functions of $a_{UV}$ with the bulk functions \protect\eqref{Num1}. The model parameters are \protect\eqref{4d2}. Note that the lightest glueball becomes tachyonic for $0.25\lesssim a_{UV}\leq a_{UV}^{max}\approx0.3$ (figure \protect\ref{fig20}).
 }
  \label{fig21}
 \end{figure}

  \begin{figure}[t]
 \begin{center}
  \includegraphics[width=.49\textwidth]{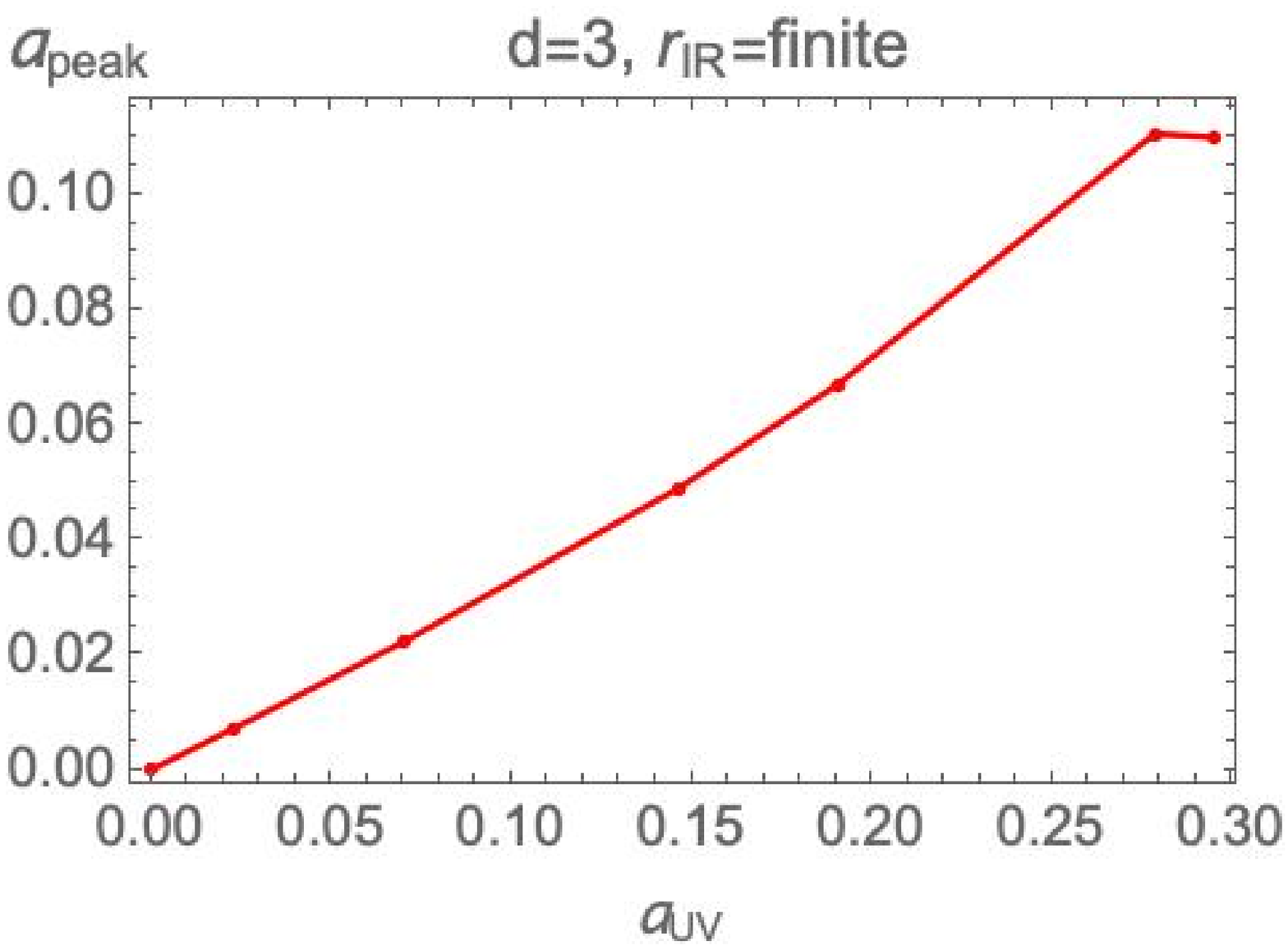}\hfil\hfil
  \includegraphics[width=.49\textwidth]{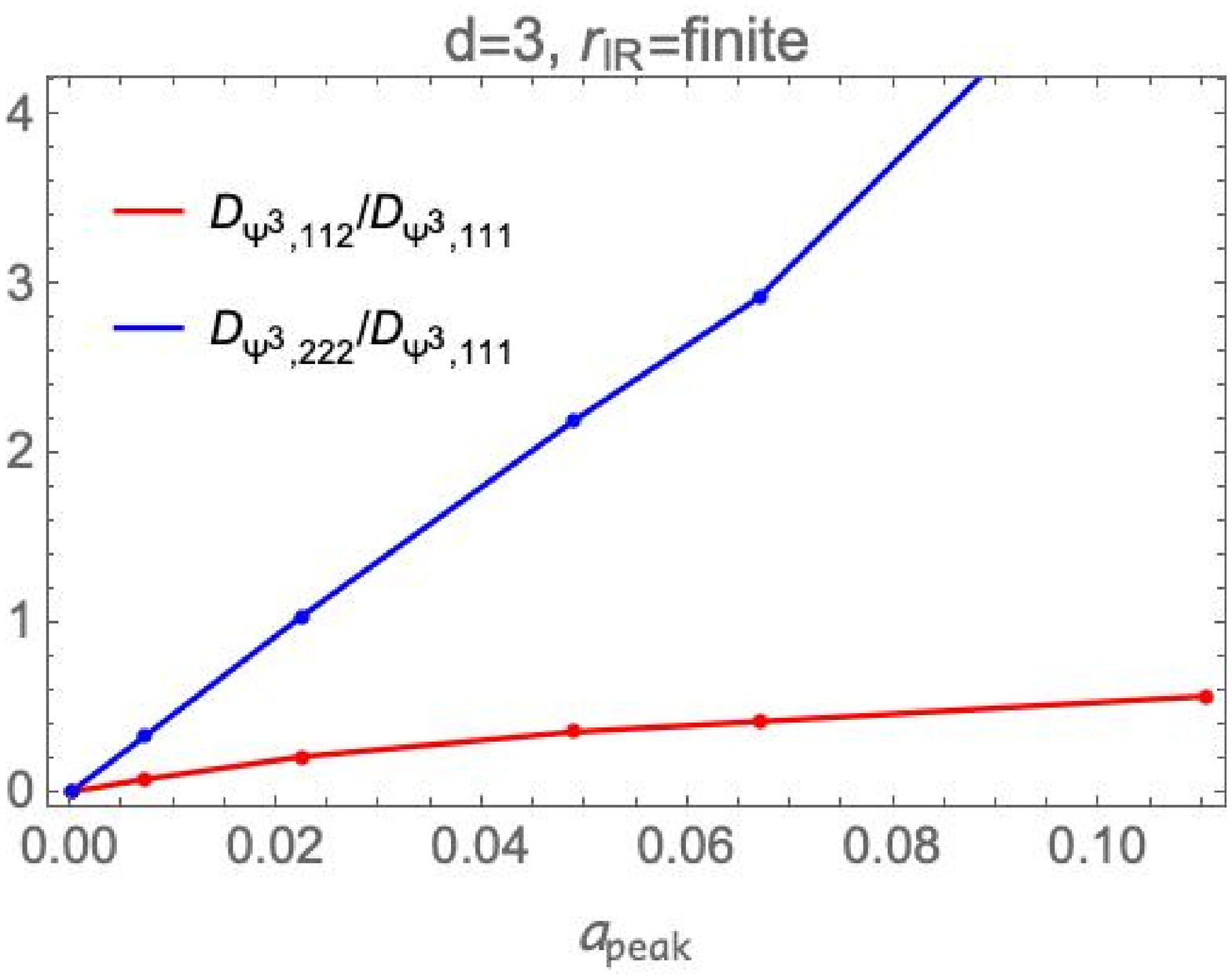}\hfil\hfil
 \end{center}
 \caption{Plots of $a_\text{peak}$ defined in \protect\eqref{tn8} as a function of $a_{UV}$ (\textbf{left}) and ratios $D_{{\Psi^3},112}/D_{{\Psi^3},111}, D_{{\Psi^3},222}/D_{{\Psi^3},111}$ (\textbf{right}) as functions of $a_\text{peak}$. The bulk functions are \protect\eqref{Num1}. The model parameters are \protect\eqref{4d2}.
 }
  \label{fig22}
 \end{figure}

\subsection{Steep potentials}

We use the bulk functions \eqref{Num1} with the model parameters \eqref{4d2}. In the absence of the axion source, the lightest spin-0 glueball is a CP-even state while the second lightest spin-0 glueball is a CP-odd state. CP-violating three point couplings are zero for  vanishing $\theta$-angle,
\be
D_{\Psi^3,112}=D_{\Psi^3,222}=0,
\quad
\text{for $a_{UV}=0$.}
\label{tn1}\ee
On the other hand, CP-conserving couplings such as $D_{\Psi^3,111}$ are nonvanishing.

In the presence of a non-zero axion source, these CP-violating couplings become nonvanishing.
The values of the CP-violating cubic couplings normalized by $D_{\Psi^3,111}$ are plotted in figure \ref{fig21}.
We observe that the ratios $D_{\Psi^3_{112}}/D_{\Psi^3_{111}}$ and $D_{\Psi^3_{222}}/D_{\Psi^3_{111}}$ are almost linear in $a_{UV}$,
\be
D_{\Psi^3_{112}}/D_{\Psi^3_{111}} \approx 2 \,a_{UV},
\quad
D_{\Psi^3_{222}}/D_{\Psi^3_{111}} \approx 15 \,a_{UV}.
\label{tn8-2}\ee
Contrary to the naive expectation mentioned in the beginning of the section, we do not find the suppression of CP-violation by the running effect.

As another measure of the CP-violation, we introduce $a_\text{peak}$ as
\be
a_\text{peak} = \left.a(A)\right|_{A=A_\text{peak}},
\label{tn8}\ee
where $A_\text{peak}$ is the value of $A$ where the square of the lightest spin-0 glueball wave-function, $\left|\psi_{\zeta A}^{(1)}\right|^2 + \left|\psi_{{\cal S} A}^{(1)}\right|^2$, becomes  maximum. This is the region in the bulk where the bulk the radial  wavefunction is peaked, and it gives a rough measure of the energy scale in the dual field theory which is most relevant  for glueball interactions.

In figure \ref{fig22}, the value of $a_\text{peak}$ as a function of $a_{UV}$ (left) and the CP-violating cubic couplings as functions of $a_\text{peak}$ are plotted. We observe that $a_\text{peak}$ is almost linear in $a_{UV}$, and the right panel of figure \ref{fig22} looks similar to figure \ref{fig21}.

\subsection{Soft potentials}

 \begin{figure}[t]
 \begin{center}
  \includegraphics[width=.49\textwidth]{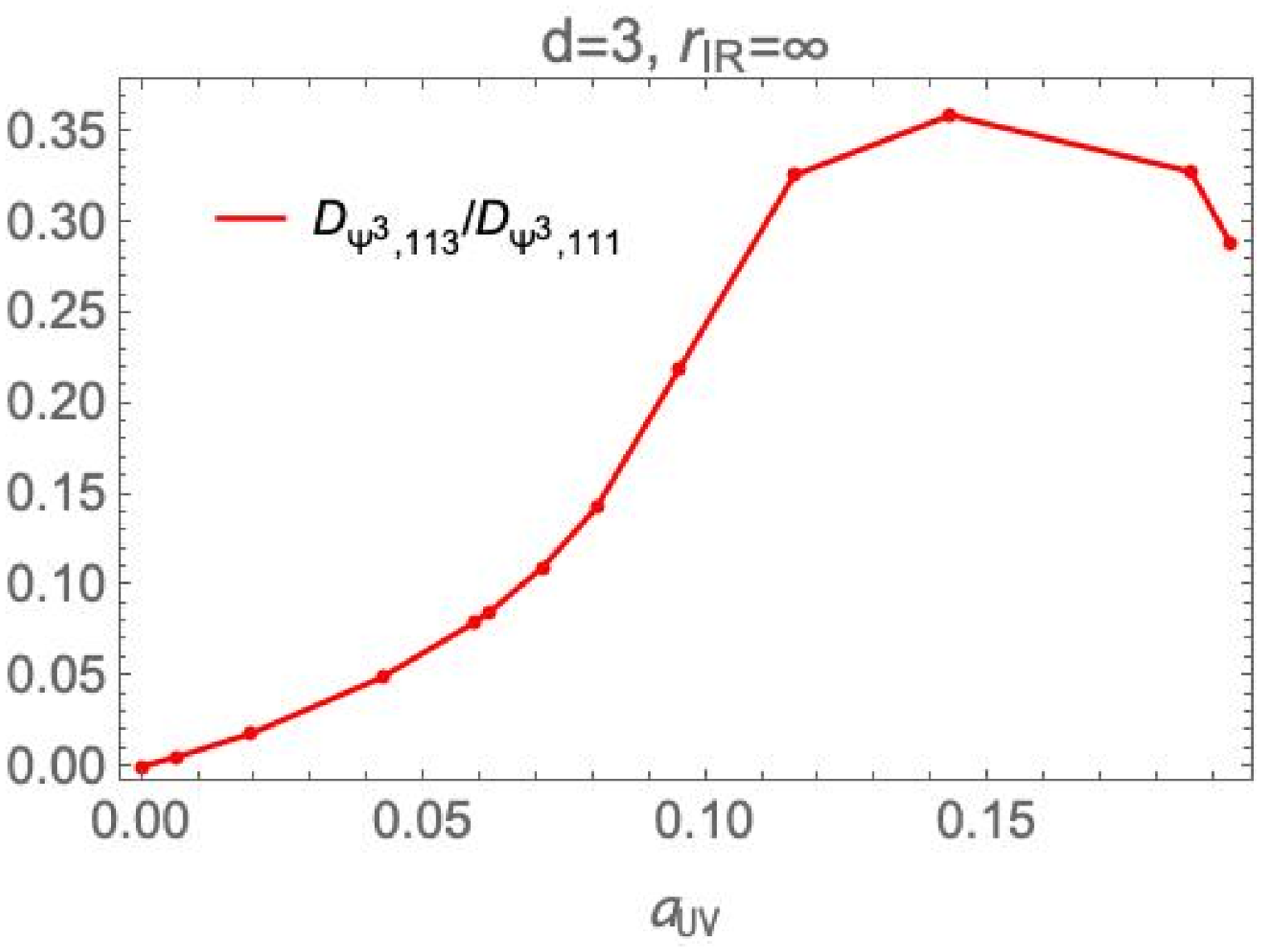}\hfil\hfil
  \includegraphics[width=.49\textwidth]{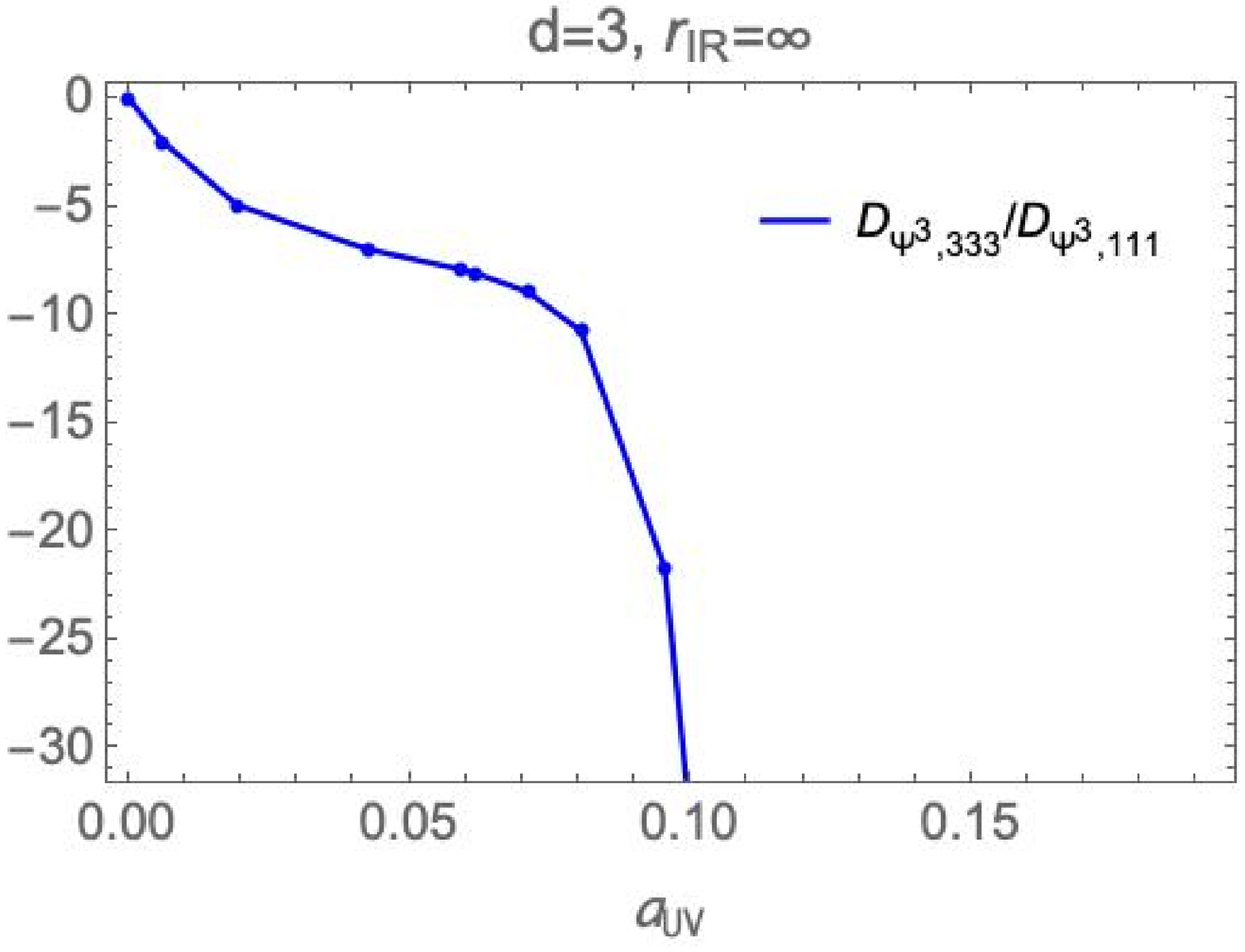}\hfil\hfil
 \end{center}
 \caption{Plots of $D_{{\Psi^3},113}/D_{{\Psi^3},111}$ (\textbf{left}) and $D_{{\Psi^3},333}/D_{{\Psi^3},111}$ (\textbf{right}) as functions of $a_{UV}$ with the bulk functions \protect\eqref{P1}. The model parameters are \protect\eqref{4d1}.
 }
  \label{fig11}
 \end{figure}

 \begin{figure}[t]
 \begin{center}
  \includegraphics[width=.49\textwidth]{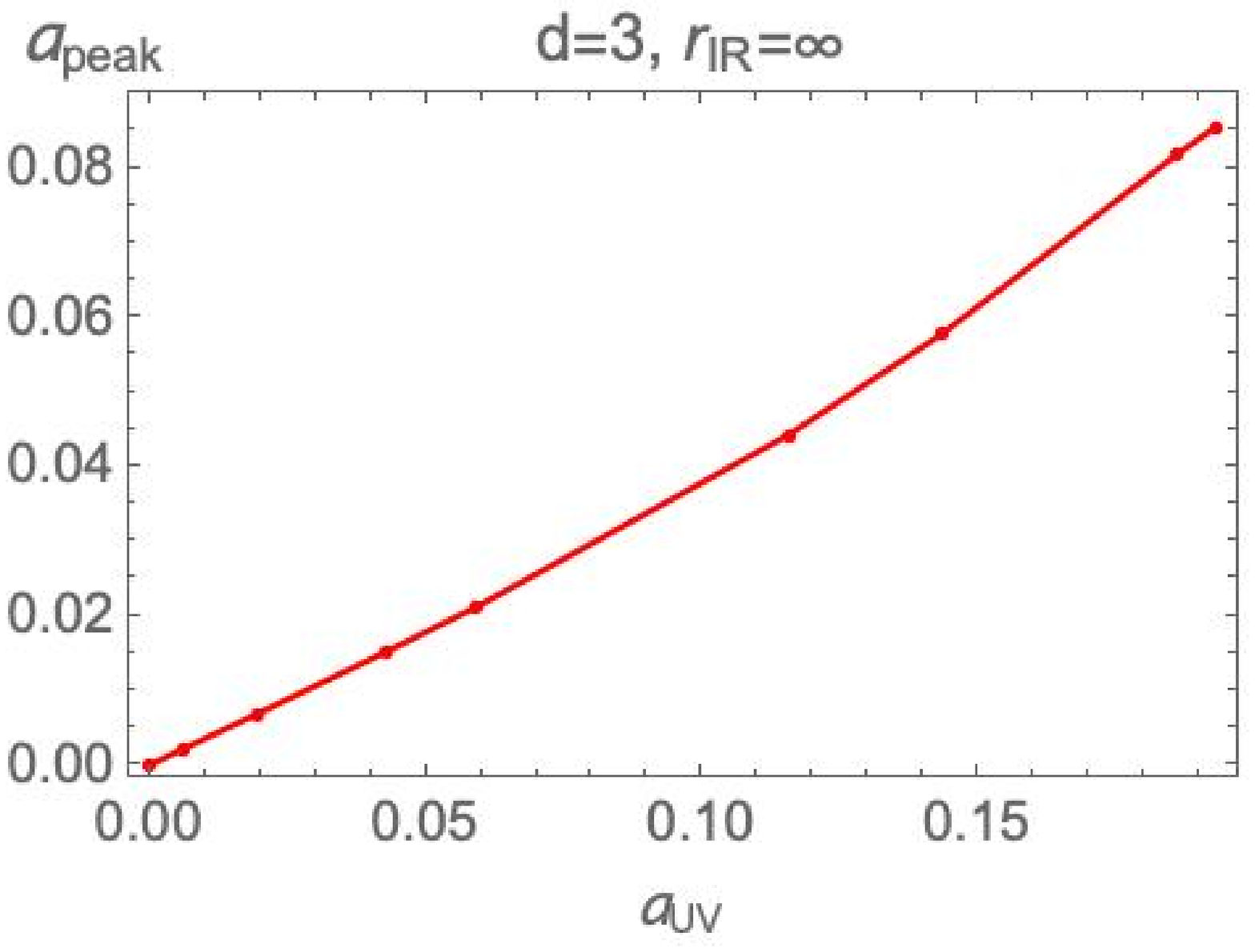}\hfil\hfil \\
  \includegraphics[width=.49\textwidth]{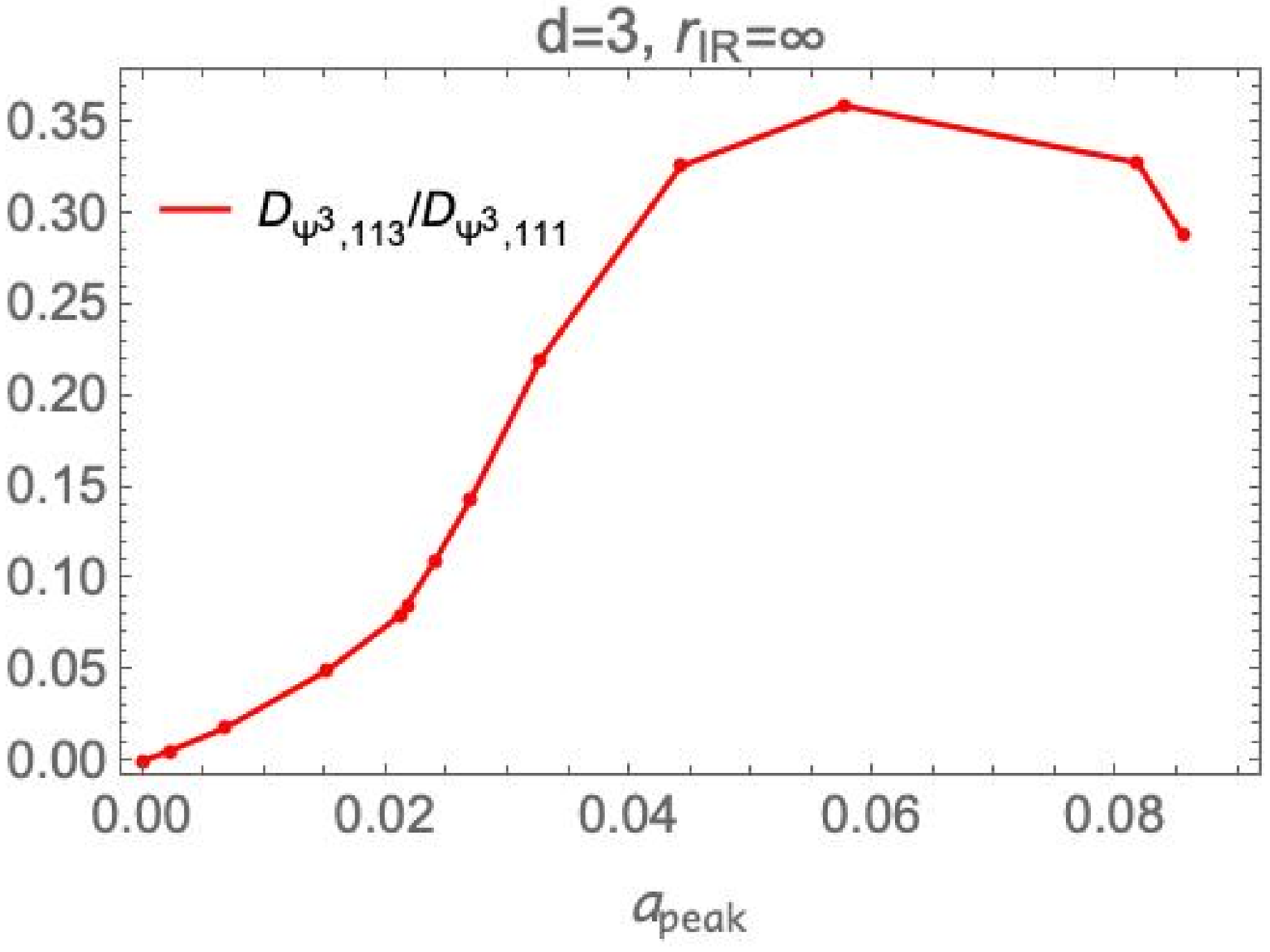}\hfil\hfil
  \includegraphics[width=.49\textwidth]{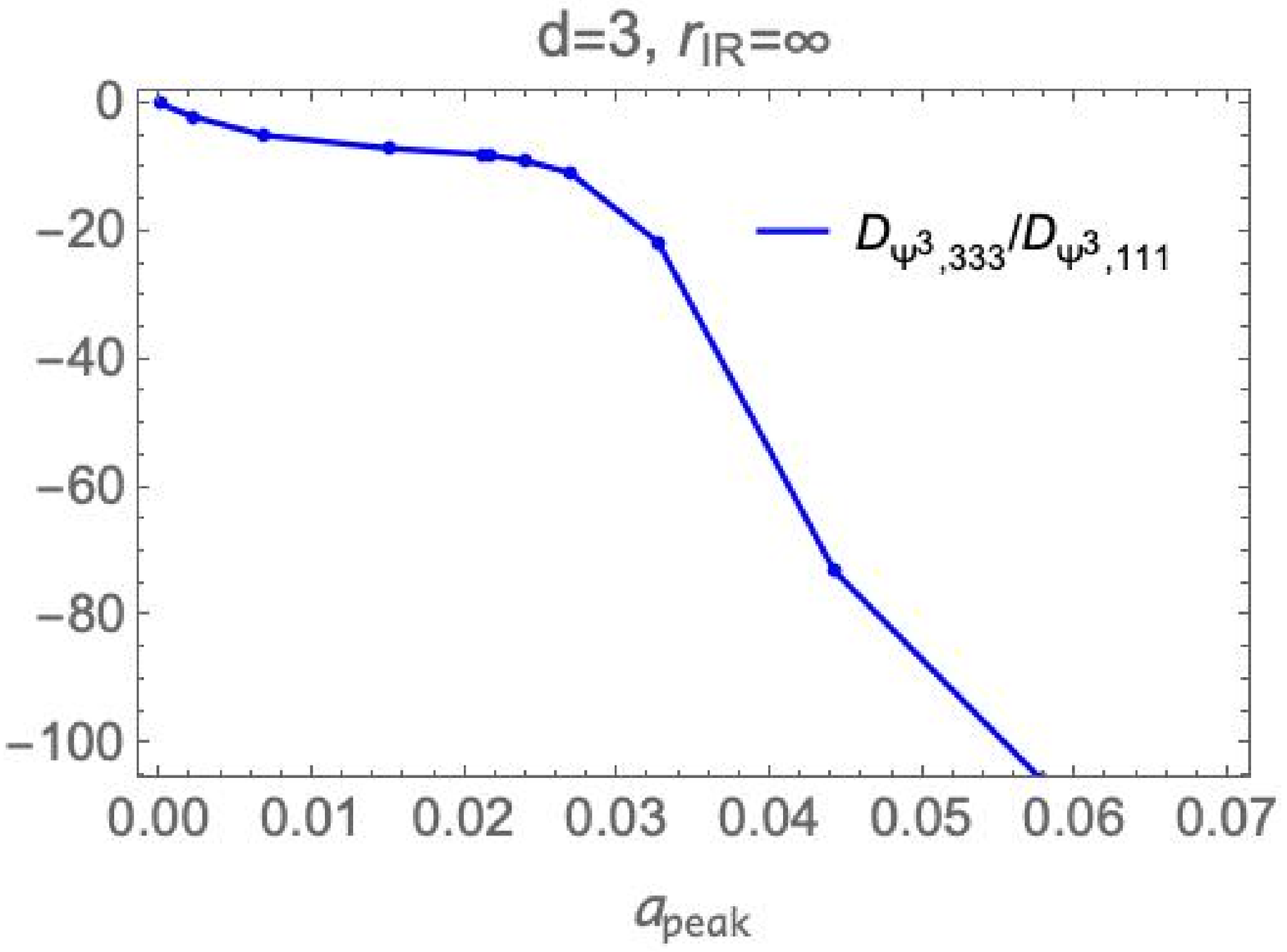}\hfil\hfil
 \end{center}
 \caption{Plots of $a_\text{peak}$ as a function of $a_{UV}$ (\textbf{upper}), $D_{{\Psi^3},113}/D_{{\Psi^3},111}$ (\textbf{lower left}) and $D_{{\Psi^3},333}//D_{{\Psi^3},111}$ (\textbf{lower right}) as functions of $a_\text{peak}$. The bulk functions are \protect\eqref{P1}. The model parameters are \protect\eqref{4d1}.
 }
 \label{fig11-2}
 \end{figure}

Here, we use the bulk functions \eqref{P1} with the model parameters \eqref{4d1}.
In this background, the lightest spin-0 glueball is a CP-even state while the third lightest spin-0 glueball is a CP-odd state. There is no CP-violation in the absence of the axion source:
\be
D_{{\Psi^3},113}=D_{{\Psi^3},333}=0,
\quad
\text{for $a_{UV}=0$.}
\label{tn2}\ee
In figure \ref{fig11}, we plot the values of the CP-violating couplings in the presence of the axion source $a_{UV}\neq0$. The couplings are normalized by $D_{\Psi^3,111}$.
We observe that, for small $a_{UV}$, the ratios $D_{\Psi^3,113}/D_{\Psi^3,111}$ and $D_{\Psi^3,333}/D_{\Psi^3,111}$ are proportional to $a_{UV}$:
\be
D_{\Psi^3_{113}}/D_{\Psi^3_{111}} \approx a_{UV},
\quad \text{for $a_{UV}\lesssim0.06$},
\label{tn3}\ee
$$
D_{\Psi^3_{333}}/D_{\Psi^3_{111}} \approx -200 \,a_{UV} ,
\quad \text{for $a_{UV}\lesssim0.02$}.
$$
Again we do not observe the suppression of the CP-violating couplings.
As $a_{UV}$ is increased, the ratios are no longer linear functions of $a_{UV}$.

The value of $a_\text{peak}$ as a function of $a_{UV}$ is plotted in an upper panel of figure \ref{fig11-2}.
The CP-violating couplings as functions of $a_\text{peak}$ are plotted in lower panels. The lower panels of figure \ref{fig11-2} are qualitatively similar to the panels in figure \ref{fig11}.

\section*{Acknowledgements}
\addcontentsline{toc}{section}{Acknowledgements}

We would like to thank A. Cotrone, M. Jarvinen, E. Shuryak, L. Witkowski and I. Zahed for useful conversations.

This work was supported in part by European Union's Seventh Framework Programme under grant agreements (FP7-REGPOT-2012-2013-1) no 316165 and the Advanced ERC grant SM-grav, No 669288. YH thanks the hospitality of the Kavli Institute for Theoretical Physics (supported by NSF PHY-1748958) where part of this work was carried out.

\newpage
\appendix

\begin{appendix}
\renewcommand{\theequation}{\thesection.\arabic{equation}}
\addcontentsline{toc}{section}{Appendices}
\section*{APPENDIX}

\section{Geometry of field space}\label{geometry}
It is convenient to summarize the geometry of the field space of the action \eqref{A2}.
The field space metric $G_{IJ}$ is defined in \eqref{thi3}.
The nonzero components of the Christoffel symbol constructed from $G_{IJ}$ are
\be
\Gamma^\f_{aa}=-\frac{1}{2}\de_\f Y,
\quad
\Gamma^a_{\f a}=\Gamma^a_{a\f}=\frac{1}{2}\frac{\de_\f Y}{Y}.
\label{a-thi5}\ee
The field space Ricci scalar and Riemann tensor are
\be
R_\text{fs} = \frac{(\de_\f Y)^2}{2Y^2} - \frac{\de_\f^2 Y}{Y}, \quad
R_{IJKL} = \frac{R_\text{fs}}{2} \left( G_{IK}G_{JL} - G_{IL}G_{JK} \right).
\label{a-thi6}\ee
The field space Ricci scalars corresponding to \eqref{Num1} and \eqref{P1} are
\be
R_{\text{fs}} = -\frac{\g^2}{2},
\quad
-(d-1) \left( 1 - \frac{Y_c^2}{\left(Y_c + Y_\infty e^{\sqrt{2(d-1)}\f}\right)^2} \right),
\label{Num1-3}\ee
respectively.

The vielbein along a background field trajectory is
\be
e_\sigma^I = \frac{1}{\sigma'}\left( \f', a'\right),
\quad e_{\sigma I} =  \frac{1}{\sigma'}\left( \f', Y a'\right),
\label{a-thi7}\ee
where a prime stands for a $r$-derivative, and $\sigma'$ is
\be
\sigma' \equiv \sqrt{\f'^2+Y a'^2}= \sqrt{2(d-1)(A'^2-A'')}= e^A \sqrt{S^2+\frac{T}{Y}}.
\label{a-re4}\ee
Here (\ref{FE7-ii}, \ref{th2}, \ref{a1}, \ref{a6}) is used.
The basis orthogonal to $e_\sigma^I$  is
\be
e_s^I = \frac{\sqrt{Y}}{\sigma'} \left( a', -\frac{\f'}{Y}\right),
\quad
e_{s I} = \frac{\sqrt{Y}}{\sigma'} \left( a', -\f' \right).
\label{a-thi9}\ee
The vectors $e_\sigma^I$ and $e_s^I$ satisfy
\be
G_{IJ} e_\sigma^Ie_\sigma^J = G_{IJ} e_s^I e_s^J=1, \quad
G_{IJ} e_\sigma^Ie_s^J=0.
\label{a-thi10}\ee

Motivated by the cosmology \cite{Garcia-Saenz:2019njm}, ``Hubble parameter" $H$ is defined as
\be
H\equiv \frac{\dot{\left(e^{A}\right)}}{e^A} = e^{-A}A'
= -\frac{W}{2(d-1)},
\label{a-thi13}\ee
and  ``slow-roll parameters" $\epsilon$ and $\eta$ are
\be
\epsilon \equiv -\frac{\dot{H}}{H^2}
= 1 - \frac{A''}{A'^2}
= \frac{1}{2(d-1)} \frac{\sigma'^2}{A'^2}
= 2(d-1) \frac{S^2+\frac{T}{Y}}{W^2},
\label{a-thi14}\ee
\be
\eta \equiv \frac{\dot{\epsilon}}{H\epsilon}
= \frac{1}{\epsilon} \frac{d\epsilon}{dA}
= \frac{2A''^2-A'A'''}{A'^2 (A'^2-A'')}
= 2(d-1) \left[
2 \frac{ S^2 + \frac{T}{Y} }{W^2}
- \frac{d}{d-1}
- \frac{2S \de_\f V}{W\left(S^2+\frac{T}{Y}\right)}
\right].
\label{a-thi24}\ee
The bending parameter $\eta_\perp$ is defined as
\be
{\cal D}_r e_\sigma^I
= e^{A} H \eta_\perp e_s^I, \quad
{\cal D}_r e_s^I
=-e^{A} H \eta_\perp e_\sigma^I,
\label{a-thi17}\ee
where the action of the covariant derivative ${\cal D}_r$ on the field space vector $A^I$ is
\be
{\cal D}_r A^I \equiv
\p_r A^I+ \Gamma^I_{JK} (\p_r \phi^J) A^K.
\label{a-thi17-2}\ee
Using \eqref{a-re4}, \eqref{FE7-ii} and \eqref{a6}, $\eta_\perp$ is written as
\be
\eta_\perp =
\frac{\sqrt{Y}a'}{A'\sigma'^2}
\left(
\f'' - \frac{a''}{a'}\f' - \frac{\de_\f Y}{Y}\f'^2 -\frac{\de_\f Y}{2} a'^2
\right)
=\sqrt{Y} a' \frac{e^{2A} \de_\f V}{A'\sigma'^2}
\label{a-thi18}\ee
$$
= -2(d-1) {\rm sign}(Q) \sqrt{\frac{T}{Y}} \frac{\de_\f V}{W \left( S^2+\frac{T}{Y}\right)}
= -2(d-1) e^{-A}\sqrt{Y} \frac{da}{dr} \frac{\de_\f V}{W \left( S^2+\frac{T}{Y}\right)}
=\sqrt{Y} \frac{da}{dA} \frac{\de_\f V}{S^2+\frac{T}{Y}}.
$$

The projection of the second order covariant derivative of the potential along the entropic direction is
\be
V_{;ss}\equiv e^I_s e^J_s V_{;IJ}
= e^I_s e^J_s \left( \de_I \de_J  V - \Gamma^K_{IJ} \de_K V\right)
= \frac{1}{\sigma'^2}
\left[ \frac{1}{2} (\de_\f V) \f'^2 \frac{\de_\f Y }{Y}
+ (\de_\f^2 V) Y a'^2
\right],
\label{a-thi15}\ee
$$
= \frac{1}{Y S^2+T}
\left(
\frac{\de_\f V}{2}S^2 \de_\f Y + T \de_\f^2 V
\right).
$$

Similarly, the projection of the third order covariant derivative of the potential along the entropic direction is
\be
V_{;sss}= e^I_s e^J_s e^K_s V_{;IJK}=
\label{a-thi16}\ee
$$
=\frac{{\rm sign}(Q)}{2}\sqrt{\frac{T}{Y}}\frac{1}{\left(S^2+\frac{T}{Y}\right)^{3/2}}
\left[
S^2\left\{ \de_\f V_0\left(\frac{\de_\f^2 Y }{Y}-2\frac{(\de_\f Y)^2}{Y^2}\right)+3\de_\f^2 V \frac{\de_\f Y}{Y} \right\}
+ 2\frac{T}{Y} \de_\f^3 V
\right].
$$
The parameters $\lambda_\perp$ and $\mu_s$ are defined as
\be
\lambda_\perp
\equiv\frac{\eta_\perp'}{A' \eta_\perp}=
d + \frac{\sigma'^2}{2(d-1)A'^2} - \frac{\f'}{2A'} \left( \frac{4e^{2A}\de_\f V}{\sigma'^2} - \frac{2\de_\f^2 V}{\de_\f V} + \frac{\de_\f Y}{Y}\right)
\label{a-thi23}\ee
$$
=d
+ 2(d-1) \frac{S^2+\frac{T}{Y}}{W^2}
+ (d-1) \frac{S}{W} \left( \frac{4 \de_\f V_0}{S^2+\frac{T}{Y}} - \frac{2\de_\f^2 V}{\de_\f V} + \frac{\de_\f Y}{Y} \right),
$$
\be
\mu_s
\equiv\frac{(m_s^2)'}{A' m_s^2}
=-2(d-1)\frac{S}{W} \frac{d}{d\f}\left(\log m_s^2\right).
\label{a-thi23-2}\ee

Finally, the derivative of $R_\text{fs}$ along the entropic direction is
\be
R_{\text{fs},s}=e^I_s \p_I R_\text{fs}=
-\frac{\sqrt{Y}a'}{\sigma'} \left[
\left(\frac{\de_\f Y}{Y}\right)^3
-2\frac{(\de_\f Y)(\de_\f^2 Y)}{Y^2}
+\frac{\de_\f^3 Y}{Y}
\right]
\label{a-thi23-3}\ee
$$
= -{\rm sign}(Q) \sqrt{\frac{T}{YS^2+T}}
\left[
\left(\frac{\de_\f Y}{Y}\right)^3
-2\frac{(\de_\f Y)(\de_\f^2 Y)}{Y^2}
+\frac{\de_\f^3 Y}{Y}
\right].
$$

\section{Conformal coordinate system}\label{conformal}
The conformal coordinate system is related to the domain wall coordinate system through
\be
{du \over dr} = e^A.
\label{a-th2}\ee
We fix an integration constant in such a way that the UV boundary is at $r_{UV}=0$.
By using the conformal coordinate $r$, the bulk equations of motion are
\be
2(d-1) (A''-A'^2) + \f'^2 + Y a'^2 = 0,
\label{a1-2}\ee
\be
\f'' + (d-1)A' \f' - e^{2A}\de_\f V - \frac{\de_\f Y}{2}a'^2 = 0,
\label{a3-2}\ee
\be
\pa_r(Y e^{(d-1)A}a')=0,
\label{a4-2}\ee
where a prime stands for a $r$ derivative.
The energy scale $\mu$ of a dual QFT is roughly identified as
\be
\mu  \leftrightarrow e^{A(r)}.
\label{a-th11}\ee

The derivative of $A$ and $\f$ with respect to $r$ is given by
\be
{dA\over dr} = -{e^A\over2(d-1)} W,
\quad
{d\f\over dr} = e^A S,
\quad
{da\over dr} = {\rm sign}(Q) e^A \frac{\sqrt{T}}{Y},
\label{sa2}\ee
from \eqref{FE7-ii} and \eqref{a6}.

\section{Asymptotics of the wave-functions}\label{wavefunction}
In this appendix, we study IR and UV solutions of the Schr\"odinger equations \eqref{tr21}.
\subsection{IR solutions}
We consider the two types IR asymptotics, \eqref{a32} (steep potentials) and \eqref{qc1}-\eqref{soft}   (soft potentials).

\noindent \textbf{Steep potentials.}
From (\ref{sub8-2}, \ref{a-thi18}, \ref{tr16}), the quantities which appear in \eqref{tr21} are calculated as
\be
\frac{d^2 B_{\zeta A}}{dA^2} + \left( \frac{dB_{\zeta A}}{dA} \right)^2 - \eta_\perp^2
= \left[\frac{1}{2}\left(d-1-\frac{1}{\delta}\right)\right]^2+\ldots,
\ee
\be
\frac{d^2 B_{ {\cal S} A}}{dA^2} + \left( \frac{dB_{ {\cal S} A}}{dA} \right)^2  + m_{{\cal S}A}^2
= \left[\frac{1}{2}\left( (d-1)[1+b(\g-\g_\text{min})] - \frac{1}{\delta} \right)\right]^2
+\ldots,
\ee
and
$$
2\eta_\perp \frac{dB_{\zeta A}}{dA}+\frac{d\eta_\perp}{dA}
= \mathcal{O}\left(e^{\frac{(d-1)b}{2}(\g-\g_\text{min})A}\right),
$$
$$
\Phi = \Phi_{IR} + \mathcal{O}\left(e^{\frac{(d-1)b}{2}(\g-\g_\text{min})A}\right)
= \Phi_{IR} + \mathcal{O}\left((r_0-r)^{\frac{(d-1)b\delta}{2}(\g-\g_\text{min})}\right),
$$
where $\Phi_{IR}=\left.\Phi\right|_{A=-\infty}$. We choose $\Phi_{IR}=0$ in the following.
We observe that the mixing term in \eqref{tr21} is neglected in the IR.
The wave-functions in the IR are
\be
\frac{\psi_{\zeta A}}{A'} \sim
c_1 e^{\left(d-1+\frac{1}{\delta}\right)\frac{A}{2}} + c_2 e^{-\left(d-1-\frac{3}{\delta}\right)\frac{A}{2}},
\label{sc14}\ee
\be
\frac{\psi_{{\cal S} A}}{A'} \sim d_1 e^{\left((d-1)[1+b(\g-\g_\text{min})]+\frac{1}{\delta}\right)\frac{A}{2}} + d_2 e^{\left(-(d-1)[1+b(\g-\g_\text{min})]-\frac{3}{\delta}\right)\frac{A}{2}}.
\label{s33}\ee
For $\delta\geq3/(d-1)$, we should impose $c_2=0$ for the normalizability \eqref{cu3}. On the other hand, for $\delta<3/(d-1)$, both the solutions are acceptable, and we should impose an extra boundary condition in order to obtain a discrete spectrum.
Similarly, the solution corresponding to $d_1$ is always acceptable, while the solution corresponding to $d_2$ is acceptable only for $\delta (1+b(\g-\g_\text{min}))<3/(d-1)$.

\noindent \textbf{Soft potentials.}

From (\ref{s27}, \ref{a-thi18}, \ref{tr16}), we obtain
\be
\frac{d^2 B_{\zeta A}}{dA^2} + \left( \frac{dB_{\zeta A}}{dA} \right)^2 - \eta_\perp^2 = \frac{(d-1)^2}{4}+\ldots,
\quad
\frac{d^2 B_{ {\cal S} A}}{dA^2} + \left( \frac{dB_{ {\cal S} A}}{dA} \right)^2  + m_{{\cal S}A}^2 = \frac{(d-1)^2}{4}+\ldots,
\label{sc5}\ee
and
\be
2\eta_\perp \frac{dB_{\zeta A}}{dA}+\frac{d\eta_\perp}{dA}
= \mathcal{O}\left((-A)^{-\frac{d+1}{2}P}\right),
\quad
\Phi = \text{const.} + \mathcal{O}\left(r^{-\frac{d+1}{2}\frac{P}{1-P}}\right).
\label{sc5-2}\ee
The mixing term in \eqref{tr21} can be neglected in the IR.
By using \eqref{sc5}, in the IR, the solutions of \eqref{tr21} are
\be
\psi_{\zeta A} \simeq
c_1 e^{\frac{d-1}{2}A}
+c_2 e^{-\frac{d-1}{2}A},
\quad
\psi_{{\cal S} A} \simeq
d_1 e^{\frac{d-1}{2}A}
+d_2 e^{-\frac{d-1}{2}A}.
\label{sc6}\ee
We should choose $c_2=d_2=0$ in order to satisfy the normalizability condition \eqref{cu3}.

\subsection{UV solutions}
Contrary to the IR solutions, UV solutions are universal for two types of bulk potentials.
Using \eqref{th2} and \eqref{UV23}, we obtain
\be
{du\over dr} = e^{-u/\ell} + \ldots,
\label{g7}\ee
from which we find
\be
r = \ell \,e^{u/\ell} + r_{UV} +\ldots,
\label{g8}\ee
where $r_{UV}$ is the integration constant, $\left.r\right|_{u=-\infty}\equiv r_{UV}$.
We take $r_{UV}=0$.
By substituting \eqref{g8} into \eqref{UV23}, we observe
\be
\f(r)= \f_- \ell^{\Delta_-} \left( r\over\ell \right)^{\Delta_-}
+ {C d |\f_-|^{\Delta_+\over\Delta_-} \ell^{\Delta_+} \over (\Delta_+-\Delta_-) \Delta_-}  \left( r\over\ell \right)^{\Delta_+}
 + \ldots
\label{g9}\ee
$$
e^{A(r)} = {\ell\over r} + \ldots,
\quad
a(r)= a_{UV} + {Q \ell^d \over d Y_0} \left(r\over \ell\right)^d + \ldots
$$

From \eqref{re8}, we observe
\be
\Phi = \text{const.}+\mathcal{O}\left( e^{-(d-\Delta_-) A}\right),
\label{g13}\ee
which becomes constant in the UV.

The quantities which appears in \eqref{tr21} are calculated as
\be
\frac{d^2 B_{\zeta A}}{dA^2} + \left( \frac{dB_{\zeta A}}{dA} \right)^2 - \eta_\perp^2 = 
\frac{(d-2\Delta_-)^2}{4} +\ldots,
\quad
\frac{d^2 B_{ {\cal S} A}}{dA^2} + \left( \frac{dB_{ {\cal S} A}}{dA} \right)^2  + m_{{\cal S}A}^2 =
\frac{d^2}{4}+\ldots,
\label{g14-2}\ee
and
$$
2\eta_\perp \frac{dB_{\zeta A}}{dA}+\frac{d\eta_\perp}{dA}
= \mathcal{O}\left( e^{-(d-\Delta_-) A}\right).
$$

Up to a constant rotation, the solutions of Schr\"{o}dinger equation \eqref{tr21} are
\be
\psi_{\zeta A} =c_1 e^{\frac{1}{2}(d-2\Delta_-)A} + c_2 e^{-\frac{1}{2}(d-2\Delta_-)A} + \ldots,
\label{s13}\ee
\be
\psi_{{\cal S}A}
=d_1 e^{{d\over2}A} + d_2 e^{-\frac{d}{2}A}
+\ldots,
\label{s13-2}\ee
and the normalization condition is \eqref{cu3}. Using $1/A'^2\propto e^{-2A}$ in the UV, we observe that
$d_1=0$ is required for the normalizability assuming $d\geq2$.
Regarding $\psi_{\zeta A}$, we need to impose $c_1=0$ if $\Delta_-<\frac{d}{2}-1$ is satisfied.

\section{Universality of asymptotic glueball spectra}\label{universality}

In this appendix we will compute the asymptotic behavior of glueball masses using the WKB approximation.
In YM theory in particular, due to the string picture behind, we expect all glueballs to have the same asymptotics. This has been used in  \cite{iQCD} to fix the axion part of the action in Improved Holographic QCD.

We start from the Schr\"odinger form differential equation \eqref{re9}. To derive the asymptotic spectrum we replace $\p^\mu \p_\mu $ by $m_n^2$:
\be
\left[
-m_n^2
- \frac{d^2}{dr^2}
+\begin{pmatrix}
V_{\zeta\zeta} & V_{\zeta{\cal S}} \\
V_{\zeta{\cal S}} & V_{{\cal S}{\cal S}}
\end{pmatrix}
\right]
\begin{pmatrix}
\tilde{\psi}_\zeta \\
\tilde{\psi}_{\cal S}
\end{pmatrix}
=0,
\label{a-re9}\ee
where we defined
\be
\begin{pmatrix}
V_{\zeta\zeta} & V_{\zeta{\cal S}} \\
V_{\zeta{\cal S}} & V_{{\cal S}{\cal S}}
\end{pmatrix}
\equiv
{\cal R}^{-1}
\begin{pmatrix}
V_\zeta - \left(A' \eta_\perp \right)^2 & 2 A' \eta_\perp \left(B_\zeta' + \frac{(A' \eta_\perp)'}{2A' \eta_\perp}\right) \\
2 A' \eta_\perp \left(B_\zeta' + \frac{(A' \eta_\perp)'}{2A' \eta_\perp}\right) & V_{\cal S} - \left(A' \eta_\perp \right)^2
\end{pmatrix}
{\cal R}.
\label{w17}\ee
Here ${\cal R}$, $V_{\zeta, {\cal S}}$, $\Phi$ and $\eta_\perp$ are defined in \eqref{re7} and \eqref{re8}.
The explicit forms are,
\be
V_{\zeta\zeta}=
\frac{V_\zeta-V_{\cal S}}{2} \cos{2\Phi} - 2 A' \eta_\perp \left(B_\zeta' + \frac{(A' \eta_\perp)'}{2A' \eta_\perp}\right) \sin{2\Phi} + \frac{V_\zeta+V_{\cal S}}{2} -\left(A' \eta_\perp \right)^2,
\label{w28}\ee
$$
V_{\zeta{\cal S}}=2 A' \eta_\perp \left(B_\zeta' + \frac{(A' \eta_\perp)'}{2A' \eta_\perp}\right) \cos{2\Phi} + \frac{V_\zeta-V_{\cal S}}{2} \sin{2\Phi},
$$
$$
V_{{\cal S}{\cal S}}= -\frac{V_\zeta-V_{\cal S}}{2} \cos{2\Phi} + 2 A' \eta_\perp \left(B_\zeta' + \frac{(A' \eta_\perp)'}{2A' \eta_\perp}\right) \sin{2\Phi} + \frac{V_\zeta+V_{\cal S}}{2} - \left(A' \eta_\perp \right)^2.
$$

\subsection{WKB approximation with a mixing term}

According to \cite{WKB} the correct WKB ansatz for our case is
\be
 \tilde{\psi}_\zeta =  X_{\zeta}(r) \, e^{iU(r)}
\sp
\tilde{\psi}_{\cal S} =  X_{\cal S}(r) \, e^{iU(r)}.
\label{w13a}\ee
Substituting into (\ref{a-re9}) and expanding in powers of $\hbar$ (that here we have set to one) we obtain to the first two leading order equations
\be
\left(\begin{matrix} U'^2+V_{\zeta\zeta}-m^2 & V_{\z\ss}\\
V_{\z\ss}& U'^2+V_{\ss\ss}-m^2\end{matrix}\right)\left(\begin{matrix} X_{\z}\\
X_{\ss}\end{matrix}\right)=\left(\begin{matrix} 0\\
0\end{matrix}\right),
\label{w14a}\ee
\be
2X_{\z,\ss}' U'+X_{\z,\ss}U''=0.
\label{w15a}\ee
In order for (\ref{w14a}) to have a nontrivial solution we must have
\be
\left(U'^2+V_{\zeta\zeta}-m^2\right)\left( U'^2+V_{\ss\ss}-m^2\right)-V_{\z\ss}^2=0
\label{w16a}\ee
with solutions
\be
U_{\pm}'^2=m^2-{V_{\z\z}+V_{\ss\ss}\over 2}\pm\sqrt{\left({V_{\z\z}-V_{\ss\ss}\over 2}\right)^2+V_{\z\ss}^2}
\label{w20}\ee
$$
=m^2 - \frac{V_\z + V_\ss}{2} + (A'\eta_\perp)^2 \pm \sqrt{\left(\frac{V_\z - V_\ss}{2}\right)^2+4(A'\eta_\perp)^2 \left(B_\zeta' + \frac{(A' \eta_\perp)'}{2A' \eta_\perp}\right)^2},
$$
where \eqref{w28} is used in the second line.
Then (\ref{w15a}) gives
\be
X_{\z,\ss}={C_{\z,\ss}\over \sqrt{U'}},
\label{w60}\ee
where $C_{\z,\ss}$ are integration constants.
It is clear that we have two distinct towers of bound states associated with $U_{\pm}$. These towers become distinct at larges masses as they (generically) will have different slopes but are not distinct at low masses (unless the mixing is tiny).

The S\"ommerfeld quantization conditions become
\be
n_{\pm}\pi = \int^{r^{\pm}_2}_{r^{\pm}_1} dr~U'_{\pm},
\label{w18a}\ee
where $r_{1,2}^{\pm}$ are the turning points in $U_{\pm}$.
When
\be
V_{\z\ss}^2\ll \left({V_{\z\z}-V_{\ss\ss}}\right)^2
\label{w612}\ee
then
\be
U_{+}'^2=m^2-V_{\z\z}\sp U_{-}'^2=m^2-V_{\ss\ss}
\label{w622}\ee
and we recover the standard towers of $0^{++}$ and $0^{+-}$ glueballs.

The S\"ommerfeld quantization conditions \eqref{w18a} are written as
\be
n_{\pm}\pi = \left(\int^{r^{\pm}_a}_{r^{\pm}_1}dr+\int^{r^{\pm}_b}_{r^{\pm}_a}dr+\int^{r^{\pm}_2}_{r^{\pm}_b}dr\right) U'_{\pm},
\label{w18b}\ee
where we introduced $r^{\pm}_{a, b}$ in such a way that $U'_{\pm}$ are approximated by UV asymptotic forms for $r_1^\pm\leq r\leq r_a^\pm$ while $U'_{\pm}$ are approximated by IR asymptotic forms for $r_b^\pm\leq r\leq r_2^\pm$.
In the intermediate region $r_a^\pm \leq r \leq r_b^\pm$, $U'_{\pm}$ are approximated by $U'_{\pm}\simeq m$ for large $m$.
In the following, we first study the UV and IR asymptotic forms of $U'_{\pm}$.
Then, by solving \eqref{w18b}, we obtain asymptotic glueball spectra. As we shall see, the asymptotic spectra do not depend on the precise value of $r_{a,b}^\pm$ provided they are chosen in the right region.

\subsection{UV asymptotics}
The UV asymptotics is universal for the steep and soft potentials.
From (\ref{g9}, \ref{re7}), we obtain
\be
V_\z \simeq \frac{(d+1-2\Delta_-)(d-1-2\Delta_-)}{4r^2},
\quad
V_\ss \simeq \frac{d^2-1}{4r^2},
\label{w632}\ee
$$
A' \eta_\perp \simeq -\frac{Q}{\sqrt{Y_0}} \left(\frac{d}{\Delta_-}-1\right) \frac{r^{d-\Delta_- -1}}{\f_-},
\quad
B_\zeta' \simeq \frac{-d+2\Delta_-+1}{2r}\;,
$$
as $r\to 0$.
Equation \eqref{w20} becomes
{
\be
U_+'^2 = m^2 - V_\z \left(1 + \mathcal{O}\left(r^{2(d-\Delta_-)}\right)\right),
\quad
U_-'^2 = m^2 - V_\ss \left(1 + \mathcal{O}\left(r^{2(d-\Delta_-)}\right)\right).
\label{w642}\ee
}
The turning points $r_1^\pm$ are calculated as
\be
r_1^+ \simeq \frac{\sqrt{(d+1-2\Delta_-)(d-1-2\Delta_-)}}{2m},
\quad
r_1^- \simeq \frac{\sqrt{d^2-1}}{2m}.
\label{w652}\ee

\subsection{Steep potentials}
In the IR, from (\ref{th20-2}, \ref{re7}, \ref{re8}, \ref{a-thi18}), we obtain
\be
V_\zeta \simeq \frac{3}{4} \left(\frac{\delta}{\delta_\text{min}}\right)^2 \frac{1-2\left(\frac{\delta_\text{min}}{\delta}-1\right)}{(r_0-r)^2},
\label{sc13}\ee
$$
V_{\cal S} \simeq
\frac{3}{4} \left(\frac{\delta}{\delta_\text{min}}\right)^2 \frac{ (4-2\left(\frac{\delta_\text{min}}{\delta}-1\right)+3b(\g-\g_\text{min}))\left( 1 + b(\g - \g_\text{min}) \right) }{(r_0-r)^2},
$$
$$
A'\eta_\perp \simeq
-\frac{Q \ell^{d-1}}{\sqrt{2\ell^2 V_\infty Y_\infty}} \frac{(\frac{2d}{d-1}-b^2)^{\frac{3}{2}}}{\frac{d-1}{2}b^2-1} \left( \frac{\ell}{r_0-r} \right)^{1-\frac{b}{b^2-\frac{2}{d-1}}\left(\g-\g_\text{min}\right)},
\quad
B_\zeta' \simeq - \frac{\delta}{2} \frac{d-1}{r_0-r},
$$
where $\delta_\text{min}$ is defined in \eqref{a34-2}.
Equation \eqref{w20} becomes{
\be
U_+'^2 = m^2 - V_\z \left(1 + \mathcal{O}\left(\left( r_0-r \right)^{\frac{2b(\g-\g_\text{min})}{b^2-\frac{2}{d-1}}}\right)\right),
\quad
U_-'^2 = m^2 - V_\ss \left(1 + \mathcal{O}\left(\left( r_0-r \right)^{\frac{2b(\g-\g_\text{min})}{b^2-\frac{2}{d-1}}}\right) \right).
\label{w22}\ee
}

The turning points $r_2^\pm$ are
\be
r_2^+ {\simeq} r_0 - \frac{\sqrt{3}}{2m} \frac{\delta}{\delta_\text{min}} \sqrt{1-2\left(\frac{\delta_\text{min}}{\delta}-1\right)},
\label{w61}\ee
$$
r_2^- {\simeq} r_0 -  \frac{\sqrt{3}}{2m} \frac{\delta}{\delta_\text{min}} \sqrt{\left(4-2\left(\frac{\delta_\text{min}}{\delta}-1\right)+3b(\g-\g_\text{min})\right)(1+b(\g-\g_\text{min}))}.
$$

Using (\ref{w652}, \ref{w22}, \ref{w61}), \eqref{w18b} becomes
\be
n_+\pi \simeq \left(\int^{r_a^+}_{r_1^+}dr + \int^{r_2^+}_{r_b^+}dr \right) \sqrt{m^2 - V_\z }
+(r^{+}_a - r^{+}_b ) m
\label{w62}\ee
$$
= \frac{\sqrt{(d+1-2\Delta_-)(d-1-2\Delta_-)}}{4} \int^{1}_{\frac{(d+1-2\Delta_-)(d-1-2\Delta_-)}{4(r_a^+)^2m^2}} dy \frac{\sqrt{1-y}}{y^{3/2}} +
$$
$$
+ \frac{3}{4}\frac{\delta}{\delta_\text{min}}\sqrt{1-\frac{2}{3}\frac{\delta_\text{min}}{\delta}}\int^1_{\frac{3}{4m^2} \left(\frac{\delta}{\delta_\text{min}}\right)^2 \frac{1-2\left(\frac{\delta_\text{min}}{\delta}-1\right)}{(r_0-r_b^+)^2}} dz \frac{\sqrt{1-z}}{z^{3/2}}
+(r^{+}_a - r^{+}_b ) m
$$
$$
=mr_0 + \mathcal{O}(m^{0}),
$$
where we changed the variable from $r$ to $y=\frac{(d+1-2\Delta_-)(d-1-2\Delta_-)}{4m^2r^2}$ and $z=\frac{3}{4m^2} \left(\frac{\delta}{\delta_\text{min}}\right)^2 \frac{1-2\left(\frac{\delta_\text{min}}{\delta}-1\right)}{(r_0-r)^2}$.
From \eqref{w62}, we find
\be
m_{n_+} = \frac{\pi}{r_0} n_+ + \ldots
\label{a-w63}\ee
Performing the similar calculation for glueballs with the quantum number $n_-$, we obtain
\be
m_{n_-} = \frac{\pi}{r_0} n_-  + \ldots
\label{a-w64}\ee
Note that the IR end-point $r_0$ depends on $(b, \g)$ and $a_{UV}$. For example, from figures \ref{fig42} and \ref{fig42-2}, we observe that $r_0$ is a monotonically increasing function of $a_{UV}$.

The ratio of slopes of the two towers is
\be
\lim_{n\to\infty} \frac{m_{n_-}}{m_{n_+}}=1.
\label{a-w65}\ee

\subsection{Soft potentials}
We parametrize $Y$ generically as
\be
Y=Y_\infty e^{\g \f}
\label{w1}\ee
instead of $Y_\infty e^{\sqrt{2(d-1)}\f}$. At the end of the calculation, we shall observe that for $\frac{1}{d+1}<P<1$, $\g=\g_\text{min}=\sqrt{2(d-1)}$ realizes the asymptotic universality of the slopes of the glueball spectrum.

By using (\ref{re7}, \ref{s9}, \ref{s19-2}, \ref{s15}), we obtain (see \eqref{s29-5} for a lower bound on $P$)
\be
V_\zeta \simeq
\begin{cases}
v \,r^{\frac{2P}{1-P}}+\ldots,
\quad v\equiv\dfrac{d-1}{8(1-P)^2}\dfrac{G^2}{\ell^2}\left(\dfrac{1}{\ell}\right)^{\frac{2P}{1-P}},
\quad \text{for $\frac{1}{d+1}<P<1$,}
\\ \\
\dfrac{V_\infty \f_\star^2 e^{2A_{IR}}}{4} \exp\left(
2 e^{A_{IR}} \sqrt{\dfrac{2V_\infty}{d-1}} r
\right),
\quad \text{for $P=1$},
\\ \\
\dfrac{d-1}{8(P-1)^2} \left(\dfrac{\ell}{G}\right)^{\frac{2}{P-1}} \dfrac{1}{(r_0-r)^{2\left(1+\frac{1}{P-1}\right)}},
\quad \text{for $1<P$},
\end{cases}
\label{w23}\ee
and
\be
V_{\cal S} {\simeq} \left(\sqrt{\dfrac{2}{d-1}}\g-1\right)^2 V_\zeta,
\label{w24}\ee
where the constant $G>0$ is defined in \eqref{s16-2}, and $A_{IR}, 0<\f_\star$ are the integration constants appearing in \eqref{s19} and \eqref{s9}.

Using \eqref{a-thi18}, we obtain
\be
A'\eta_\perp\simeq
\begin{cases}
-Q\ell^{d-1} \dfrac{(d-1)G^{1-\frac{P}{2}}}{2(1-P)\sqrt{V_\infty Y_\infty}} \left(\dfrac{r}{\ell}\right)^{\frac{2-P}{2(1-P)}} \dfrac{e^{-\frac{G}{2}\left(\frac{r}{\ell}\right)^{\frac{1}{1-P}}(\g-\g_\text{min})}}{r},
\quad \text{for $\frac{1}{d+1}<P<1$},
\\ \\
-Q\ell^{d-1} \sqrt{\dfrac{(d-1)\f_\star}{2Y_\infty}} e^{A_{IR}} \exp\left(
e^{A_{IR}} \sqrt{\dfrac{V_\infty}{2(d-1)}} r
-\dfrac{\f_\star}{2}(\g-\g_\text{min})e^{e^{A_{IR}}\sqrt{\frac{2V_\infty}{d-1}}r}
\right)
\quad \text{for $P=1$},
\\ \\
-Q \ell^{d-1} \dfrac{d-1}{2(P-1)\sqrt{V_\infty Y_\infty}} \left(\dfrac{G(r_0-r)}{\ell}\right)^{\frac{P-2}{2(P-1)}} \dfrac{e^{-\frac{\g-\g_\text{min}}{2}\left(\frac{\ell}{G(r_0-r)}\right)^{\frac{1}{P-1}}}}{r_0-r},
\quad \text{for $1<P$}.
\end{cases}
\label{s16-3}\ee

In the following, we calculate the S\"ommerfeld quantization conditions for the three cases, $\frac{1}{d+1}<P<1$, $P=1$, and $P>1$.

\subsubsection{$\frac{1}{d+1}<P<1$}
Using (\ref{w20}, \ref{w23}, \ref{w24}, \ref{s16-3}) and
\be
B_\zeta' \simeq - \frac{1}{\ell}\sqrt{\frac{d-1}{2}}\frac{G}{2(1-P)}\left(\frac{r}{\ell}\right)^{\frac{P}{1-P}},
\label{w51-2}\ee
we obtain
{
\be
U_+'^2 = m^2 - V_\z \left(1 + \mathcal{O}\left(r^{-\frac{P}{1-P}}e^{-G\left(\frac{r}{\ell}\right)^{\frac{1}{1-P}}(\g-\g_\text{min})}\right)\right),
\label{w52}\ee
$$
U_-'^2 = m^2 - V_\ss \left(1 + \mathcal{O}\left(r^{-\frac{P}{1-P}}e^{-G\left(\frac{r}{\ell}\right)^{\frac{1}{1-P}}(\g-\g_\text{min})}\right)\right).
$$
}
The turning points $r_2^\pm$ are
\be
r_2^+ \simeq
\left(\dfrac{m^2}{v}\right)^{\frac{1-P}{2P}},
\quad
r_2^- \simeq \left(\dfrac{m^2}{\left(\sqrt{\frac{2}{d-1}\g-1}\right)^2v}\right)^{\frac{1-P}{2P}}.
\label{w25-1}\ee
Then, \eqref{w18b} becomes
\be
n_+\pi \simeq \left(\int^{r_a^+}_{r_1^+}dr + \int^{r_2^+}_{r_b^+}dr \right) \sqrt{m^2 - V_\z }
+(r^{+}_a - r^{+}_b ) m
\label{w8}\ee
$$
= m^{\frac{1}{P}} {1-P\over 2P ~v^{1-P\over 2P}} \int^{1}_{v(r_b^+)^{\frac{2P}{1-P}}/m^2} dy ~y^{1-3P\over 2P}~\sqrt{1- y} + \mathcal{O}(m)
$$
$$
=\frac{m^{\frac{1}{P}}}{v^{1-P\over 2P}} {\Gamma\left[{1+P\over 2P}\right]\Gamma\left[{3\over 2}\right]
\over \Gamma\left[{2P+1\over 2P}\right]} +\mathcal{O}(m),
$$
where $V_\z$ is in \eqref{w23}, and we changed the variable from $r$ to $y=v r^{\frac{2P}{1-P}}/m^2$.
Solving for $m$ we obtain
\be
m_{n_+}\simeq \left(2\sqrt{\pi} \,{\Gamma\left[{1+2P\over 2P}\right]
\over \Gamma\left[{1+P\over 2P}\right]}\right)^{P}~v^{1-P\over 2}~n_+^{P}.
\label{w82}\ee
Similarly, we obtain
\be
m_{n_-} \simeq\left(2\sqrt{\pi} \,{\Gamma\left[{1+2P\over 2P}\right]
\over \Gamma\left[{1+P\over 2P}\right]}\right)^{P}~v^{1-P\over 2} \left(\sqrt{\frac{2}{d-1}}\g-1\right)^{1-P\over P}~n_-^{P},
\label{w83}\ee
and finally
\be
\lim_{n\to\infty}{m_{n_+} \over m_{n_-}}=\left(\sqrt{\frac{2}{d-1}}\g-1\right)^{1-P\over P}.
\label{w84}\ee
The value of $\gamma$ compatible with asymptotic glueball universality is
\be
\g= \sqrt{2(d-1)}.
\label{w5}\ee
This is the exponent used in \eqref{qc1-ii}.

\subsubsection{$P=1$}
Using (\ref{w20}, \ref{w23}, \ref{w24}, \ref{s16-3}) and
\be
B_\zeta' \simeq -\frac{e^{A_{IR}}\sqrt{V_\infty} \f_\star}{2} \exp\left(e^{A_{IR}}\sqrt{\frac{2V_\infty}{d-1}}~r\right),
\label{w40-2}\ee
we obtain
\be
U_+'^2=m^2 - V_\z \left( 1 + \mathcal{O}\left[\exp\left(
-e^{A_{IR}} \sqrt{\dfrac{2V_\infty}{d-1}} r
- \f_\star (\g-\g_\text{min})e^{e^{A_{IR}}\sqrt{\frac{2V_\infty}{d-1}}r}
\right)\right] \right),
\label{w25-3}\ee
$$
U_-'^2=m^2 - V_\ss \left( 1 + \mathcal{O}\left[\exp\left(
-e^{A_{IR}} \sqrt{\dfrac{2V_\infty}{d-1}} r
- \f_\star (\g-\g_\text{min})e^{e^{A_{IR}}\sqrt{\frac{2V_\infty}{d-1}} r}
\right)\right]\right).
$$
The turning points $r_2^\pm$ are
\be
r_2^+ \simeq
\sqrt{\dfrac{d-1}{2V_\infty}} \dfrac{e^{-A_{IR}}}{2} \log\left(\dfrac{4 e^{-2A_{IR}}m^2}{V_\infty \f_\star^2}\right),
\label{w25-2}\ee
$$
r_2^- \simeq \sqrt{\dfrac{d-1}{2V_\infty}} \dfrac{e^{-A_{IR}}}{2} \log\left(\left(\sqrt{\dfrac{2}{d-1}}\g-1\right)^{-2}\dfrac{4 e^{-2A_{IR}}m^2}{V_\infty \f_\star^2}\right).
$$

The quantization condition \eqref{w18b} is
\be
n_+\pi \simeq \left(\int^{r_a^+}_{r_1^+}dr + \int^{r_2^+}_{r_b^+}dr \right) \sqrt{m^2 - V_\z }
+(r^{+}_a - r^{+}_b ) m
\label{w41}\ee
$$
=\sqrt{\frac{d-1}{2V_\infty}}\frac{e^{-A_{IR}} m }{2} \int^1_{\frac{V_\infty \f_\star^2 e^{2A_{IR}}}{4 m^2}} dy \frac{\sqrt{1-y}}{y}
+\mathcal{O}(m)
$$
$$
=\sqrt{\frac{d-1}{2V_\infty}} e^{-A_{IR}} m \log(m) +\mathcal{O}(m),
$$
where $V_\z$ is in \eqref{w23}, and we changed variable in the integral from $r$ to $y=\frac{V_\infty \f_\star^2 e^{2A_{IR}}}{4 m^2} \exp\left(2e^{A_{IR}} \sqrt{\frac{2V_\infty}{d-1}} r \right)$.

Solving \eqref{w41} for large $m$, we obtain
\be
m_{n_+} = 2 e^{A_{IR}} \sqrt{\frac{2V_\infty}{d-1}} \frac{\pi}{2}\frac{n_+}{\log n_+} + \ldots
\label{w42}\ee
Similarly, we obtain
\be
m_{n_-}= 2 e^{A_{IR}} \sqrt{\frac{2V_\infty}{d-1}} \frac{\pi}{2}\frac{n_-}{\log n_-} + \ldots,
\label{w43}\ee
and finally
\be
\lim_{n\to\infty} \frac{m_{n_-}}{m_{n_+}} = 1.
\label{w43-2}\ee

\subsubsection{$P>1$}
Using (\ref{w20}, \ref{w23}, \ref{w24}, \ref{s16-3}) and
\be
B_\zeta' \simeq - \sqrt{\frac{d-1}{2}}\frac{1}{2(P-1)}\left( \frac{\ell}{G(r_0-r)} \right)^{\frac{1}{P-1}} \frac{1}{r_0-r},
\label{w30-2}\ee
we obtain
\be
U_+'^2 = m^2 - V_\z \left( 1 + \mathcal{O}\left(\left( r_0-r \right)^{\frac{P}{P-1}} e^{-\left(\frac{\ell}{G(r_0-r)}\right)^{\frac{1}{P-1}}(\g-\g_\text{min})}\right)\right),
\label{w30-3}\ee
$$
U_-'^2 = m^2 - V_\ss \left( 1 + \mathcal{O}\left(\left( r_0-r \right)^{\frac{P}{P-1}} e^{-\left(\frac{\ell}{G(r_0-r)}\right)^{\frac{1}{P-1}}(\g-\g_\text{min})}\right) \right).
$$

The turning points $r_2^\pm$ are
\be
r_2^+ \simeq
r_0 - \left( \dfrac{\sqrt{\frac{d-1}{2}}}{2(P-1)m} \right)^{\frac{P-1}{P}} \left(\dfrac{\ell}{G}\right)^{\frac{1}{P}},
\quad
r_2^- \simeq r_0 - \left( \dfrac{\g-\sqrt{\frac{d-1}{2}}}{2(P-1)m} \right)^{\frac{P-1}{P}} \left(\dfrac{\ell}{G}\right)^{\frac{1}{P}}.
\label{w25-4}\ee

The quantization condition \eqref{w18b} is
\be
n_+\pi \simeq \left(\int^{r_a^+}_{r_1^+}dr + \int^{r_2^+}_{r_b^+}dr \right) \sqrt{m^2 - V_\z }
+(r^{+}_a - r^{+}_b ) m
\label{w31}\ee
$$
= \frac{\sqrt{(d+1-2\Delta_-)(d-1-2\Delta_-)}}{4} \int^{1}_{\frac{(d+1-2\Delta_-)(d-1-2\Delta_-)}{4(r_a^+)^2m^2}} dy \frac{\sqrt{1-y}}{y^{3/2}} +
$$
$$
+ \frac{(d-1)^{\frac{P-1}{2P}}}{2^{\frac{5}{2}-\frac{3}{2P}}P} \left(\frac{\ell m(P-1)}{G}\right)^{\frac{1}{P}} \int^1_{\frac{d-1}{8(P-1)^2m^2} \left(\frac{\ell}{G}\right)^{\frac{2}{P-1}} \frac{1}{(r_0-r_b^+)^{2\left(1+\frac{1}{P-1}\right)}}} dz \frac{\sqrt{1-z}}{z^{\frac{3}{2}-\frac{1}{2P}}}
+(r^{+}_a - r^{+}_b ) m
$$
$$
=m r_0 + \mathcal{O}\left(m^{0}\right)
$$
where $V_\z$ is in \eqref{w23},  we changed the integration variable from $r$ to $y=\frac{(d+1-2\Delta_-)(d-1-2\Delta_-)}{4m^2r^2}$ and $z=\frac{d-1}{8(P-1)^2m^2} \left(\frac{\ell}{G}\right)^{\frac{2}{P-1}} \frac{1}{(r_0-r)^{2\left(1+\frac{1}{P-1}\right)}}$.
Solving \eqref{w31} for $m$ asymptotically, we obtain
\be
m_{n_+}= \frac{\pi}{r_0} n_+ +\ldots
\label{w32}\ee
Similarly, we obtain
\be
m_{n_-}= \frac{\pi}{r_0} n_- +\ldots,
\label{w33}\ee
and finally
\be
\lim_{n\to\infty} \frac{m_{n_-}}{m_{n_+}} = 1.
\label{w33-2}\ee

\section{IR asymptotic solutions in soft potentials}\label{IR}
In this appendix, we consider the IR asymptotic solutions in soft potentials \eqref{qc1}-\eqref{qc1-ii}.
We first  assume $0\leq P$ corresponding to a confining holographic QFT~\cite{iQCD}.\footnote{At the end of the appendix, we shall see a sharper bound on $P$ \eqref{s29-5} requiring the consistency of the expansion.}

We start from \eqref{a8}, which can be solved algebraically for $S$ and $T$ as
\be
S=2{V\over dW/d\f}+{d W^2\over 2(d-1)dW/d\f},
\quad
T=2Y\left( {d\over 4(d-1)}W^2-{S^2\over 2}+V\right)=Y\left(\frac{dW}{d\f}S-S^2\right).
 \label{a31}\ee
Removing $S$ and $T$ in \eqref{a8}, we obtain the second order differential equation for $W$:
\be
2(dW^2+4(d-1)V)^2~\frac{d^2W}{d\f^2}+2(d-1)\left[(dW^2+4(d-1)V){\p_\f Y\over Y}+4(d-1)\p_\f V\right]\left(\frac{dW}{d\f}\right)^3-
\label{aa4}\ee
$$
-(dW^2+4(d-1)V)\frac{dW}{d\f}\left[(dW^2+4(d-1)V){\p_\f Y\over Y}+2\left(4(d-1)\p_\f V+dW\frac{dW}{d\f}\right)\right]=0.
$$
We solve \eqref{aa4} asymptotically to derive the IR expression of $W$. Since \eqref{aa4} is a second order differential equation, the solution has the two integration constants.
As in the non-axionic case, one of the integration constants corresponds to the singular solutions with $W\sim S\sim e^{\sqrt{d\over 2(d-1)}\f}$.
These generic solutions violate the Gubser bound \cite{Gubser} and are not acceptable.
Potentially regular solutions have the following structure:
\be
W = W_\infty \f^{P_w} e^{w\f}+ \ldots, \quad
S\simeq S_\infty \f^{P_s} e^{s\f} + \ldots,
\label{qc1-2}\ee
for $\f\to\infty$, where $W_\infty$ and $S_\infty$ are positive.

By using \eqref{qc1} and \eqref{qc1-2}, at the leading order in the IR, \eqref{aa4} becomes
\be
w W_\infty \f^{P_w} e^{w \f}
\bigg[
16 (d-1) \sqrt{2(d-1)} \left\{ (d+1) - \sqrt{2(d-1)} w \right\} V_\infty^2 \f^{2P} e^{2\sqrt{2\over d-1}\f}+
\label{qc3}\ee
$$
+ d \sqrt{2(d-1)} \left\{ d-2(d-1)w^2 \right\} W_\infty^4 \f^{4P_w} e^{4w\f}+
$$
$$
+ 8d \left\{ (d-1) w \left(1+\sqrt{2(d-1)}w\right)-d\sqrt{2(d-1)}\right\} V_\infty W_\infty^2 \f^{P+2P_w} e^{\sqrt{2\over d-1}\f + 2 w \f}
\bigg]=0.
$$
This indicates the relation
\be
w= \sqrt{1\over 2(d-1)},
\quad\quad
P_w = {P\over2},
\label{qc8}\ee
and \eqref{qc3} becomes
\be
d(d-1) W_\infty \left(4V_\infty - W_\infty^2\right)^2 \f^{5P/2} e^{{5\over\sqrt{2(d-1)}}\f}=0.
\label{qc9}\ee
From \eqref{qc9}, we obtain
\be
W = 2 \sqrt{V_\infty} \f^{P_w} e^{w\f}+ \ldots
\label{qc10}\ee
From \eqref{a31}, the IR behaviors of $S$ and $T$ are given by
\be
S = \sqrt{2V_\infty\over d-1} \f^{P/2} e^{{1\over\sqrt{2(d-1)}}\f}+ \ldots
\label{qc11}\ee
\be
T\simeq T_\infty \f^{P} e^{\sqrt{2\over d-1}d\f},
\quad
T_\infty=0.
\label{qc12}\ee
Summarizing so far, the leading IR solutions are given by (\ref{qc10}, \ref{qc11}, \ref{qc12}).

Next, we consider the subleading order solution,
\be
W =
\bigg(2 \sqrt{V_\infty} \f^{P\over2} + W_{{P\over2}-1} \f^{{P\over2}-1} + {\cal O} \left( \f^{{P\over2}-2} \right)\bigg) e^{{1\over\sqrt{2(d-1)}}\f}.
\label{s1}\ee

By substituting \eqref{s1} into \eqref{aa4}, we obtain the condition
\be
V_\infty^{3\over2}
\left(2d(d-1) W_{{P\over2}-1}^2 - \sqrt{2 (d-1) V_\infty} (1+(3d-1)P) W_{{P\over2}-1} +2P(1+(d-1)P)V_\infty\right) =0,
\label{s2-1}\ee
from $\mathcal{O}\left(\f^{{5P\over2}-2} e^{{5\over\sqrt{2(d-1)}}\f}\right)$ term.
We obtain two solutions $W_{{P\over2}-1}= W_{{P\over2}-1}^{(S)}, \, W_{{P\over2}-1}^{(L)}$:
\be
W_{{P\over2}-1}^{(S)} \equiv P \sqrt{2V_\infty \over d-1},
\quad
W_{{P\over2}-1}^{(L)} \equiv {P (d-1)+1 \over d} \sqrt{V_\infty\over2(d-1)}.
\label{s2-2}\ee
Next, from \eqref{a31} and \eqref{s1}, the asymptotic expansion of $S$ is given by
\be
S=
\begin{cases}
\Bigg(
\sqrt{\dfrac{2V_\infty}{d-1}} \f^{P\over2} + P \dfrac{d\sqrt{V_\infty}}{d-1} \f^{{P\over2}-1}
\Bigg) e^{{1\over\sqrt{2(d-1)}}\f}
\quad \text{for $W_{{P\over2}-1} = W_{{P\over2}-1}^{(S)}$}
\\ \\
\Bigg(
\sqrt{\dfrac{2V_\infty}{d-1}} \f^{P\over2} + \dfrac{2d-1-P(d-1)}{2d(d-1)} \sqrt{V_\infty}  \f^{{P\over2}-1}
\Bigg) e^{{1\over\sqrt{2(d-1)}}\f}
\quad \text{for $W_{{P\over2}-1} = W_{{P\over2}-1}^{(L)}$},
\end{cases}
\label{s4-2}\ee

Finally, by using \eqref{a31}, \eqref{s1}, \eqref{s2-2} and \eqref{s4-2}, we obtain
\be
T=
\begin{cases}
0
\quad \text{for $W_{{P\over2}-1} = W_{{P\over2}-1}^{(S)}$}
\\ \\
\dfrac{P(d+1)-1}{d} {\dfrac{2V_\infty Y_\infty }{\sqrt{2(d-1)}}} \f^{P-1}
e^{\sqrt{2\over d-1}d\f}
\quad \text{for $W_{{P\over2}-1} = W_{{P\over2}-1}^{(L)}$}.
\end{cases}
\label{s6}\ee
We observe that the solution $W_{{P\over2}-1}^{(L)}$ would not be a good one because the axion vev is fixed independently of the axion source (see also the discussion in \cite{axion-rg}).
In the following, we use $W_{{P\over2}-1}^{(S)}$.

We add the effect of the axion vev to the solutions found in (\ref{s1}, \ref{s4-2}, \ref{s6}).
From \eqref{s1} and \eqref{s4-2} with $W_{{P\over2}-1}=W_{{P\over2}-1}^{(S)}$ in \eqref{s2-2}, we observe
\be
{d\over d-1} {W\over S}= d \sqrt{2\over d-1} - {d P\over \f}.
\label{s20}\ee
By solving \eqref{a8}, we obtain
\be
T= D \,\f^{-d P} e^{\sqrt{2\over d-1}d\f},
\label{s21}\ee
where $D$ is an integration constant associated with the axion. The mass dimension of $D$ is $2$. Next, we investigate the effect of the integration constant $D$ on $W$ and $S$.
We denote
\be
W= \left.W\right|_{D=0} + \delta W\sp
S= \left.S\right|_{D=0} + \delta S,
\label{s23}\ee
where $\delta W$ and $\delta S$ are assumed to be small.
From \eqref{a8} with \eqref{s21}, we obtain
\be
\delta S - \frac{d(\delta W)}{d\f} = - {D\over Y_\infty} \sqrt{d-1\over2V_\infty}  \f^{-\left(d+{1\over2}\right)P}e^{{1\over\sqrt{2(d-1)}}\f}.
\label{s22}\ee
By substituting \eqref{s23} and \eqref{s22} into the third equation of \eqref{a8}, we observe
\be
-2 \sqrt{2V_\infty \over d-1}  \f^{\left(d+{1\over2}\right)P} \frac{d(\delta W)}{d\f} + {2d\sqrt{V_\infty} \over d-1} \f^{\left(d+{1\over2}\right)P} \delta W + {D \over Y_\infty} e^{{1\over\sqrt{2(d-1)}}\f} = 0.
\label{s24}\ee
For $\f\to\infty$, the solution satisfying the Gubser bound is
\be
\delta W = - {D\over 2\sqrt{V_\infty} Y_\infty}  \f^{-\left(d+{1\over2}\right)P} e^{{1\over\sqrt{2(d-1)}}\f}.
\label{s25}\ee
By using \eqref{s22} and \eqref{s25}, we find
\be
\delta S = - {D(2d-1) \over 2\sqrt{2(d-1)V_\infty} Y_\infty}  \f^{-\left(d+{1\over2}\right)P} e^{{1\over\sqrt{2(d-1)}}\f}.
\label{s26}\ee

Combining all the above, we obtain (\ref{s27}) in the main text.
Note that the corrections including $D$ to $W$ and $S$ are not exponentially suppressed compared to the solutions \eqref{s1} and \eqref{s4-2}, but are suppressed by the power of $\f$.

We notice that ${\cal O}(D^2)$ contribution to the scalar function $T$ can be calculated by substituting \eqref{s27} into the second equation of \eqref{a8}:
\be
{1\over T}\frac{dT}{d\f}
= d \,\sqrt{2\over d-1} - {d P\over \f}
+ {D d\over V_\infty Y_\infty} \sqrt{d-1\over 2} \f^{-(d+1)P}.
\label{s29-2}\ee
By integrating \eqref{s29-2}, we obtain
\be
T = D\, \f^{-d P}\exp\left(
d \sqrt{2\over d-1} \f
+ {D d\over V_\infty Y_\infty} \sqrt{d-1\over 2} {1\over1-(d+1)P} \f^{1-(d+1)P}
\right).
\label{s29-3}\ee
From \eqref{a8} with (\ref{s23}, \ref{s29-3}), we observe
\be
\delta W, \delta S \sim  D \f^{-\left(d+{1\over2}\right)P}  e^{\sqrt{1\over2(d-1)}\f +{D d\over V_\infty Y_\infty} \sqrt{d-1\over 2} {1\over1-(d+1)P} \f^{1-(d+1)P}}.
\label{s29-4}\ee
Therefore, if we require that the correction coming from axion vev does not change the leading asymptotic behavior in \eqref{s1} and \eqref{s4-2}, we obtain a bound on $P$ as
\be
1-(d+1)P <0.
\label{s29-5}\ee
In the following, we assume this inequality.

Finally, we derive the IR asymptotic behavior of $A$, $\f$, and $a$.
From (\ref{a6}, \ref{s27}), we obtain
\be
{dA\over d\f}
= - {1\over\sqrt{2(d-1)}} + {P \over 2} {1\over \f}
- {D\over V_\infty Y_\infty} \sqrt{d-1\over 8} \f^{-(d+1)P}
+\ldots
\label{s18}\ee
Hence, the IR asymptotics of the scale factor $A$ is
\be
A= A_{IR} - {\f \over\sqrt{2(d-1)}} + {P \over 2} \log \f
- {D\over V_\infty Y_\infty} \sqrt{d-1\over 8} {1\over P(d+1)-1} \f^{1- (d+1)P}
+\ldots,
\label{s19}\ee
where $A_{IR}$ is an integration constant.

From \eqref{a6} and \eqref{th2} \eqref{s27}, we obtain
\be
{d\f\over dr} = e^A S =
e^{A_{IR}} \f^{P} \sqrt{\dfrac{2V_\infty}{d-1}}
\bigg( 1
- {D\over V_\infty Y_\infty} \sqrt{d-1\over 8} {1\over P(d+1)-1} \f^{1- (d+1)P} +\ldots
\bigg),
\label{s8}\ee
where we assumed \eqref{s29-5}.
By solving \eqref{s8}, we obtain the expression of $\f$ as the function of $r$:
\be
\f= \begin{cases}
\left(
(1-P) e^{A_{IR}} \sqrt{\dfrac{2V_\infty}{d-1}} (r-r_0) 
\right)^{\frac{1}{1-P}}
+\ldots,
\quad
\text{for $P\neq1$},\\ \\
\f_\star e^{e^{A_{IR}}  \sqrt{\frac{2V_\infty}{d-1}}  r} + \ldots,
\quad
\text{for $P=1$}.
\end{cases}
\label{s9}\ee
where $\f_\star>0$ and $r_0$ are integration constants.
We observe that the IR end-point corresponds to $r\to\infty$ for $P\leq1$ and $r\to\text{(finite)}=r_0$ for $P>1$.
By combining \eqref{s19} and \eqref{s9}, the behavior of the scale factor in the IR is
\be
e^A \sim \begin{cases}
\exp\left[
- \dfrac{1}{\sqrt{2(d-1)}}
\left(
(1-P) e^{A_{IR}} \sqrt{\dfrac{2V_\infty}{d-1}} (r-r_0)
\right)^{1\over 1-P}
\right],
\quad
\text{for $P\neq1$},\\ \\
\exp\left(
-\dfrac{1}{\sqrt{2(d-1)}} \f_\star e^{e^{A_{IR}} \sqrt{\frac{2V_\infty}{d-1}}  r}
\right),
\quad
\text{for $P=1$}.
\end{cases}
\label{s19-2}\ee

By integrating \eqref{a5}, the IR asymptotic solution of the axion field is
\be
a = - {\rm sign} (Q) \int^\infty_\f d\f {\sqrt{T}\over Y S} \simeq
- {\rm sign} (Q) {\sqrt{D}\over Y_\infty} \sqrt{d-1 \over 2V_\infty} \int^\infty_\f d\f  \f^{-{d+1\over2}P} e^{-\sqrt{d-1\over2}\f}
\label{s15}\ee
$$
\simeq - \frac{{\rm sign} (Q)}{Y_\infty} \sqrt{D\over V_\infty} \f^{-{d+1\over2}P} e^{-\sqrt{d-1\over2}\f},
$$
where \eqref{qc1} and \eqref{s27} are used. Furthermore, using \eqref{s19}, we obtain
\be
a \simeq
- \frac{{\rm sign} (Q)}{Y_\infty} \sqrt{D\over V_\infty}
\bigg(\sqrt{2(d-1)}(A_{IR}-A)\bigg)^{-dP} e^{-(d-1)(A_{IR}-A)}.
\label{s16}\ee
By using \eqref{s16} and the identification \eqref{th11}, we obtain the energy dependence of the axion field, \eqref{s17}.

\section{Derivation of the quadratic fluctuation equations}\label{quadratic}
In this appendix, we provide a derivation of the quadratic fluctuation equations \eqref{re6}.
Under a $(d+1)$-dimensional diffeomorphism ($\delta x^\mu=\xi^\mu, \delta r=\xi^{d+1}$), the fluctuations defined in \eqref{pt2-2} transform as
\be
h_{\mu\nu} \to h_{\mu\nu}  -\partial_\mu \xi_\nu - \partial_\nu \xi_\mu - 2\eta_{\mu\nu} A' \xi^{d+1}\sp
B_\mu \to B_\mu - \xi'_\mu - \partial_\mu \xi^{d+1}.
\label{pt4}\ee
\be
\phi \to \phi  -{\xi^5}' - A' \xi^{d+1}\sp
\delta\f \to \delta\f -\f' \,\xi^{d+1} \sp
\delta a\to \delta a - a' \xi^{d+1}.
\label{pt5}\ee
Notice that, $\delta a$ is a gauge invariant quantity if the axion does not change along with the flow, but this is not true in the presence of the non-trivial axion RG flow.

By substituting \eqref{pt2-2} into \eqref{A2}, we obtain
\be
S= M_p^{d-1} \int d^dx dr e^{(d-1)A}\sqrt{-\tilde{g}} \left[\tilde{R} + d(d-1) \tilde{g}^{ab} \de_a A \de_b A - \frac{1}{2}\tilde{g}^{ab}\de_a\f\de_b \f -{1\over 2}Y \tilde{g}^{ab}\de_a a \pa_b a - e^{2A} V \right].
\label{pt8}\ee
Here $\tilde{R}$ is the Ricci curvature constructed from $\tilde{g}_{ab}$ defined in \eqref{pt2-2}.

We expand the action \eqref{pt8} to quadratic order in the fluctuations defined in \eqref{pt2-2}.
The result is
\be
S^{(2)} = M_p^{d-1} \int d^dx dr \bigg[e^{(d-1)A} \bigg\{
\mathcal{L}_\text{ein}^{(2)}
- \frac{1}{4} h'_{\rho\sigma} h'^{\rho\sigma}
+\frac{1}{4} h'^2
- \frac{1}{4} F_{\mu\nu} F^{\mu\nu} -
\label{pt19}\ee
$$
  - (\de^\mu \phi) (\de^\nu h_{\mu\nu} - \de_\mu h)
  + \left(\frac{h'}{2}+\phi'\right) \left( \f' \delta \f + Y a' \delta a \right)
  + 2\phi \left( \f' \delta \f' + Y a'  \delta a'  \right) +
$$
$$
+ B^{\mu} \left( \f' \de_\mu \delta \f
+ Y a' \de_\mu \delta a  \right)
+  \phi(\de_\f Y)  a'^2 \delta \f
- \frac{1}{2}  \bigg(\de_\mu \delta\f \de^\mu \delta \f +Y \de_\mu \delta a \de^\mu \delta a \bigg) -
$$
$$
- \frac{1}{2} \bigg( (\delta\f')^2 +Y (\delta a')^2\bigg)
- (\de_\f Y) a' \delta\f \delta a'
- \left( e^{2A}  \de_\f^2 V + \frac{\de_\f^2 Y}{2}  a'^2 \right) \frac{(\delta\f)^2}{2} \bigg\} -
$$
$$
- \left(e^{(d-1)A} B^\mu \right)' \left( \de_\mu h - \de^\nu h_{\mu\nu} \right)
- \left(e^{(d-1)A}\right)' \bigg( 2B_\mu \de^\mu\phi + (\phi^2)' +\phi h'  \bigg) \bigg],
$$
where $h\equiv h^\mu_\mu$, $F_{\mu\nu}\equiv \de_\mu B_\nu - \de_\nu B_\mu$ and
\be
\mathcal{L}_\text{ein}^{(2)} =
  - \frac{1}{4} \left(\de_\mu h_{\rho\sigma} \right) \left( \de^\mu h^{\rho\sigma} \right)
  + \frac{1}{2}(\de_\mu h^{\mu\rho}) (\de^\nu h_{\nu\rho})
   - \frac{1}{2} (\de^\nu h)(\de^\mu h_{\mu\nu})
   +\frac{1}{4} (\de_\mu h) (\de^\mu h),
\label{pt19-2}\ee
is the quadratic part of the $d$-dimensional Einstein-Hilbert Lagrangian.

From \eqref{pt19}, the equations of motion for the fluctuations are
\be
h_{\mu\nu}''
+(d-1)A' h_{\mu\nu}'
+\p^2 h_{\mu\nu}
-2\de^\rho \de_{(\mu} h_{\nu)\rho}
+\de_\mu \de_\nu (h+2\phi)
- \frac{(e^{(d-1)A} \de_{(\mu} B_{\nu)})'}{e^{(d-1)A}}+
\label{pt21}\ee
$$
+\eta_{\mu\nu}\bigg(
-h''
-(d-1)A'h'
-\de^2 (h+2\phi)
+2(d-1)A'\phi'
+2(d-1)(A''+(d-1)A'^2)\phi-
$$
$$
-\frac{(e^{(d-1)A}\f'\delta\f+Y a'\delta a))'}{e^{(d-1)A}}
+2\frac{(e^{(d-1)A}\de_\mu B^\mu)'}{e^{(d-1)A}}
+\de_\rho \de_\sigma h^{\rho\sigma}
\bigg)=0,
$$
\be
\left( \de^\rho\de_\rho \eta_{\mu\nu} -\de_\mu \de_\nu \right) B^\nu
- \de^\nu h'_{\mu\nu}
+\de_\mu \bigg(-2(d-1) A' \phi
+ h'
+ \f' \delta\f
+ Y a' \delta a
\bigg)
=0,
\label{pt21-2}\ee
\be
- \de^\mu \de_\mu h + \de^\mu \de^\nu h_{\mu\nu} - (d-1) A' h' + 2(d-1)  (A''+(d-1)A'^2) \phi +
\label{pt21-3}\ee
$$
+2(d-1)A' \de_\mu B^\mu
- e^{-(d-1)A} \left(e^{(d-1)A} \f' \right)' \delta\f
+ \left(\f' \delta\f' + Y a' \delta a' \right)
+ a'^2 \delta \f \de_\f Y
=0,
$$
\be
\delta\f''
+ (d-1) A' \delta\f'
+ \de^\mu \de_\mu \delta\f - \left( e^{2A} \de_\f^2 V + \frac{1}{2}a'^2 \de_\f^2 Y \right)\delta\f -
\label{pt21-4}\ee
$$
- 2e^{-(d-1)A} \left( e^{(d-1)A} \f'  \right)' \phi
+  \f'  \left(\frac{h'}{2} - \de_\mu B^\mu - \phi'  \right)
+   \de_\f Y \left( a'^2 \phi -   a' \delta a' \right)
=0,
$$
\be
Y \delta a'' + \left[(d-1)A' Y + \f' \de_\f Y \right] \delta a'  + Y \de^\mu \de_\mu \delta a +
\label{pt21-5}\ee
$$
+ e^{-(d-1)A} \bigg( e^{(d-1)A} a' \de_\f Y  \delta\f \bigg)'
+ Y a' \left(\frac{h'}{2}- \de_\mu B^\mu - \phi'  \right)
=0,
$$
where $X_{(\mu\nu)}\equiv \frac{1}{2}(X_{\mu\nu}+X_{\nu\mu})$.
The equations \eqref{pt21}, \eqref{pt21-2}, \eqref{pt21-3}, \eqref{pt21-4} and \eqref{pt21-5} correspond to the variation with respect to $h_{\mu\nu}, B_\mu, \phi, \delta\f$, and $\delta a$, respectively.

Next, we perform the decomposition of $B_\mu$ and $h_{\mu\nu}$ as in \eqref{pt6-1} and \eqref{pt6}.
The transformation law under $(d+1)$-dimensional diffeomorphism is
\be
\psi\to \psi-A'\xi^{d+1} \sp
E\to E-\xi\sp
V_\mu^T = \xi_\mu^T\sp
h_{\mu\nu}^{TT}\to h_{\mu\nu}^{TT}.
\label{pt7}\ee
Here we take $\delta x^\mu = \xi^\mu = \xi^{T\mu} + \partial^\mu \xi$ and $\delta r=\xi^{d+1}$ where $\partial^\mu \xi^{T}_{\mu}=0$.
The equations of motion are decomposed into the scalar, vector, and tensor modes.
The equation for the transverse traceless tensor mode is
\be
{h^{TT}_{\mu\nu}}''
+(d-1)A' {h^{TT}_{\mu\nu}}'
+\p^2 {h^{TT}_{\mu\nu}}
=0.
\label{a-pt22}\ee
The equations for the scalar modes are
\be
\de_\mu \bigg(
2(d-1) \psi' - 2(d-1) A' \phi + \f' \delta\f + Y a' \delta a
\bigg)
=0,
\label{pt22-5}\ee
\be
- 2(d-1)\de^\mu \de_\mu \psi
- 2d(d-1) A' \psi'
+ 2(d-1)\bigg(A''+(d-1)A'^2\bigg) \phi
+2(d-1) A' \de^\mu \de_\mu (W - E')  -
\label{pt22-6}\ee
$$
- e^{-(d-1)A} \left(e^{(d-1)A} \f' \right)' \delta\f
+ \left(\f' \delta\f' + Y a' \delta a' \right)
+ a'^2  \de_\f Y \delta \f
=0,
$$
\be
\delta\f'' + (d-1) A' \delta\f' + \de^\mu \de_\mu \delta\f - \left( e^{2A} \de_\f^2 V + \frac{1}{2}a'^2 \de_\f^2 Y \right)\delta\f -
\label{pt22-7}\ee
$$
-2e^{2A} \de_\f V \, \phi
+ \f' \bigg( d \psi' -  \de_\mu \de^\mu (W-E') - \phi' \bigg)
-  \de_\f Y_  a' \delta a'
=0,
$$
\be
Y \delta a''
+[(d-1)A' Y + \f'(\de_\f Y) ]\delta a'
+ Y \de^\mu \de_\mu \delta a   +
\label{pt22-8}\ee
$$
+ e^{-(d-1)A} \bigg( e^{(d-1)A} a' \de_\f Y  \delta\f \bigg)'
+ Y a' \bigg(  d \psi' -  \de^\mu \de_\mu (W - E') - \phi' \bigg)
=0.
$$

From \eqref{pt22-5} and \eqref{pt22-6}, we observe
\be
\phi = \frac{1}{A'} \left( \psi' + \frac{\f' \delta \f + Y a' \delta a}{2(d-1)} \right),
\label{pt23}\ee
\be
d \psi'  -\de^\mu \de_\mu (W-E')-\phi' =
-\frac{1}{A'}
\left[  \de^\mu \de_\mu \psi
+ \psi''
-\left( 2\frac{A''}{A'} + (d-1) A' \right)\psi'
\right]
- \frac{e^{2A}\de_\f V}{(d-1)A'}\delta\f+
\label{pt24-2}\ee
$$
+\left( \frac{A''}{(d-1)A'^2} + 1 \right)\left(\f' \delta\f + Y a' \delta a\right).
$$
By substituting \eqref{pt23} and \eqref{pt24-2} into \eqref{pt22-7} and \eqref{pt22-8}, we obtain \eqref{re6}.

For convenience, we write down the expression $B_\zeta$ in \eqref{re7} in terms of the scalar functions and $\epsilon$,
\be
B_\zeta = \frac{d-1}{2}A+\frac{1}{2}\log\frac{\sigma'^2}{A'^2}
=  \frac{d-1}{2}A+\frac{1}{2}\log\frac{S^2+\frac{T}{Y}}{W^2}
=  \frac{d-1}{2}A+ \frac{1}{2}\log\epsilon,
\label{re7-5-3}\ee
where $\epsilon$ is defined in \eqref{a-thi14}.

\section{A coordinate transformation}\label{transformation}

Numerically, it turns out to be convenient to change the coordinate from $r$ to $A$.
By using $A$ as a coordinate, the equation \eqref{re6} becomes
\be
\left[
- \frac{d^2}{dA^2}
+ \begin{pmatrix}
V_{\zeta A} & 2\eta_\perp\left( \frac{dB_{\zeta A}}{dA} + \frac{1}{\eta_\perp}\frac{d\eta_\perp}{dA} +\frac{d}{dA} \right) \\
2\eta_\perp \left(\frac{dB_{\zeta A}}{dA}-\frac{d}{dA}\right) & V_{{\cal S}A}
\end{pmatrix}
\right]
\begin{pmatrix}
\tilde{\psi}_{\zeta A} \\
\tilde{\psi}_{{\cal S}A}
\end{pmatrix}=
\frac{\de^\mu \de_\mu}{A'^2}
\begin{pmatrix}
\tilde{\psi}_{\zeta A} \\
\tilde{\psi}_{{\cal S}A}
\end{pmatrix},
\label{tr13}\ee
where
\be
\tilde{\psi}_{\zeta A}= e^{-F} \psi_\zeta,
\quad
\tilde{\psi}_{{\cal S} A}= e^{-F} \psi_{\cal S},
\quad
F \equiv -\frac{A}{2}-\frac{1}{2}\log W.
\label{tr11}\ee
\be
V_{\zeta A} =
\frac{d^2 B_{\zeta A}}{dA^2} + \left( \frac{dB_{\zeta A}}{dA} \right)^2,
\quad
B_{\zeta A}=B_\zeta-F=\frac{d}{2}A+\frac{1}{2}\log\frac{S^2+\frac{T}{Y}}{W},
\label{a-tr16}\ee
\be
V_{{\cal S} A} =
\frac{d^2 B_{ {\cal S} A}}{dA^2} + \left( \frac{dB_{ {\cal S} A}}{dA} \right)^2
 +  \frac{e^{2A}}{A'^2}V_{;ss} + \eta_\perp^2  - (d-1) \epsilon R_{\text{fs}},
\quad
B_{ {\cal S} A}=\frac{d}{2}A+\frac{1}{2}\log W.
\label{tr16-2}\ee

Furthermore, $d\tilde{\psi}/dA$ terms in the equation (\ref{tr13}) can be eliminated by the redefinition,
\be
\begin{pmatrix}
\tilde{\psi}_{\zeta A} \\
\tilde{\psi}_{{\cal S}A}
\end{pmatrix}
=
{\cal R}
\begin{pmatrix}
\psi_{\zeta A} \\
\psi_{{\cal S}A}
\end{pmatrix},
\quad
{\cal R}\equiv
\begin{pmatrix}
\cos\Phi & \sin\Phi \\
-\sin\Phi & \cos\Phi
\end{pmatrix}
,\quad
\frac{d \Phi}{dA}
= \eta_\perp.
\label{a-tr19}\ee
The integration constant of $\Phi$ corresponds to the freedom to rotate the basis with a $r$-independent matrix. As a result, we obtain \eqref{tr20}.

\section{Analytic continuation}\label{continuation}
To compute cubic coupling among the glueballs, we start from the reference \cite{Garcia-Saenz:2019njm} where non-gaussianities in the multi-field inflation are calculated. By analytically continuing to the setup in holography \cite{McFadden:2009fg,McFadden:2010vh}, we shall obtain cubic interaction terms.
We focus on $d=3$ in this appendix.

\subsection{Cubic couplings in cosmology}
In cosmology, the action is
\be
S_{\text{cosmology}}=  \int d^{4}x \sqrt{-g} \left[\frac{\bar{M}_p^2}{2} R - {1\over 2} \bar{G}_{IJ} g^{ab}\de_a \bar{\phi}^I \de_b \bar{\phi}^J  - \bar{V}(\f)\right],
\label{cos1}\ee
where the background metric is
\be
ds^2 = - dt^2 + e^{2A} \delta_{ij} dx^i dx^j
= e^{2A} \left( - d\tau^2 + \delta_{ij} dx^i dx^j  \right),
\label{cos2}\ee
and the field space metric is
\be
\bar{G}_{IJ}=
\begin{pmatrix}
\bar{G}_{\f\f}&\bar{G}_{\f a}\\
\bar{G}_{a\f}&\bar{G}_{aa}
\end{pmatrix}=
\begin{pmatrix}
1&0\\
0&Y(\bar{\f})
\end{pmatrix},
\quad
\bar{\phi}^I =
\begin{pmatrix}
\bar{\f} \\
\bar{a}
\end{pmatrix}.
\label{thi3-2}\ee

By taking the comoving gauge, the metric of the spacetime with the fluctuation is
\be
ds^2 = -N^2 dt^2 + e^{2A} e^{2\bar{\zeta}} \delta_{ij} \left( dx^i + N^i dt \right) \left( dx^j + N^j dt \right)
\label{thi1}\ee
$$
= e^{2A} \left[-N^2 d\tau^2 + e^{2\bar{\zeta}} \delta_{ij} \left( dx^i + N^i dt \right) \left( dx^j + N^j dt \right)\right],
$$
where $N$ is the lapse function, and $N^i$ is the shift vector.
In this section, we use a dot for the derivative with respect to $t$.
The fluctuations $\alpha$ and $\theta$ are defined as
\be
N=1+\alpha, \quad
N^i = \delta^{ij} \frac{\de_j \theta}{e^{2A}}.
\label{thi2}\ee

Next, we define the adiabatic and entropic field fluctuations. To this end, we consider the geometry of the field space.
The nonzero components of the Christoffel symbol are
\be
\bar{\Gamma}^\f_{aa}=-\frac{1}{2}\de_\f \bar{Y},
\quad
\bar{\Gamma}^a_{\f a}=\bar{\Gamma}^a_{a\f}=\frac{1}{2}\frac{\de_\f \bar{Y}}{\bar{Y}}.
\label{thi5}\ee
The field space Ricci scalar and Riemann tensor are
\be
\bar{R}_\text{fs} = \frac{(\de_\f \bar{Y})^2}{2\bar{Y}^2} - \frac{\de_\f^2 \bar{Y}}{\bar{Y}}, \quad
\bar{R}_{IJKL} = \frac{\bar{R}_\text{fs}}{2} \left( G_{IK}G_{JL} - G_{IL}G_{JK} \right).
\label{thi6}\ee

The vielbein along the background field trajectory is
\be
\bar{e}_\sigma^I = \frac{1}{\bar{\sigma}'}\left( \bar{\f}', \bar{a}'\right),
\quad \bar{e}_{\sigma I} =  \frac{1}{\bar{\sigma}'}\left( \bar{\f}', \bar{Y} \bar{a}'\right),
\label{thi7}\ee
where a prime stands for $\tau$-derivative, and $\bar{\sigma}'=\sqrt{\bar{G}_{IJ}\bar{\phi}'^I\bar{\phi}'^J}=\sqrt{\bar{\f}'^2 + \bar{Y} \bar{a}'^2}$.
The basis orthogonal to $e_\sigma^I$  is
\be
\bar{e}_s^I = \frac{\sqrt{\bar{Y}}}{\bar{\sigma}'} \left( \bar{a}', -\frac{\bar{\f}'}{\bar{Y}}\right),
\quad
\bar{e}_{sI} = \frac{\sqrt{Y}}{\bar{\sigma}'} \left( \bar{a}', -\bar{\f}' \right).
\label{thi9}\ee
The adiabatic and entropic field fluctuations are
\be
\bar{e}_{\sigma I} \bar{\phi}^I =\frac{\bar{\f}' \delta \bar{\f} + \bar{Y} \bar{a}' \delta \bar{a}}{\bar{\sigma}'}, \quad
\bar{{\cal F}} \equiv \bar{e}_{s I} \bar{\phi}^I
= \sqrt{\bar{Y}}\frac{\bar{a}' \delta \bar{\f} - \bar{\f}'\delta \bar{a}}{\bar{\sigma}'}.
\label{thi11}\ee
Note that $\bar{e}_{\sigma I} \bar{\phi}^I =0$ in the comoving gauge.

The fluctuations in \eqref{thi2} are related to the other fluctuations,
\be
\alpha=\frac{\dot{\bar{\zeta}}}{\bar{H}}+...,\quad
\theta = -\frac{\bar{\zeta}}{\bar{H}} + \bar{\chi} + ...,\quad
\label{thi12}\ee
where $\bar{\chi}$ and $\bar{H}$ are defined as
\be
e^{-A}\de^2 \bar{\chi} = \bar{\epsilon} {\bar{\zeta}}' +\frac{\bar{\sigma}'\bar{\eta}_\perp}{\bar{M}_p^2}\bar{\cal F},
\quad
\bar{H}\equiv \frac{\dot{\left(e^{A}\right)}}{e^A} = e^{-A}A',
\label{thi13}\ee
The slow-roll parameters are
\be
\bar{\epsilon} \equiv -\frac{\dot{\bar{H}}}{\bar{H}^2} = 1 - \frac{A''}{A'^2}, \quad
\bar{\eta} \equiv \frac{\dot{\bar{\epsilon}}}{\bar{H}\bar{\epsilon}} = \frac{2A''^2-A'A'''}{A'^2 (A'^2-A'')}.
\label{thi14}\ee

The projection of the covariant derivative of the potential along the entropic direction is
\be
\bar{V}_{;ss} \equiv \bar{e}^I_s \bar{e}^J_s \bar{V}_{;IJ}
= \bar{e}^I_s \bar{e}^J_s \left( \de_I \de_J \bar{V} - \bar{\Gamma}^K_{IJ} \de_K \bar{V}\right)
= \frac{1}{\bar{\sigma}'^2}
\left[ \frac{1}{2} (\de_{\bar{\f}} \bar{V}) \bar{\f}'^2 \frac{\de_{\bar{\f}} \bar{Y} }{\bar{Y}}
+ (\de_{\bar{\f}}^2 \bar{V}) \bar{Y} \bar{a}'^2
\right],
\label{thi15}\ee
\be
\bar{V}_{;sss}= \bar{e}^I_s \bar{e}^J_s \bar{e}^K_s \bar{V}_{;IJK}.
\label{thi16}\ee

The bending parameter $\bar{\eta}_\perp$ is defined as
\be
{\cal D}_\tau \bar{e}_\sigma^I
= e^{A} \bar{H} \bar{\eta}_\perp \bar{e}_s^I, \quad
{\cal D}_\tau \bar{e}_s^I
=-e^{A} \bar{H} \bar{\eta}_\perp e_\sigma^I,
\label{thi17}\ee
where the action of ${\cal D}_\tau$ is ${\cal D}_\tau A^I \equiv \p_\tau A^I+ \bar{\Gamma}^I_{JK} (\p_\tau \bar{\phi}^J) A^K.$
The explicit form of $\eta_\perp$ is
\be
\bar{\eta}_\perp =
\frac{\sqrt{\bar{Y}}\bar{a}'}{A'\bar{\sigma}'^2}
\left(
\bar{\f}'' - \frac{\bar{a}''}{\bar{a}'}\bar{\f}' - \frac{\de_{\bar{\f}} \bar{Y}}{\bar{Y}}\bar{\f}'^2 -\frac{\de_{\bar{\f}} \bar{Y}}{2} \bar{a}'^2
\right)
=\sqrt{\bar{Y}} \bar{a}' \frac{e^{2A} \de_{\bar{\f}} \bar{V}}{A'\bar{\sigma}'^2},
\label{thi18}\ee

From (3.5) in \cite{Garcia-Saenz:2019njm}, the cubic couplings are
\be
\left(S^{(3)}\right)_{\text{cosmology}}
=\int dt d^3x \left({\cal L}^{(3)}\right)_{\text{cosmology}}
= \int d\tau d^3x e^A \left({\cal L}^{(3)}\right)_{\text{cosmology}},
\label{thi19-0}\ee
\be
\left({\cal L}^{(3)}\right)_{\text{cosmology}}=\bar{M}_p^2 e^{A}
\left[
\bar{\epsilon} (\bar{\epsilon} - \bar{\eta}) \bar{\zeta}'^2 \bar{\zeta}
+\bar{\epsilon} (\bar{\epsilon}+\bar{\eta}) \bar{\zeta} (\de \bar{\zeta})^2
+\frac{\frac{\bar{\epsilon}}{2}-2}{e^{2A}}(\de \bar{\zeta})(\de\bar{\chi}) \de^2\bar{\chi}
+\frac{\bar{\epsilon}}{4e^{2A}}\de^2\bar{\zeta} (\de\bar{\chi})^2
\right]+
\label{thi19}\ee
$$
+e^{3A} \bigg[
\frac{1}{2} \bar{m}_s^2 (\bar{\epsilon}+\bar{\mu}_s) \bar{\zeta} \bar{\cal F}^2
+(2\bar{\epsilon} -\bar{\eta} -2 \bar{\lambda}_\perp) \frac{\bar{\sigma}' \bar{\eta}_\perp}{e^{2A}}  \bar{\zeta} \bar{\zeta}' \bar{\cal F}
+\frac{\bar{\sigma}' \bar{\eta}_\perp}{e^{3A} \bar{H}} (\de \bar{\zeta})^2 \bar{\cal F} -
$$
$$
- \frac{\bar{\sigma}' \bar{\eta}_\perp}{e^{3A} \bar{H}} \bar{\zeta}'^2 \bar{\cal F}
- \frac{\bar{H}^2 \bar{\eta}_\perp^2 - \bar{\epsilon} \bar{M}_p^2 \bar{H}^2 \bar{R}_{\text{fs}}}{e^A \bar{H}} \bar{\zeta}' \bar{\cal F}^2
- \frac{\bar{V}_{;sss} - 2e^{-A}\bar{\sigma}' \bar{H} \bar{\eta}_\perp \bar{R}_{\text{fs}} + \bar{\epsilon} \bar{M}_p^2 \bar{H}^2 \bar{R}_{\text{fs},s}}{6} \bar{\cal F}^3 +
$$
$$
+ 2e^{-2A}\bar{\epsilon}\bar{\zeta} \left( \bar{\cal F}'^2 + (\de \bar{\cal F})^2 \right) - e^{-3A} \bar{\cal F}' \de \bar{\cal F} \de \bar{\chi} \bigg]
+{\cal D}+{\cal E}.
$$
where
\be
{\cal D}=
\frac{d}{dt} \bigg\{
-\frac{e^A}{2\bar{H}} \bar{\zeta} (\de \bar{\cal F})^2
+ \frac{e^A \bar{M}_p^2}{\bar{H}}(1-\bar{\epsilon}) \bar{\zeta} (\de \bar{\zeta})^2
-9\bar{H} \bar{M}_p^2 e^{3A} \bar{\zeta}^3
-\frac{e^{3A}}{2\bar{H}} \left( \bar{m}_s^2 +4\bar{H}^2 \bar{\eta}_\perp^2\right) \bar{\zeta} \bar{\cal F}^2
\label{thi20}\ee
$$
-\frac{\bar{M}_p^2}{4e^A \bar{H}^3} (\de \bar{\zeta})^2 \de^2 \bar{\zeta}
- \frac{\bar{\zeta} \bar{p}_\zeta^2}{4\bar{\epsilon} \bar{H} e^{3A} \bar{M}_p^2}
+\frac{\dot{\bar{\sigma}} \bar{\eta}_\perp}{\bar{\epsilon} \bar{H} \bar{M}_p^2} \bar{\cal F} \bar{\zeta} \bar{p}_\zeta
- \frac{\bar{\zeta}}{8e^{3A} \bar{H} \bar{M}_p^2} \left( \de^{-2}\bar{p}_{\zeta,ij} \de^{-2} \bar{p}_{\zeta,ij} - \bar{p}_\zeta^2 \right) +
$$
$$
+\frac{\bar{\zeta}}{4e^A \bar{H}^2}\left( \bar{\zeta}_{,ij}\de^{-2}\bar{p}_{\zeta,ij} - \de^2 \bar{\zeta} \bar{p}_\zeta \right)
-\frac{1}{2\bar{H}e^{3A}}\bar{\zeta} \bar{p}_{\cal F}^2
\bigg\},
$$
and
\be
{\cal E}=
\frac{e^{3A}}{\bar{H}}{\cal E}_\zeta
\left[
\dot{\bar{\zeta}}\bar{\zeta}
- \frac{1}{4e^{2A}\bar{H}}
\bigg\{
(\de \bar{\zeta})^2
- \frac{\de_i\de_j}{\de^2}(\de_i\bar{\zeta}\de_j\bar{\zeta})
-2\bar{H} \bigg(\de\bar{\zeta}\de\bar{\chi} - \frac{\de_i\de_j}{\de^2}(\de_i\bar{\zeta}\de_j\bar{\chi})\bigg)
\bigg\}
\right]
+ \frac{e^{3A}}{\bar{H}}  {\cal E}_{\cal F} \bar{\zeta} \dot{\bar{\cal F}}.
\label{thi21}\ee
Here we have defined
\be
\bar{m}_s^2= \bar{V}_{;ss} -\bar{H}^2\bar{\eta}_\perp^2 + \bar{\epsilon} \bar{H}^2 \bar{M}_p^2 \bar{R}_{\text{fs}},
\label{thi22}\ee
\be
\bar{\lambda}_\perp \equiv
\frac{\dot{\bar{\eta}}_\perp}{\bar{H} \bar{\eta}_\perp}
=\frac{\bar{\eta}_\perp'}{A'\bar{\eta}_\perp},
\quad
\bar{\mu}_s\equiv \frac{\dot{\bar{m}}_s^2}{\bar{H} \bar{m}_s^2}
=\frac{(\bar{m}_s^2)'}{A' \bar{m}_s^2},
\label{thi23}\ee
\be
\bar{p}_\zeta \equiv 2 e^{2A} \bar{M}_p^2 \left( \bar{\epsilon} \bar{\zeta}' + \frac{\bar{\sigma}' \bar{\eta}_\perp}{\bar{M}_p^2} \bar{\cal F} \right),
\quad
\bar{p}_{\cal F} \equiv e^{3A} \dot{\bar{{\cal F}}}
= e^{2A} \bar{{\cal F}}',
\label{thi24}\ee
\be
{\cal E}_\zeta \equiv 2\bar{M}_p^3 \left[ e^{-4A} \left(e^{2A} \bar{\epsilon} \bar{\zeta}' \right)' - e^{-2A}\bar{\epsilon} \,\de^2 \bar{\zeta} \right]
+ 2 e^{-4A} \left(e^{2A} \bar{\sigma}' \bar{\eta}_\perp \bar{\cal F}\right)',
\label{thi25}\ee
\be
{\cal E}_{\cal F}\equiv
e^{-4A} \left(e^{2A} \bar{\cal F}' \right)'
- e^{-2A}\de^2 \bar{\cal F}
+\bar{m}_s^2 \bar{\cal F}
-2 e^{-2A} \bar{\sigma}' \bar{\eta}_\perp \bar{\zeta}',
\label{thi26}\ee
\be
\bar{R}_{\text{fs},s}\equiv
e_s^I \de_I \bar{R}_{\text{fs}},
\label{thi26-2}\ee
\be
\de^2\equiv \delta^{ij}\de_i \de_j,
\quad
(\de ...)^2 \equiv \delta^{ij} (\de_i ...) (\de_j ...).
\label{thi26-3}\ee
Notice that ${\cal D}$ is the total derivative term, and ${\cal E}$ vanishes after imposing the equation of motion (The equations of motion is ${\cal E}_\zeta={\cal E}_{\cal F}=0$), and therefore we can omit these terms.

\subsection{Analytic continuation to our spacetime}
The cubic coupling for our spacetime \eqref{pt2-2} is obtained by the replacement \cite{McFadden:2009fg,McFadden:2010vh}:
\be
\bar{M}_p^2 \to -2M_p^2,
\quad
\bar{G}_{IJ}\to -M_p^2 G_{IJ},
\quad
\bar{V} \to  M_p^2 V,
\quad
\bar{Y}\to Y,
\label{thi30}\ee
$$
\bar{\Gamma}^I_{JK} \to \Gamma^I_{JK},
\quad
\de^2 \to \eta^{\mu\nu} \de_\mu \de_\nu,
\quad
(\de ...)^2 \to \eta^{\mu\nu} (\de_\mu ...) (\de_\nu ...),
$$
$$
\bar{R}_{\text{fs}} \to - \frac{R_{\text{fs}}}{M_p^2},
\quad
\bar{V}_{;ss} \to - V_{;ss},
\quad
\bar{H} \to H,
\quad
\bar{\zeta}\to \zeta,
\quad
\bar{\epsilon} \to \epsilon,
\quad
\bar{\eta} \to \eta,
\quad
\bar{\chi} \to \chi,
$$
$$
\bar{m}_s^2 \to -m_s^2,
\quad
\bar{\mu}_s \to \mu_s,
\quad
\bar{\eta}_\perp \to \eta_\perp,
\quad
\bar{\lambda}_\perp \to \lambda_\perp.
$$
$$
\bar{\cal F}^2 \to -M_p^2 {\cal F}^2,
\quad
\bar{\sigma}' \bar{F} \to - M_p^2 \sigma' {\cal F},
\quad
\bar{V}_{;sss} \bar{\cal F}^3 \to M_p^2 V_{;sss} {\cal F}^3,
\quad
\bar{R}_{\text{fs},s} \bar{\cal F}^3 \to R_{\text{fs},s} {\cal F}^3,
$$
and a prime stands for $r$-derivative instead of $\tau$-derivative.
Here $m_s^2$ is \eqref{re7}, $\zeta, {\cal F}$ are
\be
\zeta \equiv \frac{\sigma'}{A'}\psi - \frac{\f' \delta\f+Y a' \delta a}{\sigma'}
=e^{-A}\psi_\zeta,
\quad
{\cal F} \equiv \sqrt{Y}\frac{a' \delta \f - \f'\delta a}{\sigma'}
=-e^{-A} \psi_{\cal S},
\label{a-re1}\ee
and $\chi$ satisfies
\be
e^{-A} \de^2 \chi = \epsilon \zeta' + \frac{\sigma' \eta_\perp}{2} {\cal F}.
\label{thi30-2}\ee

After the replacement, we obtain the action at the cubic order in the fluctuations $S^{(3)}$ as
\be
S^{(3)}
= \int^{r_{IR}}_{r_{UV}} dr \int d^3x \,e^A {\cal L}^{(3)}
= -\int^{\infty}_{-\infty} dA \int d^3x \, \frac{e^A}{A'} {\cal L}^{(3)},
\label{a-thi19-2}\ee
\be
\frac{{\cal L}^{(3)}}{M_p^2}
=-2 e^{A}
\left[
\epsilon (\epsilon - \eta) \zeta'^2 \zeta
+\epsilon (\epsilon + \eta) \zeta (\de \zeta)^2
+\frac{\frac{\epsilon}{2}-2}{e^{2A}}(\de \zeta)(\de \chi) \de^2 \chi
+\frac{\epsilon}{4e^{2A}}\de^2 \zeta (\de \chi)^2
\right]+
\label{a-thi19-3}\ee
$$
+e^{3A} \bigg[
\frac{1}{2} m_s^2 (\epsilon + \mu_s) \zeta {\cal F}^2
-(2\epsilon -\eta -2 \lambda_\perp) \frac{\sigma' \eta_\perp}{e^{2A}}  \zeta \zeta' {\cal F}
+ \frac{\sigma' \eta_\perp}{e^{3A} H}  {\cal F} \left(\zeta'^2 - (\de \zeta)^2 \right) +
$$
$$
+ \frac{H^2 \eta_\perp^2 - 2 \epsilon H^2 R_{\text{fs}}}{e^A H} \zeta' {\cal F}^2
- \frac{V_{;sss} + 2e^{-A} \sigma' H \eta_\perp R_{\text{fs}} -2 \epsilon H^2 R_{\text{fs},s}}{6} {\cal F}^3 -
$$
$$
- 2 e^{-2A}\epsilon \zeta \left( {\cal F}'^2 + (\de {\cal F})^2 \right)
+ e^{-3A} {\cal F}' \de {\cal F} \de \chi \bigg]
$$
where we have omitted the terms corresponding to ${\cal D}$ and ${\cal E}$.

From \eqref{a-re1}, \eqref{re1}, \eqref{tr11}, \eqref{a-tr19},
\be
\begin{pmatrix}
\zeta \\
{\cal F}
\end{pmatrix}
= e^{-A}
\begin{pmatrix}
1 & 0 \\
0 & -1
\end{pmatrix}
\begin{pmatrix}
\psi_\zeta \\
\psi_{\cal S}
\end{pmatrix}
=\frac{e^{-\frac{3}{2}A} }{\sqrt{W}}
\begin{pmatrix}
1 & 0 \\
0 & -1
\end{pmatrix}
\begin{pmatrix}
\tilde{\psi}_{\zeta A} \\
\tilde{\psi}_{{\cal S}A}
\end{pmatrix}
=\frac{e^{-\frac{3}{2}A} }{\sqrt{W}}
\begin{pmatrix}
1 & 0 \\
0 & -1
\end{pmatrix}
\begin{pmatrix}
\cos\Phi & \sin\Phi \\
-\sin\Phi & \cos\Phi
\end{pmatrix}
\begin{pmatrix}
\psi_{\zeta A} \\
\psi_{{\cal S}A}
\end{pmatrix}.
\label{a-cu1}\ee

Depending on the structure of the interaction, the cubic terms are
\be
{\cal L}^{(3)}=
{\cal L}_{\Psi^3}^{(3)}
+ {\cal L}_{\Psi \left(\de\Psi\right)^2}^{(3)}
+ {\cal L}_{\Psi (\de \Psi)(\de \chi)}^{(3)}
+ {\cal L}_{\left(\de^2\Psi\right)\left(\de \chi\right)^2}^{(3)},
\label{thi29-2}\ee
where
\be
\frac{{\cal L}_{\Psi^3}^{(3)}}{M_p^2} =
-2 e^{A} \epsilon (\epsilon - \eta) \zeta'^2 \zeta
+ \frac{e^{3A}}{2} m_s^2 (\epsilon + \mu_s) \zeta {\cal F}^2
-e^{A}(2\epsilon -\eta -2 \lambda_\perp) \sigma' \eta_\perp  \zeta \zeta' {\cal F}
+ \frac{\sigma' \eta_\perp}{H}  {\cal F} \zeta'^2 +
\label{thi29-3}\ee
$$
+ e^{2A} \frac{H^2 \eta_\perp^2 - 2 \epsilon H^2 R_{\text{fs}}}{H} \zeta' {\cal F}^2
- \frac{V_{;sss} + 2e^{-A} \sigma' H \eta_\perp R_{\text{fs}} -2 \epsilon H^2 R_{\text{fs},s}}{6} e^{3A} {\cal F}^3
- 2 e^{A}\epsilon \zeta {\cal F}'^2
$$
$$
\equiv -\frac{A'}{e^A} C_{\Psi^3, ijk} \psi_i^{(3)}\psi_j^{(3)}\psi_k^{(3)},
$$
\be
\frac{{\cal L}_{\Psi \left(\de\Psi\right)^2}^{(3)}}{M_p^2}=
-2 e^{A} \epsilon (\epsilon + \eta) \zeta (\de \zeta)^2
- \frac{\sigma' \eta_\perp}{H}  {\cal F} (\de \zeta)^2
- 2 e^A \epsilon \zeta (\de {\cal F})^2
\equiv -\frac{A'}{e^A} C_{\Psi \left(\de\Psi\right)^2, ijk} \psi_i^{(3)}\left(\de\psi_j^{(3)}\right)\left(\de\psi_k^{(3)}\right),
\label{thi29-4}\ee
\be
\frac{{\cal L}_{\Psi (\de \Psi)(\de \chi)}^{(3)}}{M_p^2}=
-2\frac{\frac{\epsilon}{2}-2}{e^{A}}(\de \zeta)(\de \chi) \de^2 \chi
+  {\cal F}' \de {\cal F} \de \chi
\equiv -\frac{A'}{e^A} C_{\Psi (\de \Psi)(\de \chi),ijk} \psi_i^{(3)}\left(\de\psi_j^{(3)}\right)\left(\frac{\de}{\de^2}\psi_k^{(3)}\right),
\label{thi29-5}\ee
\be
\frac{{\cal L}_{\left(\de^2\Psi\right)\left(\de \chi\right)^2}^{(3)}}{M_p^2}=
-\frac{\epsilon}{2e^{A}}\de^2 \zeta (\de \chi)^2
\equiv -\frac{A'}{e^A}  C_{\left(\de^2\Psi\right)\left(\de \chi\right)^2,ijk} \left(\de^2\psi_i^{(3)}\right)\left(\frac{\de}{\de^2}\psi_j^{(3)}\right)\left(\frac{\de}{\de^2}\psi_k^{(3)}\right),
\label{thi29-6}\ee
where $\psi^{(3)}_i$ is the $3$d part of the KK decomposition \eqref{pt32}.

The effective action describing the cubic interaction of three dimensional glueball is
\be
S=-\int d^3x\int^{\infty}_{-\infty} dA  \, \frac{e^A}{A'}
\left(
{\cal L}_{\Psi^3}^{(3)}
+ {\cal L}_{\Psi \left(\de\Psi\right)^2}^{(3)}
+ {\cal L}_{\Psi (\de \Psi)(\de \chi)}^{(3)}
+ {\cal L}_{\left(\de^2\Psi\right)\left(\de \chi\right)^2}^{(3)}
\right)
\label{a-thi29-7}\ee
$$
=\int d^3x \bigg[
D_{\Psi^3, ijk} \psi_i^{(3)}\psi_j^{(3)}\psi_k^{(3)}
+ D_{\Psi \left(\de\Psi\right)^2, ijk} \psi_i^{(3)}\left(\de\psi_j^{(3)}\right)\left(\de\psi_k^{(3)}\right)+
$$
$$
+ D_{\Psi (\de \Psi)(\de \chi),ijk} \psi_i^{(3)} \left(\de\psi_j^{(3)}\right)\left(\frac{\de}{\de^2}\psi_k^{(3)}\right)
+ D_{\left(\de^2\Psi\right)\left(\de \chi\right)^2, ijk} \left(\de^2\psi_i^{(3)}\right)\left(\frac{\de}{\de^2}\psi_j^{(3)}\right)\left(\frac{\de}{\de^2}\psi_k^{(3)}\right)
\bigg]
$$
where
\be
D_{\Psi^3, ijk} = \int^{\infty}_{-\infty} dA\, C_{\Psi^3, ijk},
\quad
D_{\Psi \left(\de\Psi\right)^2, ijk}= \int^{\infty}_{-\infty} dA\, C_{\Psi \left(\de\Psi\right)^2, \,ijk} ,
\label{a-thi29-8}\ee
\be
D_{\Psi (\de \Psi)(\de \chi),\ell mn} =
\int^{\infty}_{-\infty} dA\, C_{\Psi (\de \Psi)(\de \chi),\ell mn} ,
\quad
D_{\left(\de^2\Psi\right)\left(\de \chi\right)^2,\ell mn} =
\int^{\infty}_{-\infty} dA\, C_{\left(\de^2\Psi\right)\left(\de \chi\right)^2,\ell mn}
\label{a-thi29-9}\ee

\end{appendix}


\end{document}